\documentclass[a4paper,11pt]{article}
\usepackage[top=1in,bottom=1in,left=1in,right=1in]{geometry}
\usepackage{array}
\usepackage{caption}
\usepackage{subcaption}
\usepackage{booktabs} %
\usepackage{multirow} %
\usepackage{hyperref} %
\usepackage{xurl} %
\usepackage{xcolor}   %
\usepackage{colortbl} %
\definecolor{blueD}{RGB}{38,68,143}
\definecolor{bluetint}{RGB}{232,238,251}
\definecolor{blueband}{RGB}{244,247,253}
\definecolor{tealD}{RGB}{13,102,92}
\definecolor{tealtint}{RGB}{233,246,242}
\definecolor{tealband}{RGB}{242,250,248}
\definecolor{orgD}{RGB}{209,110,22}
\definecolor{orgtint}{RGB}{251,239,219}
\definecolor{orgband}{RGB}{253,247,237}

\usepackage{cite}     %
\usepackage[english]{babel}
\usepackage[utf8]{inputenc}
\usepackage{amsmath}
\usepackage{amssymb}
\usepackage{amsthm}
\usepackage{algorithm}
\usepackage{graphicx}
\graphicspath{ {images/} }
\usepackage{float}

\newlength{\figwidth}
\newcommand{\widefig}[1]{\noindent\makebox[\linewidth][c]{\resizebox{\figwidth}{!}{#1}}}
\newcommand{\widetab}[1]{\noindent\makebox[\linewidth][c]{\resizebox{\figwidth}{!}{#1}}}
\usepackage{tikz}
\usetikzlibrary{positioning, calc, shapes.geometric, arrows.meta, backgrounds, fit, decorations.pathreplacing, shadows.blur, shadows, matrix, patterns}
\usepackage{pgfplots}
\pgfplotsset{compat=1.16}
\usepgfplotslibrary{fillbetween}
\usepackage{algcompatible}
\usepackage{setspace}
\usepackage{lineno}
\usepackage[affil-it]{authblk}

\makeatletter
\providecommand{\Require}{\REQUIRE}
\providecommand{\Ensure}{\ENSURE}
\providecommand{\State}{\STATE}
\providecommand{\For}{\FOR}
\providecommand{\EndFor}{\ENDFOR}
\providecommand{\If}{\IF}
\providecommand{\Else}{\ELSE}
\providecommand{\ElsIf}{\ELSIF}
\providecommand{\EndIf}{\ENDIF}

\makeatother

\theoremstyle{definition}

\usepackage{titlesec}
\titleformat{\section}
{\normalfont\Large\bfseries\color{black}}{\thesection}{1em}{}

\makeatletter
\newcommand{\thickhline}{%
  \noalign {\ifnum 0=`}\fi \hrule height 1pt
  \futurelet \reserved@a \@xhline
}
\newcolumntype{"}{@{\hskip\tabcolsep\vrule width 1pt\hskip\tabcolsep}}
\makeatother

\title{UBio-MolFM: Enabling Biomolecular Dynamics at DFT Accuracy and \texorpdfstring{$10^5$}{10\textasciicircum 5} Atoms with One Untuned Potential}
\hypersetup{unicode=true,
            pdftitle={UBio-MolFM: Enabling Biomolecular Dynamics at DFT Accuracy and 10⁵ Atoms with One Untuned Potential},
            pdfauthor={IQuest Research, UBio Team}}

\author[1,$*$]{Lin Huang}
\author[1,$*$]{Frank Peng}
\author[1]{JiaJun Cheng}
\author[1]{Zion Wang}
\author[1]{Hao Yin}
\author[1]{Hao Li}
\author[1]{Ji Zhang}
\author[1]{Jack Jia}
\author[1]{Junping Zhao}
\author[1]{Arthur Jiang}
\author[1,$\dagger$]{Jia Zhang}
\affil[1]{IQuest Research, UBio Team}
\affil[$*$]{These authors contributed equally to this work.}
\affil[$\dagger$]{Correspondence: \textit{jialrs.z@iquestlab.com}}

\date{}

\begin{document}

\maketitle
\thispagestyle{empty}

\begin{abstract}
  \noindent
Ion conduction, membrane permeation and metal recognition hinge on electronic structure, yet first-principles simulation reaches only hundreds of atoms. \textbf{UBio-MolFM} lifts that ceiling: a foundation model trained on 160 million quantum-chemical labels, its receptive field spanning non-covalent distances at near-linear cost. The barrier is cost, not principle. One untuned potential keeps force error near 20 meV/\AA{} past a thousand atoms, reproduces water's X-ray structure and ion hydration, and holds an RNA Mg$^{2+}$ site without ion-specific parameters. Cyclosporine~A pays $3.5\text{ kcal/mol}$ in water for its permeable conformer, gated by one kinetically asymmetric hydrogen bond that a fixed-charge model flattens. In a 108{,}964-atom KcsA channel on one GPU, the relaxed four-ion column is anhydrous in all five replicas, in direct contact in four---the knock-on geometry ten fixed-charge simulations never form. It remains orders of magnitude costlier. Where electronic structure decides the answer, first-principles simulation is in reach.
\end{abstract}

\newpage
Many of biology's most consequential events---signal transduction, ion conduction, enzyme catalysis, ligand binding, membrane permeation---hinge on electronic effects---polarization, charge transfer, transient transition states---that fixed atomic charges and harmonic forces cannot reproduce. Resolving them from first principles has stayed out of reach: density functional theory (DFT) scales roughly cubically and stops at a few hundred atoms, while the classical force fields that reach whole proteins, nucleic acids and membranes discard exactly that detail. Closing the gap demands more than low error on static benchmarks: a potential is useful for biology only if it reproduces the \emph{emergent} thermodynamics of the condensed phase---liquid densities, ionic solvation, the balances governing permeation---under unbiased finite-temperature dynamics.

Machine-learning force fields (MLFFs) bridge this gap, regressing potential energy surfaces on quantum-chemical data. Pretrained foundation models---MACE-OMol~\cite{mace,omol25}, UMA~\cite{wood2025family}---transfer across drug-like chemistry without per-system retraining, setting strong baselines. Yet training and evaluation lean static and small-molecule (SPICE~\cite{spice} caps near 110 atoms, OMol25~\cite{omol25} near 350), and low benchmark error does not guarantee \emph{thermodynamic consistency}---the stable densities and non-aggregating solvation that emerge only in long condensed-phase trajectories, and that the pretrained baselines tested here do not reproduce (Extended Data Fig.~\ref{fig:density_baselines_si}). A different line reaches biological scale by building the training set around the target: GEMS~\cite{gems} drives nanosecond-aggregate dynamics of a $25{,}257$-atom solvated protein at PBE0/MBD quality, pairing bottom-up fragments with top-down ones cut from classical trajectories of that system---three models for three systems, with reduced thermal stability and much larger prediction errors once the system-specific half is ablated. Its authors name a chemically transferable universal potential as future work.

Among classical force fields, the corrections that restore part of that detail stay inside a fixed functional form: the Li--Merz 12-6-4 model adds a pairwise ion-induced-dipole term but leaves every charge fixed~\cite{li2014ion}; genuinely polarizable models---AMOEBA~\cite{amoeba2013} and the classical Drude oscillator~\cite{drude2013}---carry induced dipoles self-consistently at roughly fourfold the cost on GPUs~\cite{huang2018drude}, yet fix bonding topology at parameterization time, so bond making and breaking lie outside them. The ion--carbonyl and ion--ion balance that sets selectivity-filter occupancy is sensitive to exactly such choices: in K$^+$ channels mechanism assignments shift with the water model and ion parameters alone~\cite{kopfer2014knockon,bosio2026}. Quantum-mechanical fidelity at the $10^5$-atom mesoscale therefore poses three coupled challenges: training data beyond small molecules; a receptive field capturing non-local interactions (transmembrane gradients, concerted polarization~\cite{wang2026scalable}) without stacking short cutoffs, which degrade under distance scaling~\cite{qu2026allscaip}; and an equivariant backbone whose high-order tensor products stay tractable on $10^5$-atom graphs.

\begin{figure}[H]
  \centering
  \vspace*{-45pt}
  \resizebox{\linewidth}{!}{\input{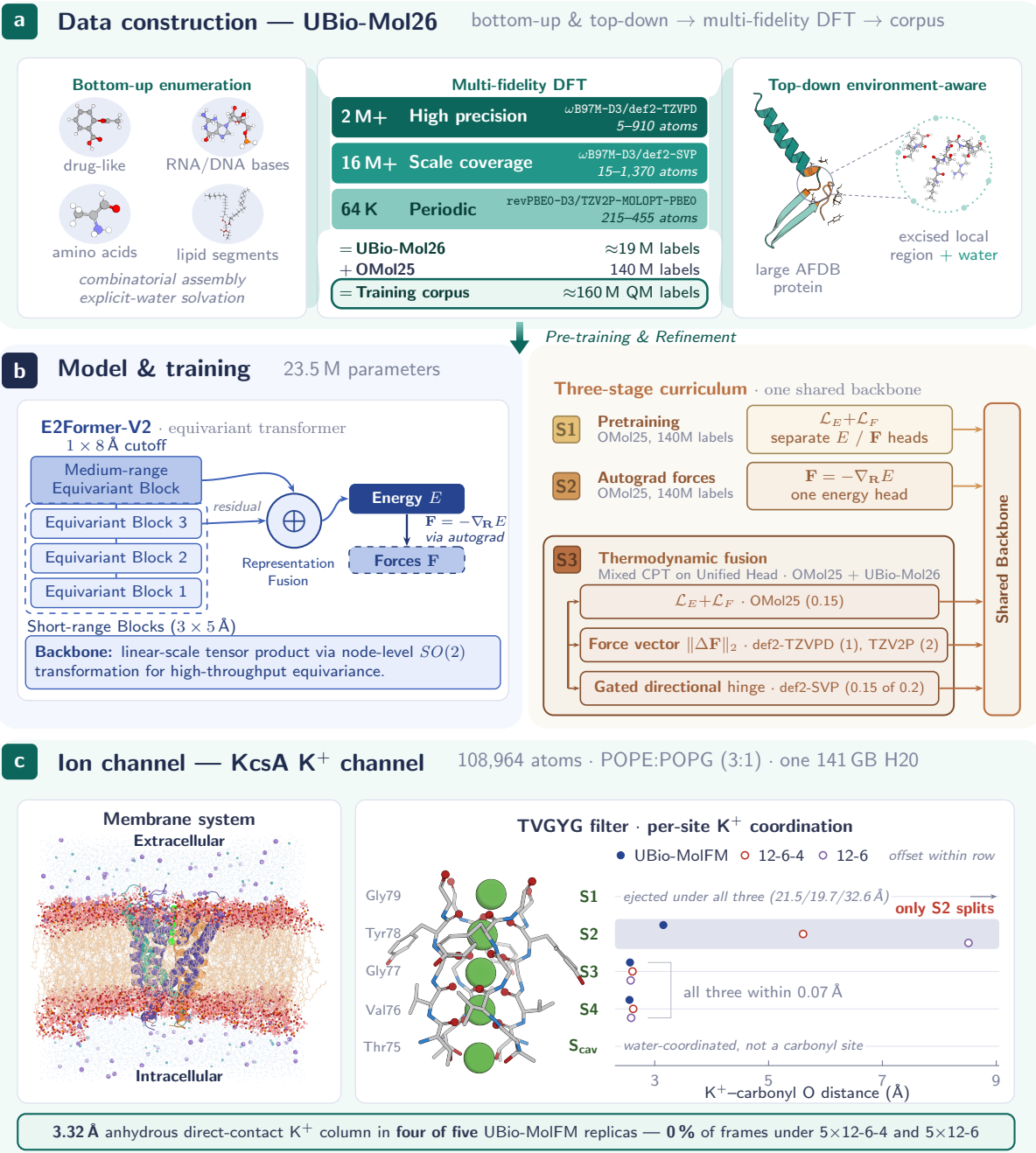}}
  \setlength{\abovecaptionskip}{2pt}\captionsetup{font={footnotesize,stretch=0.95}}
  \caption{\textbf{The UBio-MolFM framework.}
    \textbf{(a) Data.} UBio-Mol26 pairs bottom-up enumeration of the explicitly solvated building blocks shown with top-down, environment-aware sampling of local regions excised from AlphaFold proteins together with their surrounding water. Structures are labelled at the three DFT fidelities listed---the third \emph{periodic}---for $\approx$19\,M labels, $\approx$160\,M with OMol25.
    \textbf{(b) Model \& training.} E2Former-V2 is an $\mathrm{SO}(3)$-equivariant transformer; on it we configure a hierarchical hybrid cutoff---this work's choice, not part of the backbone---in which three short-range all-atom blocks (5\,\AA) feed a medium-range block (8\,\AA) with heavy-atom neighbourhoods through a residual skip, both merging at a fusion node that concatenates their channels and projects the result back, feeding the energy head; node-level $\mathit{SO}(2)$ transformations replace dense tensor products for linear-scale equivariance. Three stages over one shared backbone: S1, OMol25 pretraining with separate energy and force heads; S2, still OMol25 but with that force head retired in favour of autograd, $\mathbf{F}=-\nabla_{\mathbf{R}}E$; S3, the first mixed continued pretraining on OMol25 and UBio-Mol26 together, whose \emph{four} data branches---sampling ratio in parentheses---fall into the three objective families shown. A low-weight auxiliary \texttt{svp} energy head is omitted from the schematic; it and the \texttt{omol25} branch's tenfold oversampling above 200 atoms are in Methods.
    \textbf{(c) KcsA channel.} Unbiased $NPT$ simulation of the KcsA tetramer in the bilayer shown. From an ion-loaded start that relaxes to four ions on axis, the TVGYG filter sustains under UBio-MolFM a dehydrated, contiguous direct-contact K$^+$ column in four of five replicas and in the fifth until $700$~ps---the $3.32$~\AA{} on the strip is the five-replica mean, seed 126 included---and in none of the ten replicas of two matched fixed-charge treatments from the same equilibrated system. Per-site K$^{+}$--carbonyl distances (right) pool all five replicas per potential over each trajectory's second half; they coincide at the deep sites S3/S4 and split only at S2, the Gly77--Tyr78 cage, where our $3.15$~\AA{} is raised by seed 126 alone---the other four sit at $2.60$~\AA{} (Extended Data Table~\ref{tab:validation_summary}). S$_\text{cav}$ separates differently---the S$_\text{cav}$--S4 break of Fig.~\ref{fig:transmembrane_channel}e: its cavity ion contacts the innermost filter carbonyls under UBio-MolFM and sits a water layer below them in four of five 12-6-4 replicas; the fifth reaches contact by $1$~ns, making the difference kinetic rather than structural (Methods).}\label{fig:framework}
\end{figure}

Here we present UBio-MolFM, a thermodynamically consistent foundation model: one untuned potential that carries DFT accuracy to proteins, nucleic acids and membranes and drives unbiased dynamics past $10^5$ atoms on one GPU (Fig.~\ref{fig:framework}). Three components make it possible (Methods). \textbf{Data}---UBio-Mol26, a bio-specific quantum-chemical corpus built for this work---$\approx$19~million labels at three DFT fidelities, $\approx$160~million with OMol25~\cite{omol25}---whose periodic condensed-phase tier calibrates the pressure and density response thermodynamic consistency demands (Extended Data Fig.~\ref{fig:density_baselines_si}). \textbf{Model}---an unmodified E2Former-V2 backbone~\cite{huang2026e2former,li2025e2former} under a hierarchical hybrid cutoff reaching non-covalent distances, the range this architecture family needs at this fragment size~\cite{wang2026scalable}; we confine that reach to heavy-atom neighbourhoods, curbing H--H pair growth without losing accuracy. \textbf{Training}---a three-stage curriculum over one shared backbone, forces from autograd on a predicted energy ($\mathbf{F}=-\nabla_{\mathbf{R}}E$) for the self-consistency stable dynamics needs; quantum-mechanical references let it encode the many-body polarization fixed-charge models only approximate. Together these remove three long-standing costs: per-system parameterization, a per-system quantum-chemical dataset, and the memory barrier keeping $10^5$-atom quantum-derived trajectories off one GPU. What remains is one set of weights.

What that potential reports about condensed-phase biology is the substance of this work; three of the four systems below have no precomputed answer. Liquid water and physiological electrolytes settle at their measured density and scattering structure under unbiased dynamics. In the Mg$^{2+}$-dependent BWYV RNA pseudoknot the metal keeps a perfect octahedral inner-sphere shell with no ion- or RNA-specific parameter, matching the Li--Merz 12-6-4 model built for this problem and bettering it wherever an external reference can adjudicate. In explicit water, Cyclosporine~A's membrane-permeable conformer carries a $3.5\,\text{kcal/mol}$ toll gated by one kinetically asymmetric hydrogen bond that a fixed-charge model, on a matched protocol, flattens away entirely. And in a 108{,}964-atom KcsA channel on one GPU, the relaxed four-ion column is anhydrous in every quantum-derived replica and in direct contact---the knock-on geometry proposed for permeation---in four of five, against no frame of ten fixed-charge ones, a divergence tracing to one backbone carbonyl. Across all four tiers the weights are identical---the signature of transferability rather than per-system fit.

\section{Results}
\label{sec:results}

A potential fitted to quantum-chemical forces and energies is told nothing about a liquid's density, a hydration shell's geometry, or how an ion channel holds its ions. Whether that knowledge emerges, and survives to $10^5$ atoms, is what the tiers below test.

\subsection{DFT-Level Force Accuracy Holds Past a Thousand Atoms}

\paragraph{A size-resolved benchmark.}
We measure accuracy against MACE-OMol~\cite{mace,omol25}, UMA-S-1p2~\cite{wood2025family} and DPA-4~\cite{li2026dpa4} on three tiers of increasing size---the baselines' own training distribution (\emph{OMol-Bio-10k}), held-out biomolecular fragments (\emph{TZVPD}) and an extreme extrapolation past every model's training-size cap (\emph{TZVP})---comparing UBio-MolFM after OMol25 pretraining (S2) and after biomolecular fine-tuning (S3). Tiers, sizes, reference fidelities and metric definitions are in Table~\ref{tab:rel_energy_force} and \S\ref{sec:eval_benchmarks}; absolute energies are comparable only on the first tier (Methods), so the larger two rest on forces and two offset-free measures.

\begin{table}[htbp]
  \centering
  \caption{\textbf{Microscopic accuracy across the three benchmark tiers.} Three pretrained baselines (MACE-OMol, UMA-S-1p2, DPA-4 with its OMol head) and UBio-MolFM trained on OMol25 only (S2) or OMol25\,$+$\,UBio-Mol26 (S3), evaluated against quantum-chemical references on OMol-Bio-10k (OMol25 validation, $\omega$B97M-V/\texttt{def2-TZVPD}), the UBio-Mol26 TZVPD held-out set ($\omega$B97M-D3/\texttt{def2-TZVPD}, mixed basis) and the extreme-size TZVP relaxation/dynamics trajectories; atom ranges are given under each tier. \textbf{E}$^{\dagger}$, per-atom energy MAE (meV/atom), reported for OMol-Bio-10k only; \textbf{F}, force MAE (meV/\AA). TZVPD lists force alone, because a constant cross-functional offset between reference levels of theory makes its absolute energies incomparable. For the TZVP trajectories, \textbf{relE} is the relative-energy MAE (referenced to each trajectory's first frame) and \textbf{$\Delta$E} that of the frame-to-frame change ($E_i-E_{i-1}$, predicted and reference differenced independently), both meV/atom. MACE-OMol uses float64, all others float32. Lower is better; per column, best in \textbf{bold}, second-best \underline{underlined}. S3 cuts the best baseline's force error by $47$, $47$, $21$ and $64\%$ on TZVPD protein, DNA, RNA and lipid, and by $32$--$51\%$ across the TZVP classes.}\label{tab:rel_energy_force}
  \scriptsize\setlength{\tabcolsep}{2.5pt}\renewcommand{\arraystretch}{1.25}
  \widetab{%
  \begin{tabular}{l !{\color{black!12}\vrule} cc !{\color{black!12}\vrule} cccc !{\color{black!12}\vrule} *{15}{c}}
    \toprule
    & \multicolumn{2}{c!{\color{black!12}\vrule}}{\cellcolor{bluetint}\textcolor{blueD}{\begin{tabular}[t]{@{}c@{}}\textbf{OMol-Bio-10k}\\[-2pt]{\tiny 309--350 atoms}\end{tabular}}}
    & \multicolumn{4}{c!{\color{black!12}\vrule}}{\cellcolor{tealtint}\textcolor{tealD}{\begin{tabular}[t]{@{}c@{}}\textbf{UBio-Mol26 TZVPD}\\[-2pt]{\tiny 390--909 atoms}\end{tabular}}}
    & \multicolumn{15}{c}{\cellcolor{orgtint}\textcolor{orgD}{\begin{tabular}[t]{@{}c@{}}\textbf{TZVP (extreme)}\\[-2pt]{\tiny 1{,}215--1{,}555 atoms}\end{tabular}}} \\
    \textbf{Model}
      & \multicolumn{2}{c!{\color{black!12}\vrule}}{\cellcolor{blueband}}
      & \multicolumn{1}{c}{\cellcolor{tealband}Protein} & \multicolumn{1}{c}{\cellcolor{tealband}DNA}
      & \multicolumn{1}{c}{\cellcolor{tealband}RNA} & \multicolumn{1}{c!{\color{black!12}\vrule}}{\cellcolor{tealband}Lipid}
      & \multicolumn{3}{c}{\cellcolor{orgband}Prot.\ opt} & \multicolumn{3}{c}{\cellcolor{orgband}DNA opt}
      & \multicolumn{3}{c}{\cellcolor{orgband}RNA opt} & \multicolumn{3}{c}{\cellcolor{orgband}Lipid opt} & \multicolumn{3}{c}{\cellcolor{orgband}Prot.\ MD} \\
    \cmidrule(lr){2-3}\cmidrule(lr){4-4}\cmidrule(lr){5-5}\cmidrule(lr){6-6}\cmidrule(lr){7-7}\cmidrule(lr){8-10}\cmidrule(lr){11-13}\cmidrule(lr){14-16}\cmidrule(lr){17-19}\cmidrule(lr){20-22}
      & E$^{\dagger}$ & F & F & F & F & F
      & relE & F & $\Delta$E & relE & F & $\Delta$E & relE & F & $\Delta$E & relE & F & $\Delta$E & relE & F & $\Delta$E \\
    \midrule
    \multicolumn{22}{@{}l}{\textit{Pretrained baselines}}\\
    MACE-OMol    & \underline{1.66} & \underline{6.88} & 43.8 & 40.5 & 66.5 & 33.1 & 0.77 & 39.3 & 0.016 & 2.31 & 37.4 & 0.098 & 4.74 & 34.7 & 0.077 & 6.30 & 32.7 & 0.094 & 0.46 & 44.7 & 0.295 \\
    UMA-S-1p2    & \textbf{0.15} & \textbf{4.12} & \underline{40.8} & \underline{38.9} & \underline{59.6} & \underline{33.0} & 0.77 & \underline{38.2} & \underline{0.015} & \underline{1.33} & \underline{35.4} & \underline{0.073} & \underline{3.28} & \underline{33.5} & 0.062 & 5.80 & 32.6 & 0.087 & \underline{0.37} & 44.5 & \textbf{0.241} \\
    DPA-4 (OMol) & 16.01 & 32.57 & 44.2 & 41.4 & 68.7 & 33.3 & \underline{0.68} & 39.5 & 0.017 & \textbf{0.98} & 51.2 & 0.100 & \textbf{0.70} & 37.6 & \underline{0.058} & \textbf{5.27} & \underline{31.0} & \textbf{0.078} & 0.58 & \underline{41.6} & 0.320 \\
    \addlinespace
    \multicolumn{22}{@{}l}{\textit{UBio-MolFM (ours)}}\\
    UBio-MolFM (S2)          & 6.00 & 7.93 & 44.4 & 41.7 & 65.8 & 34.0 & 0.81 & 42.3 & 0.019 & 1.91 & 43.1 & 0.085 & 4.66 & 37.4 & 0.076 & 6.11 & 33.4 & 0.090 & 0.41 & 46.2 & 0.286 \\
    \textbf{UBio-MolFM (S3)} & 2.12 & 7.69 & \textbf{21.8} & \textbf{20.7} & \textbf{47.3} & \textbf{11.9} & \textbf{0.12} & \textbf{18.8} & \textbf{0.011} & 1.41 & \textbf{23.6} & \textbf{0.056} & 3.35 & \textbf{18.9} & \textbf{0.051} & \underline{5.49} & \textbf{21.2} & \underline{0.085} & \textbf{0.32} & \textbf{20.2} & \underline{0.269} \\
    \bottomrule
  \end{tabular}}%
\end{table}

\paragraph{Best in class from 400 atoms up.}
The ranking inverts once molecules reach biomolecular size, and holds there. On force, Stage~3 leads every pretrained baseline on every class of the held-out TZVPD fragments, and on all five categories of the extreme TZVP trajectories again (per-class margins in Table~\ref{tab:rel_energy_force}). Force accuracy, not absolute energy, is decisive here---it governs long-trajectory stability---and the margin follows from Stage 3's force-only supervision. The energetics sharpen with it: Stage~3 takes the best or second-best frame-to-frame energy tracking ($\Delta E$) on all five trajectories and the lowest relative energy on proteins (DPA-4 leads on the data-sparse nucleic-acid and lipid classes). Biomolecular \emph{energy} supervision is marginal by design (Methods), so these gains are inherited from the forces rather than fitted. 

\paragraph{The specialist still wins on small molecules.}
On OMol-Bio-10k, the baselines' home distribution, UMA-S-1p2 remains strongest, clearly on absolute energy and more narrowly on force. We do not contest that tier: UBio-MolFM stays well inside chemical accuracy per atom there, so reaching the biological mesoscale costs little on drug-like chemistry, and what it buys begins where those training distributions end.

\subsection{Bulk Water and Physiological Electrolytes Settle at Experimental Density and Structure, but Diffuse Too Slowly}
\label{subsec:hydration_solvation}

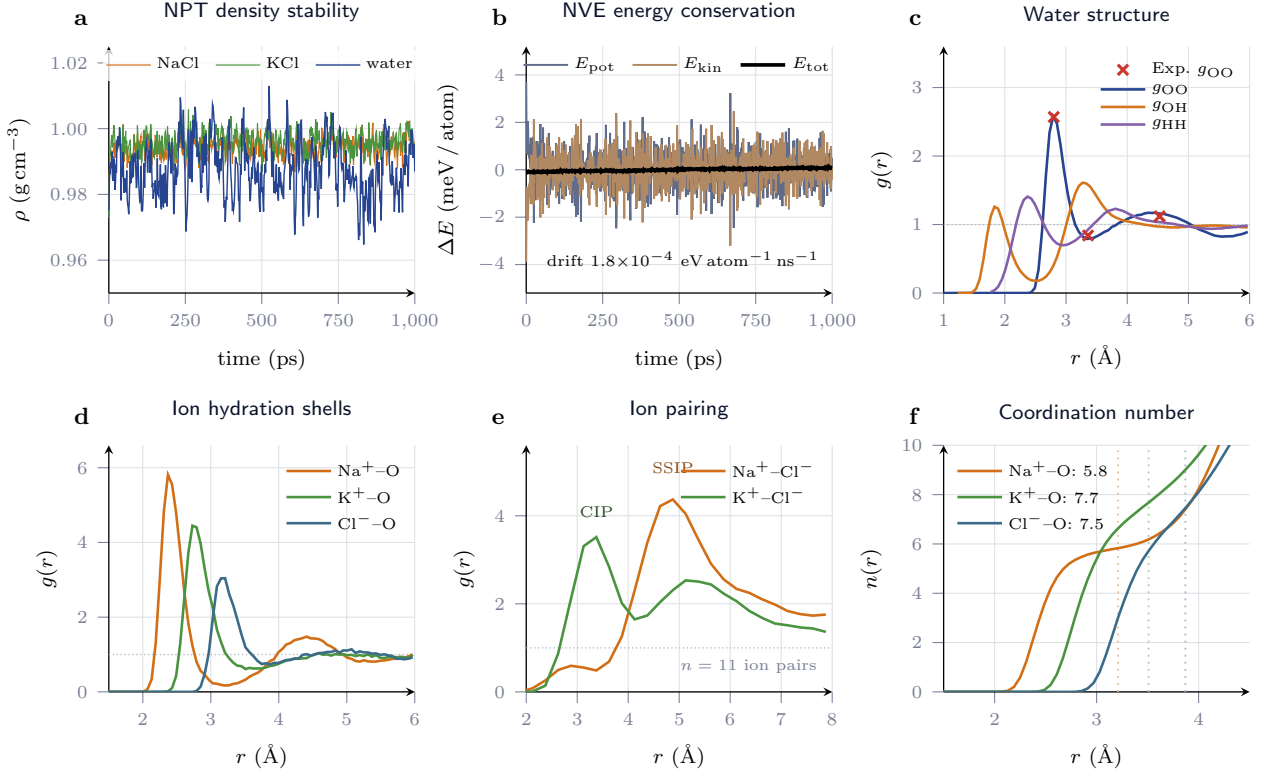
\begin{figure}[htbp]
  \centering
  \widefig{\definecolor{oursC}{RGB}{38,68,143}%
\definecolor{amberC}{RGB}{209,110,22}%
\definecolor{grnC}{RGB}{74,148,64}%
\definecolor{steelC}{RGB}{58,107,136}%
\definecolor{tealC}{RGB}{35,130,130}%
\definecolor{epotC}{RGB}{96,108,138}%
\definecolor{ekinC}{RGB}{176,136,98}%
\definecolor{expC}{RGB}{205,52,44}%
\definecolor{gohC}{RGB}{214,118,28}%
\definecolor{ghhC}{RGB}{132,96,168}%
\definecolor{axg}{RGB}{120,126,148}%
\tikzset{panellabel/.style={font=\fontsize{9}{10}\selectfont\bfseries, text=black, anchor=south west}}%
\begin{tikzpicture}[font=\sffamily]
\pgfplotsset{
  every axis/.style={
    width=5.7cm, height=4.9cm, axis lines=left,
    line width=0.5pt, tick align=outside, tick style={axg, thin},
    label style={font=\fontsize{8.1}{9.5}\selectfont},
    tick label style={font=\fontsize{7.2}{8.5}\selectfont, text=axg},
    title style={font=\fontsize{8.1}{9.5}\selectfont, text=oursC!30!black},
    every axis x label/.style={at={(ticklabel cs:0.5)}, anchor=north, font=\fontsize{8.1}{9.5}\selectfont},
    grid=major, major grid style={axg!25, line width=0.3pt},
    clip=true,
  },
  s/.style={line width=1.0pt, mark=none},
  t/.style={line width=0.75pt, mark=none},
  legstyle/.style={font=\fontsize{6.3}{7.1}\selectfont, draw=none, fill=none,
    row sep=-1.5pt, inner sep=1pt, legend cell align=left},
}

\begin{axis}[
  name=DEN, title={\textsf{NPT density stability}},
  xlabel={time (ps)}, ylabel={$\rho$ (g\,cm$^{-3}$)},
  xmin=0, xmax=1000, ymin=0.95, ymax=1.025, xtick={0,250,500,750,1000}, ytick={0.96,0.98,1.00,1.02},
  yticklabel style={/pgf/number format/fixed, /pgf/number format/precision=2, /pgf/number format/fixed zerofill},
  legend style={legstyle, at={(0.5,0.99)}, anchor=north, legend columns=3, fill=white, fill opacity=0.78, text opacity=1, /tikz/every even column/.append style={column sep=4pt}},
]
\addplot[amberC, line width=0.45pt] table[col sep=comma, x=t_ps, y=rho]{data_thermo/density_nacl.csv}; \addlegendentry{NaCl}
\addplot[grnC, line width=0.45pt] table[col sep=comma, x=t_ps, y=rho]{data_thermo/density_kcl.csv}; \addlegendentry{KCl}
\addplot[oursC, line width=0.55pt] table[col sep=comma, x=t_ps, y=rho]{data_thermo/density_water_ours.csv}; \addlegendentry{water}
\end{axis}
\node[panellabel] at ($(DEN.north west)+(-0.6cm,0.15cm)$) {a};

\begin{axis}[
  name=NVE, at={($(DEN.east)+(1.5cm,0)$)}, anchor=west,
  title={\textsf{NVE energy conservation}},
  xlabel={time (ps)}, ylabel={$\Delta E$ (meV\,/\,atom)},
  xmin=0, xmax=1000, ymin=-5.2, ymax=5.2, xtick={0,250,500,750,1000}, ytick={-4,-2,0,2,4},
  legend style={legstyle, at={(0.5,0.99)}, anchor=north, legend columns=3, /tikz/every even column/.append style={column sep=4pt}},
]
\addplot[epotC, t] table[col sep=comma, x=t_ps, y=dEpot_meV]{data_thermo/nve_plot.csv}; \addlegendentry{$E_{\text{pot}}$}
\addplot[ekinC, t] table[col sep=comma, x=t_ps, y=dEkin_meV]{data_thermo/nve_plot.csv}; \addlegendentry{$E_{\text{kin}}$}
\addplot[black, line width=1.4pt] table[col sep=comma, x=t_ps, y=dEtot_meV]{data_thermo/nve_plot.csv}; \addlegendentry{$E_{\text{tot}}$}
\node[font=\fontsize{6.3}{7.3}\selectfont, text=black, anchor=south east, align=right] at (axis cs:990,-4.6)
  {drift $1.8{\times}10^{-4}$ eV\,atom$^{-1}$\,ns$^{-1}$};
\end{axis}
\node[panellabel] at ($(NVE.north west)+(-0.6cm,0.15cm)$) {b};

\begin{axis}[
  name=WAT, at={($(NVE.east)+(1.5cm,0)$)}, anchor=west,
  title={\textsf{Water structure}},
  xlabel={$r$ (\AA)}, ylabel={$g(r)$},
  xmin=1, xmax=6, ymin=0, ymax=3.6, xtick={1,2,3,4,5,6}, ytick={0,1,2,3},
  legend style={legstyle, at={(0.98,0.97)}, anchor=north east},
]
\draw[axg!55, densely dotted, line width=0.5pt] (axis cs:1,1)--(axis cs:6,1);
\addplot[only marks, mark=x, mark size=2.8pt, expC, line width=1.2pt] coordinates {(2.80,2.57)(3.36,0.84)(4.53,1.12)}; \addlegendentry{Exp.\ $g_{\text{OO}}$}
\addplot[oursC, s] table[col sep=comma, x=r, y=gOO]{data_thermo/rdf_water.csv}; \addlegendentry{$g_{\text{OO}}$}
\addplot[gohC, s, restrict x to domain=1.2:6, unbounded coords=jump] table[col sep=comma, x=r, y=gOH]{data_thermo/rdf_water.csv}; \addlegendentry{$g_{\text{OH}}$}
\addplot[ghhC, s, restrict x to domain=1.75:6, unbounded coords=jump] table[col sep=comma, x=r, y=gHH]{data_thermo/rdf_water.csv}; \addlegendentry{$g_{\text{HH}}$}
\end{axis}
\node[panellabel] at ($(WAT.north west)+(-0.6cm,0.15cm)$) {c};

\begin{axis}[
  name=ION, at={($(DEN.south west)+(0,-2.05cm)$)}, anchor=north west,
  title={\textsf{Ion hydration shells}},
  xlabel={$r$ (\AA)}, ylabel={$g(r)$},
  xmin=1.5, xmax=6, ymin=0, ymax=6.6, xtick={2,3,4,5,6}, ytick={0,2,4,6},
  legend style={legstyle, at={(0.98,0.97)}, anchor=north east},
]
\draw[axg!55, densely dotted, line width=0.5pt] (axis cs:1.5,1)--(axis cs:6,1);
\addplot[amberC, s] table[col sep=comma, x=r, y=gNaO]{data_thermo/rdf_nacl.csv}; \addlegendentry{Na$^+$--O}
\addplot[grnC, s] table[col sep=comma, x=r, y=gKO]{data_thermo/rdf_kcl.csv}; \addlegendentry{K$^+$--O}
\addplot[steelC, s] table[col sep=comma, x=r, y=gClO]{data_thermo/rdf_nacl.csv}; \addlegendentry{Cl$^-$--O}
\end{axis}
\node[panellabel] at ($(ION.north west)+(-0.6cm,0.15cm)$) {d};

\begin{axis}[
  name=PAIR, at={($(NVE.south west)+(0,-2.05cm)$)}, anchor=north west,
  title={\textsf{Ion pairing}},
  xlabel={$r$ (\AA)}, ylabel={$g(r)$},
  xmin=2, xmax=8, ymin=0, ymax=5.6, xtick={2,3,4,5,6,7,8}, ytick={0,2,4},
  legend style={legstyle, at={(0.97,0.97)}, anchor=north east},
]
\draw[axg!55, densely dotted, line width=0.5pt] (axis cs:2,1)--(axis cs:8,1);
\addplot[amberC, s] table[col sep=comma, x=r, y=gNaCl_s]{data_thermo/rdf_pair_nacl.csv}; \addlegendentry{Na$^+$--Cl$^-$}
\addplot[grnC, s] table[col sep=comma, x=r, y=gKCl_s]{data_thermo/rdf_pair_kcl.csv}; \addlegendentry{K$^+$--Cl$^-$}
\node[font=\fontsize{6}{7}\selectfont, text=grnC!50!black, anchor=south] at (axis cs:3.38,3.75) {CIP};
\node[font=\fontsize{6}{7}\selectfont, text=amberC!65!black, anchor=south] at (axis cs:4.88,4.75) {SSIP};
\node[font=\fontsize{5.8}{6.6}\selectfont, text=axg, anchor=south east] at (axis cs:7.9,0.2) {$n=11$ ion pairs};
\end{axis}
\node[panellabel] at ($(PAIR.north west)+(-0.6cm,0.15cm)$) {e};

\begin{axis}[
  name=COORD, at={($(WAT.south west)+(0,-2.05cm)$)}, anchor=north west,
  title={\textsf{Coordination number}},
  xlabel={$r$ (\AA)}, ylabel={$n(r)$},
  xmin=1.5, xmax=4.5, ymin=0, ymax=10, xtick={2,3,4}, ytick={0,2,4,6,8,10},
  legend style={legstyle, at={(0.03,0.97)}, anchor=north west},
]
\draw[amberC!45, dotted, line width=0.7pt] (axis cs:3.21,0)--(axis cs:3.21,10);
\draw[grnC!45, dotted, line width=0.7pt] (axis cs:3.51,0)--(axis cs:3.51,10);
\draw[steelC!45, dotted, line width=0.7pt] (axis cs:3.87,0)--(axis cs:3.87,10);
\addplot[amberC, s] table[col sep=comma, x=r, y=nNaO]{data_thermo/rdf_nacl.csv}; \addlegendentry{Na$^+$--O: 5.8}
\addplot[grnC, s] table[col sep=comma, x=r, y=nKO]{data_thermo/rdf_kcl.csv}; \addlegendentry{K$^+$--O: 7.7}
\addplot[steelC, s] table[col sep=comma, x=r, y=nClO]{data_thermo/rdf_nacl.csv}; \addlegendentry{Cl$^-$--O: 7.5}
\end{axis}
\node[panellabel] at ($(COORD.north west)+(-0.6cm,0.15cm)$) {f};

\end{tikzpicture}}
  \caption{\textbf{Thermodynamic consistency and quantitative agreement with experiment for bulk water and physiological electrolytes.} Under unbiased dynamics UBio-MolFM (Stage~3; 1-ns trajectories) settles at the correct equilibrium density and conserves energy (\textbf{a,b}) and reproduces high-precision scattering references for water structure and ion hydration (\textbf{c--f}). \textbf{(a)}~Mass density during $NPT$ simulations (300~K, 1~bar) of 512-molecule water and $0.15\,\text{mol/L}$ NaCl/KCl boxes ($0.987$, $0.995$, $0.996\,\text{g\,cm}^{-3}$), all within $1\%$ of experiment; the pure-water box is smaller than the salt boxes ($\approx$4{,}000 molecules) and so shows larger finite-size fluctuations. \textbf{(b)}~$NVE$ energy conservation: per-atom deviations of kinetic, potential and total energy from their time-means. The total stays flat (drift $1.8\times10^{-4}\,\text{eV\,atom}^{-1}\,\text{ns}^{-1}$), confirming conservative forces; the early rise in kinetic energy is the box relaxing onto our surface from a classically equilibrated start---$3.9$~meV/atom converting from potential into kinetic inside the first $25$~ps, carrying it from $299$ to $328$~K---not an energy-conservation violation; over the remaining $975$~ps the box is stationary, and the residual gap between the potential and kinetic traces there is the drift itself. \textbf{(c)}~Water $g_{\text{OO}}$, $g_{\text{OH}}$, $g_{\text{HH}}$; first peak $2.79$~\AA{} at $g=2.58$ against the X-ray $2.80(1)$ and $2.57(5)$, coordination $4.3$ against $4.3(2)$. Red crosses mark experimental $g_{\text{OO}}$ peak and valley positions~\cite{skinner2014structure, chen2016ab} ($g_{\text{OH}}$, $g_{\text{HH}}$ shown for the intermolecular region only). \textbf{(d)}~Ion--water RDFs. \textbf{(e)}~Cation--anion RDFs resolving solvent-separated (SSIP) and contact (CIP) pairing; curves are coarse-binned and smoothed owing to limited $n=11$ ion-pair statistics. \textbf{(f)}~Running coordination numbers for the RDFs of \textbf{(d)} (dotted lines, first-shell cut-offs; first-shell values annotated). Dotted horizontals in \textbf{(c--e)} mark $g(r)=1$. All RDFs and coordination numbers are averaged over the equilibrated second half of each trajectory. Density failures of pretrained baselines are shown in Extended Data Fig.~\ref{fig:density_baselines_si}.}\label{fig:thermo}
\end{figure}

Under unbiased $NPT$ dynamics (300~K, 1~bar), a 512-molecule water box holds a mean density of $0.987 \pm 0.009\,\text{g\,cm}^{-3}$ over 1~ns---$1.0\%$ below the experimental $0.997$---and the $0.15\,\text{mol/L}$ NaCl and KCl boxes fall within the same margin (Fig.~\ref{fig:thermo}a). No density was supplied in training: the number is emergent, and discriminating---DPA-4 (OMol branch) collapses toward a gas-like density while MACE-OMol and UMA-S-1p2 over-densify---as does our own Stage-2 checkpoint, at $1.118\,\text{g\,cm}^{-3}$, until UBio-Mol26's periodic tier enters at Stage~3 (Extended Data Fig.~\ref{fig:density_baselines_si}; SI~S4.1). Energy conservation is equally clean: under $NVE$ the total drifts by $1.8\times10^{-4}\,\text{eV\,atom}^{-1}\,\text{ns}^{-1}$ (Fig.~\ref{fig:thermo}b), so the learned forces carry no non-physical source or sink. All condensed-phase simulations below use Stage~3.

The measured structure follows. The oxygen--oxygen radial distribution function matches the benchmark wide-$Q$ X-ray determination~\cite{skinner2013benchmark} at the nearest-neighbour maximum in position and height, recovering the tetrahedral network at the experimental coordination number~\cite{skinner2014structure}; the first minimum is slightly deeper than experiment, the mild over-structuring expected of DFT-quality water with classical nuclei~\cite{chen2016ab} (Fig.~\ref{fig:thermo}c; Extended Data Table~\ref{tab:validation_summary}). Ion hydration shells fall inside the experimental and high-level-DFT ranges, and ion--ion correlations show none of the spurious aggregation afflicting fixed-charge and some machine-learned models at physiological concentration: Na$^{+}$--Cl$^{-}$ pairs are mostly solvent-separated, the more weakly hydrated K$^{+}$ also sampling contact pairs (Fig.~\ref{fig:thermo}d--f; Extended Data Tables~\ref{tab:nacl_comparison} and~\ref{tab:validation_summary}). With only $11$ pairs sampled, that contrast is indicative and the absence of aggregation the robust observation.

Structure is not dynamics, and the one dynamical observable this trajectory yields is where agreement is weakest. The water self-diffusion coefficient---from the mean-square displacement of the molecular centres of mass---is $1.60\times10^{-5}\,\text{cm}^2\,\text{s}^{-1}$ raw, $1.88$ with the periodic finite-size correction~\cite{yeh2004}, against a measured $2.30\times10^{-5}$ at $298$~K~\cite{holz2000}: $18\%$ too slow, and a floor rather than an estimate, since the correction assumes the experimental viscosity where our more over-structured liquid would be more viscous (Methods). That shortfall is ours, not the reference level's: revPBE0-D3---the functional our periodic tier is computed at---gives $2.67\times10^{-5}$ with classical nuclei~\cite{marsalek2017}, placing our corrected value $30\%$ below it, and our raw value $24\%$ below its raw counterpart, and nuclear quantum effects would widen rather than close that gap, since they lower $D$ on \emph{ab initio} surfaces~\cite{marsalek2017}. Still, $1.88$ beats the dispersionless generalized-gradient functionals, which are near-glassy at ambient conditions, and SCAN and BLYP-D3 on a single correction convention~\cite{villard2024}. The direction matches our own structure: the over-structuring that deepens the first $g_\text{OO}$ minimum also slows exchange between hydration shells, so this is one defect, not two.

\subsection{A Structural Mg\texorpdfstring{$^{2+}$}{2+} Site in Folded RNA, Reached Without Ion-Specific Parameters}
\label{subsec:rna_metal_coordination}

\begin{figure}[htbp]
  \centering
  \widefig{\input{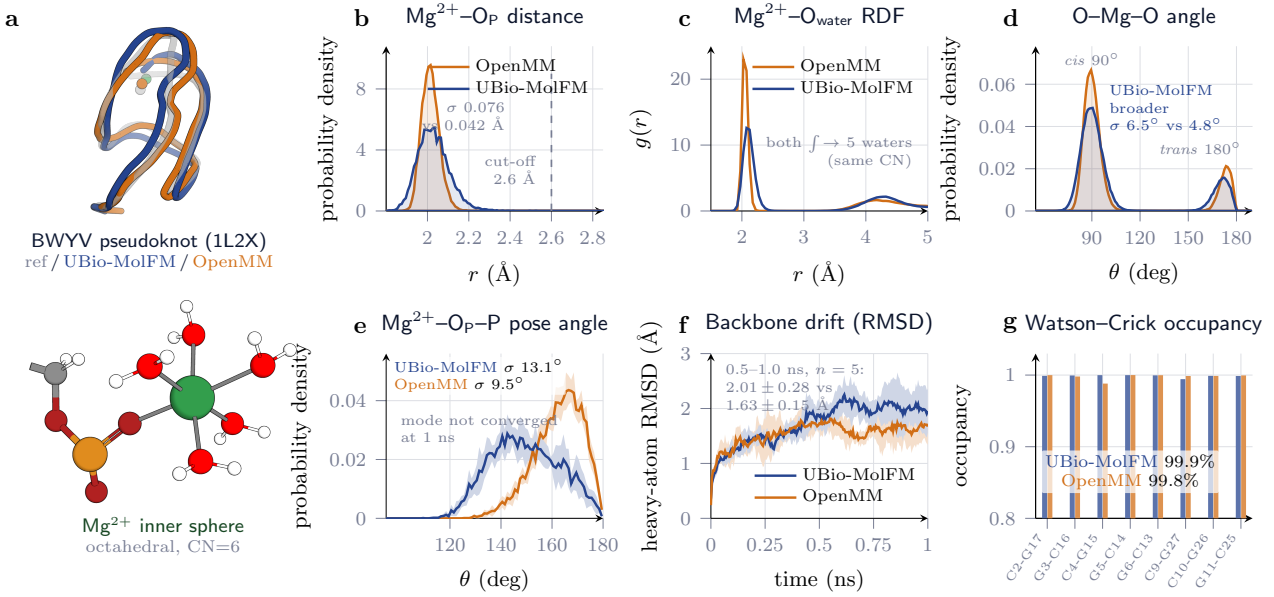}}
  \caption{\textbf{Site-specific Mg$^{2+}$ coordination in the BWYV RNA pseudoknot (1L2X): UBio-MolFM reproduces a Mg-specific classical force field.} Unbiased $NPT$ simulations of the 27-nt pseudoknot retaining the crystal's single inner-sphere Mg$^{2+}$, comparing UBio-MolFM (blue) with Amber~OL3\,$+$\,TIP3P\,$+$\,Li--Merz 12-6-4 Mg$^{2+}$ (orange) from the same equilibrated structure. \textbf{Five replicas per potential}, $1.0$~ns each; curves are five-replica means, envelopes seed-to-seed standard deviations (Methods). \textbf{(a)}~\emph{Top}: final backbones on the common reference (grey) with their bound Mg$^{2+}$; both preserve the fold and the metal site. One replica per potential is drawn---seed 42, the highest-drift classical run and the second-highest of ours, so neither is a favourable pick. \emph{Bottom}: the inner-sphere octahedron (five water O plus OP1 of the G1 phosphate), the site both potentials hold in every frame of all ten trajectories---the complete CN$=$6 shell in every classical frame and $99.8\%$ of ours, a single water briefly crossing the $2.8$~\AA{} counting cut-off in the rest. \textbf{(b)}~Mg$^{2+}$--O$_\text{P}$ distance density: one peak for both, well inside the $2.6$~\AA{} inner-sphere cut-off (dashed), but $1.8\times$ broader under UBio-MolFM ($\sigma$ $0.076$ vs $0.042$~\AA). \textbf{(c)}~Mg$^{2+}$--O$_\text{water}$ radial distribution function: UBio-MolFM's first peak ($2.08\pm0.02$~\AA) sits on the experimental Mg--O distance ($2.09\pm0.04$~\AA~\cite{ohtaki1993hydration}) while the fitted shell is shorter ($2.03$) and over-structured ($g_\text{max}$ $23.0$ vs $12.7$), both integrating to five first-shell waters. \textbf{(d)}~O--Mg--O angle distribution: both recover the octahedral \emph{cis}/\emph{trans} doublet, but UBio-MolFM's widths ($\sigma$ $6.5^\circ$, $5.1^\circ$) match AIMD of \emph{aqueous} Mg$^{2+}$ ($6.3^\circ$, $5.0^\circ$)~\cite{yu2022mgtfsi} where the fitted potential is narrower on both ($4.8^\circ$, $3.7^\circ$). \textbf{(e)}~Mg$^{2+}$--O$_\text{P}$--P pose angle, again broader under UBio-MolFM ($\sigma$ $13.1^\circ$ vs $9.5^\circ$); its \emph{mode} is the lower one ($147\pm10^\circ$ vs $166\pm3^\circ$, $p=0.015$) but is not converged, so only the width is relied on. \textbf{(f)}~Heavy-atom RMSD to the common start over the second half: ours drift modestly further ($2.01\pm0.28$ vs $1.63\pm0.15$~\AA) with about twice the replica spread, and neither ensemble unfolds---drift from the shared origin, not deviation from the crystal. \textbf{(g)}~Occupancy of the eight canonical Watson--Crick stem pairs; both hold the stems ($99.9$ vs $99.8\%$, $p=0.64$). Panels (b,d,e) carry the four width measures, with no replica of one potential overlapping any replica of the other; (c) corroborates the same picture without an experimental counterpart of its own. All values, and the convention that quoted scalars are per-replica means rather than properties of the plotted curve, are given in \S\ref{subsec:rna_metal_coordination} and Extended Data Table~\ref{tab:validation_summary}.}\label{fig:rna_mg_analysis}
\end{figure}

The Mg$^{2+}$ at the core of the Beet western yellows virus pseudoknot keeps its octahedral inner-sphere shell---five waters and one phosphate oxygen---under a potential that has never seen an ion-specific parameter: the phosphate contact holds in every frame of all five unbiased replicas and the complete CN$=$6 shell in $99.8\%$ of them, and the fold, eight Watson--Crick stem pairs and \emph{cis}/\emph{trans} angular doublet hold with it (Fig.~\ref{fig:rna_mg_analysis}a,b,d,g). This is the regime fixed-charge models handle worst---the intense polarization of a polyvalent metal buried in a folded nucleic acid---and reproducing it has historically needed heavy, ion-specific parameterization. We simulated the 27-nucleotide pseudoknot of PDB 1L2X~\cite{egli2002metal} in explicit water against the gold standard for exactly this problem---Amber~OL3~\cite{zgarbova2011ol3}/TIP3P~\cite{tip3p} with the ion-specifically fitted Li--Merz \emph{12-6-4} model~\cite{li2014ion}---five replicas each from one shared equilibrated structure (Methods). Both preserve the metal site and the stems---coordination number, topology, inner-sphere fraction and canonical pair occupancy indistinguishable---though ours drift modestly further from the shared origin in RMSD and radius of gyration; neither unfolds (Fig.~\ref{fig:rna_mg_analysis}f).

Where the two differ, the fitted potential is stiffer, and every observable carrying an external reference favours the learned one. Four independent width measures separate the two five-replica sets with no overlap. Angular breadth is the sharpest test of rigidity: UBio-MolFM reproduces the octahedral \emph{cis} and \emph{trans} thermal widths of \emph{ab initio} aqueous Mg$^{2+}$~\cite{yu2022mgtfsi} almost exactly where the fitted potential is narrower on each, seed-to-seed scatter at or below $0.06^\circ$ in all four replica sets, and Welch's $t$ on the five replicas per potential separating them at $p$ between $1.4\times10^{-12}$ and $2.3\times10^{-4}$ across the four measures. The metal--phosphate distance distribution is $1.8\times$ wider under ours at an identical mean, its pose angle on the phosphate $1.4\times$ broader, and the first-shell distance matches the experimental Mg--O value where the fitted shell falls marginally short, both integrating to five inner-shell waters (Fig.~\ref{fig:rna_mg_analysis}b--e; Extended Data Table~\ref{tab:validation_summary}). Where a Mg-specific force field returns a near-frozen octahedron, a potential fitted to no ion returns a librating one, at the width scattering and \emph{ab initio} dynamics report for the aqueous ion.

Three limits bound the claim. The fitted shell is also far more structured radially, but peak heights have no experimental counterpart, so this corroborates rather than establishes the over-rigidity. Every external reference here describes \emph{aqueous} Mg$^{2+}$ rather than an RNA-bound one---a transfer we justify but cannot bound, whose failure would move the reference \emph{toward} the fitted potential rather than ours (SI~S4.2). The pose angle's \emph{position} we set aside: not converged at $1.0$~ns, so only its width enters the argument (Methods).

\subsection{A Hidden Hydrogen-Bond Latch Gates the Permeable Conformer of Cyclosporine~A}
\label{subsec:csa_thermodynamics}

\begin{figure}[htbp]
  \centering
  \captionsetup{singlelinecheck=false}%
  \widefig{\input{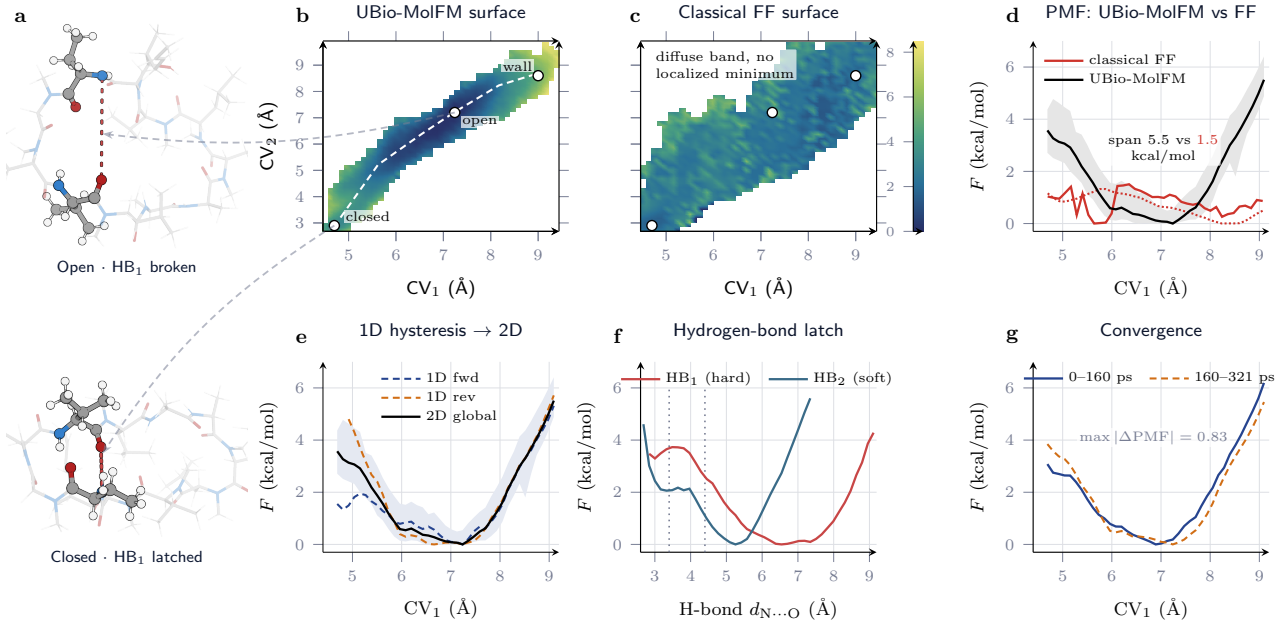}}
  \caption{\textbf{Aqueous conformational free-energy landscape of Cyclosporine~A, and its collapse under a classical force field.} Energies in kcal/mol, zeroed at each surface's minimum. $\text{CV}_1$ is the transannular backbone carbonyl--carbonyl distance (Abu2/Val5); the latch coordinate $\text{CV}_2\equiv\text{HB}_1$ is the Abu2~N$\cdots$O=C~Val5 distance. \textbf{(a)}~Representative conformations (latch residues highlighted, water omitted): expanded \emph{open} state (top, $\text{HB}_1$ ruptured) and compact \emph{closed} state (bottom, $\text{HB}_1$ engaged); dashed arrows mark their positions on the surface. \textbf{(b)}~UBio-MolFM 2D surface $F(\text{CV}_1,\text{CV}_2)$---a single diagonal funnel in which the latch closes as the ring compacts; white markers, the closed, open and over-extended stationary points; dashed line, minimum-free-energy path. \textbf{(c)}~The same surface from a classical force field (Amber/GAFF2--TIP3P) with identical system, CVs, springs and 2D-MBAR protocol (Methods), on the same colour scale: a diffuse low-free-energy band running the whole diagonal, with no localized minimum. \textbf{(d)}~PMF along $\text{CV}_1$ (HB$_1$ marginalized): UBio-MolFM (black) resolves a $5.51$-kcal/mol funnel across the well-sampled window, whereas the classical model (crimson solid, 2D-MBAR; dotted, its 1D-only estimate) fits inside $1.51$~kcal/mol on the same window---a value unchanged on the narrower $4.7$--$9.0$~\AA{} range Methods restricts the classical curve to, both of its extremes lying inside it---closed basin $1.04$ ($0.05$--$2.00$) vs $3.54$ ($2.69$--$4.41$), wall $0.89$ ($0.12$--$1.29$) vs $5.07$ ($4.43$--$6.04$), neither pair overlapping. Ranges here and in \textbf{(e)} are the full spread of $20$ window-level bootstrap resamples, not calibrated $95\%$ intervals (Methods); non-overlap therefore means no resample of either campaign reached the other. \textbf{(e)}~UBio-MolFM PMF: the 80-state 2D-MBAR estimate (black; shaded window-bootstrap range) over the well-sampled window $4.7$--$9.1\,\text{\AA}$, which contains all three stationary points; the two terminal bins, where the bootstrap bound is unstable, are omitted. Naive 1D-MBAR over forward or reverse windows alone disagrees---the latch hysteresis the 2D patches remove. \textbf{(f)}~Free energy projected onto each backbone hydrogen bond: $\text{HB}_1$ is the hard latch that must break for the ring to open (unlatch/relatch $0.43$/$3.73$~kcal/mol, against $0.70$/$1.20$ classically), $\text{HB}_2$ the soft one that follows passively ($0.11$); dotted lines, latched/unlatched thresholds. \textbf{(g)}~Convergence: PMF from the first versus second half of production, agreeing to $\max|\Delta\text{PMF}|=0.83$~kcal/mol, with the residual at the sparsely sampled basin and wall ends and inside the bootstrap range of \textbf{(e)}. The MeLeu9--MeLeu10 amide stays \emph{cis} throughout, so this surface is the branch carrying the membrane-permeable closed form; the dominant aqueous \emph{resting} conformer A1~\cite{limbach2022csa} and the all-\emph{trans} Fab-bound form lie on other branches reached by isomerizations not sampled here (Extended Data Fig.~\ref{fig:csa_conformer_si}).}\label{fig:csa_pmf}
\end{figure}

Cyclosporine~A pays a conformational toll in water before crossing a membrane. The 1{,}200-Da cyclic undecapeptide is the textbook chameleon---polar and solvent-exposed in water, compact and closed by transannular backbone hydrogen bonds in apolar media---and fixed-charge models mispredict the permeability of such beyond-rule-of-five macrocycles~\cite{witek2016csa}, a failure attributed to the polarization-dependent stabilization that switch turns on upon desolvation. On a converged, quantum-derived free-energy surface in explicit water---5{,}992 atoms, every water molecule on the ML potential---the toll is $3.54\,\text{kcal/mol}$ (window-bootstrap range $2.69$--$4.41$): the solvent-exposed conformer is the minimum, the compact latched one sits that far above it, and beyond $9\,\text{\AA}$ the surface climbs into a steric wall (Fig.~\ref{fig:csa_pmf}b). That is a Boltzmann population of $\approx$$0.3\%$ at $300$~K on this branch---a restatement of the toll, not an independent measurement, and consistent with the long-standing observation that CsA rarely populates its fully hydrogen-bonded form~\cite{witek2016csa}. What enforces it is not the size of the ring. It is one hydrogen bond.

That the surface is quantum-derived is measured rather than asserted. Single-point DFT on 100 decorrelated conformers from the sampled windows reproduces the model to $0.44\,\text{kcal/mol}$ at its energy head's calibration level---closer than the three DFT references stand to each other---and to $1.09$ at UBio-Mol26's collection level, force errors landing within the spread of the corresponding tiers of Table~\ref{tab:rel_energy_force} ($7.2$ against $7.69$~meV/\AA{} at the calibration level, $31.5$ against $21.8$ at the collection level): this macrocycle is described about as well as the benchmark predicts. The reference functionals disagree about the closed-versus-open \emph{electronic} energy across the dispersion divide by $+2.73$ and $+3.33\,\text{kcal/mol}$ (CI $[+2.32,+3.15]$ and $[+2.54,+4.17]$), in the direction that would enlarge the $3.54$ toll. We neither correct nor bound it (Methods; Extended Data Table~\ref{tab:csa_qm}).

Projecting the free energy onto each backbone hydrogen bond resolves the gate (Fig.~\ref{fig:csa_pmf}f). $\text{HB}_1$ (Abu2~N--H$\cdots$O=C~Val5) is a hard latch that must break for the ring to open, and it is kinetically asymmetric: relatching costs nearly an order of magnitude more than unlatching, where $\text{HB}_2$ follows the ring passively. That asymmetry is also why the obvious progress coordinate fails: $\text{CV}_1$ is degenerate against $\text{HB}_1$, so forward and reverse pulling disagree on the closed basin's depth (Fig.~\ref{fig:csa_pmf}e); localized two-dimensional patches on $(\text{CV}_1,\text{HB}_1)$, analysed as one global 2D-MBAR over all 80 states, remove it (Methods). The remedy is general: wherever an intuitive reaction coordinate conceals a slow latch, a handful of two-dimensional patches restores a self-consistent partition function without a full high-dimensional scan.

Repeating the protocol under a classical force field---identical system, springs, window layout and analysis (Methods)---shows what quantum accuracy buys (Fig.~\ref{fig:csa_pmf}c,d). Amber/GAFF2 with TIP3P returns no funnel. Its entire surface over the same window fits inside $1.51\,\text{kcal/mol}$ against the quantum surface's $5.51$, and resolves no minimum within that range: its largest adjacent-bin step is $72\%$ of the span against $10\%$ on the quantum surface, and its lowest bin is undetermined (Methods). Re-running the whole campaign unconstrained at our own $0.5$-fs timestep moves four quantities, three of them deepening the classical failure and the fourth the reverse---the closed basin from $1.04$ to $1.32\,\text{kcal/mol}$. None is separable from each campaign's own convergence drift, which exceeds the span it measures, so we rest nothing on them: what both protocols agree on is that the classical surface is $3$--$4\times$ flatter. We report the constrained $2$-fs one, standard practice rather than uniformly conservative (SI~S4.3). A surface that resolves no minimum cannot express the polar-versus-apolar preference CsA is known for; ours does, with the correct sign in water.

The latch is where the loss localizes: the same bond is $1.7\times$ asymmetric under the standard classical protocol, $1.0\times$ once that protocol is matched to ours, and $8.6\times$ under UBio-MolFM. What the classical Hamiltonian loses is not the hydrogen bond but its asymmetry---the property that makes it a gate rather than a contact.

What causes that flatness is measurable on fixed geometries. Evaluated as single points on the same 100 conformers---no sampling, solvent, integrator or constraints left in---the classical relative energies depart from $\omega$B97M-V by roughly $\pm$$10\,\text{kcal/mol}$ at a correlation of $0.80$, against the model's $0.44$ and $0.9996$. So what is flat is the free-energy profile, not the potential beneath it: the classical energy spread across these conformers, $107\,\text{kcal/mol}$, is \emph{wider} than DFT's own $101$, so the failure is misordering rather than compression, and ensemble averaging turns such a surface flat. Neither obvious escape closes it: refitting the AM1-BCC charges accounts for at most $12\%$, and the error is worst at the \emph{open} minimum (Extended Data Table~\ref{tab:csa_qm}; Methods). What remains is functional form, whose two candidates we do not separate: torsions never fitted to an $N$-methylated peptide backbone~\cite{yamane2022csaff}, and the absent polarization response invoked for macrocycles of this class~\cite{witek2016csa}. The claim is scoped accordingly: a property of an automatically parameterized classical model, not of a hand-parameterized one.

Both surfaces are resolved on the same single branch of CsA's conformational space---the MeLeu9--MeLeu10 amide stays \emph{cis} throughout---so the contrast between them is unaffected by that restriction (Discussion; Extended Data Fig.~\ref{fig:csa_conformer_si}).

\subsection{At \texorpdfstring{$10^5$}{1e5} Atoms, a Divergence in the KcsA Filter One Carbonyl Wide}
\label{subsec:transmembrane_channel}

In a 108{,}964-atom KcsA system---channel tetramer, POPE:POPG (3:1) bilayer, $0.15$~mol/L KCl, explicit water---the relaxed four-ion K$^+$ column is anhydrous in all five quantum-derived replicas and holds direct contact throughout in four of them---all three upper spacings simultaneously below $4$~\AA{} in $78$--$89\%$ of late frames---against no frame of ten fixed-charge replicas launched from the same equilibrated system. The two descriptions of the conducting filter the field has argued over for two decades are separated here by the potential, and the disagreement is narrow: across $10^5$ atoms it traces to one backbone carbonyl's orientation.

The regime is quantum-mechanically inaccessible, and the leading pretrained foundation models either exhaust memory or fall an order of magnitude short on throughput (Extended Data Fig.~\ref{fig:throughput}); per-system equivariant potentials~\cite{kozinsky2023scaling} reach comparable sizes only with system-specific QM training. UBio-MolFM's hybrid-cutoff stack instead sustains $\approx$$0.24$~steps/s at this size on one 141~GB H20 under activation recompute (interpolated from the size sweep; Methods)---to our knowledge the first transferable foundation model at this scale on one GPU without per-system retraining. Against five UBio-MolFM replicas we ran two matched fixed-charge treatments from the same equilibrated system, five replicas each: the induced-dipole-corrected Li--Merz 12-6-4 model~\cite{li2014ion} and the plain 12-6 Lennard-Jones model~\cite{joung2008jc}. All fifteen run $1.0$~ns (Methods; Extended Data Fig.~\ref{fig:kcsa_validation_si}a). Three things differ: the potential-energy surface, the barostat coupling, and a $500$-step minimization only our side received. A released-box control bounds the second; the third by its own size, $0.14$~\AA, which straddles the contiguity criterion at the first frame and matters nowhere after (Methods). What remains is whether QM accuracy matters and whether explicit ion-induced polarization does.

\begin{figure}[t]
  \centering
  \setlength{\abovecaptionskip}{4pt}
  \widefig{\input{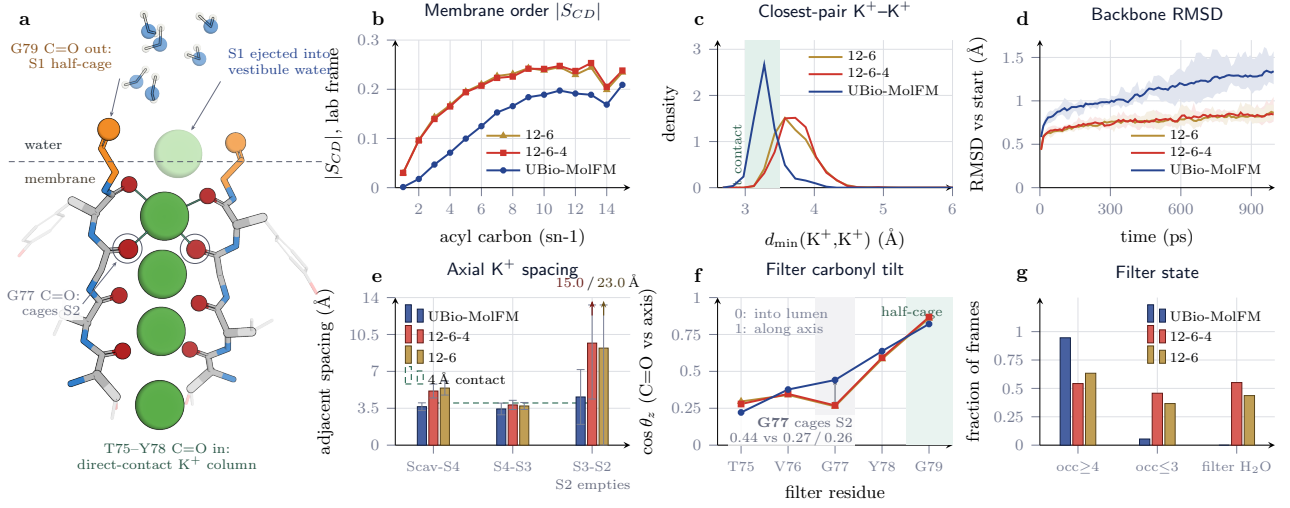}}
  \captionsetup{font={footnotesize,stretch=1.0}}
  \caption{\textbf{A 108{,}964-atom transmembrane K$^+$ channel (KcsA): a quantum-derived potential and two fixed-charge treatments diverge at the selectivity filter.} Five replicas each of UBio-MolFM (blue), the induced-dipole-corrected Li--Merz 12-6-4 model (crimson) and the plain 12-6 Lennard-Jones model (goldenrod), all from the identical system and equilibrated configuration under one $NPT$ protocol (310~K, 1~atm; the UBio-MolFM replicas start after a $500$-step minimization on their own potential, which the classical references do not receive---a $0.14$~\AA{} offset at the first frame, straddling the $4$~\AA{} contiguity criterion and quantified in Methods); values are in \S\ref{subsec:transmembrane_channel} and Extended Data Table~\ref{tab:validation_summary}. \textbf{(a)}~Diagonal two-subunit cutaway of the filter at the ion-loaded start (K$^+$ green, front and back monomers hidden, resolved vestibule waters pale blue). T75--Y78 carbonyls point into the lumen and coordinate the column; the G79 carbonyls (orange) point outward, leaving S1 without a ceiling---the ``half-cage'' from which the excess ion is expelled under all three potentials. Teal cylinders, the measured S2 contacts identifying that cage as Y78$+$G77 (SI~S4.4); rings, the two G77 oxygens; dashed line, extracellular phosphate plane, to scale. \textbf{(b)}~Membrane chain order $|S_{CD}|$ along the sn-1 acyl chain in the laboratory frame; all three give a fluid $L_\alpha$ bilayer, the two classical treatments indistinguishable. UBio-MolFM lies below both, but this frame conflates collective tilt with conformational disorder and the difference is the former, so nothing is read off this panel; what the released-box control does establish about the bilayer, unfavourably, is in Methods. \textbf{(c)}~Closest K$^+$--K$^+$ distance in the filter (late-trajectory frames, pooled); shaded band, direct ionic contact. Replica means $3.32\pm0.13$~\AA{} against $3.74\pm0.10$ and $3.71\pm0.18$. \textbf{(d)}~Backbone RMSD to the equilibrated start, all replicas (mean; shaded min--max). \textbf{(e)}~The three upper adjacent axial K$^+$ spacings (mean\,$\pm$\,s.d.): UBio-MolFM holds all three below $4$~\AA{} in four of five replicas, whereas both classical columns break at S3--S2 (below the vacated S2, marked; opening to $9.7$ and $9.2$~\AA) and at S$_\text{cav}$--S4 ($5.1$, $5.4$~\AA). Both classical S3--S2 bars are truncated at $13.3$~\AA{} for legibility (true upper bounds $15.0$ and $23.0$~\AA). \textbf{(f)}~Filter-carbonyl tilt $\cos\theta_z$ from T75 to G79, quantifying the half-cage of (a): all three reproduce the signature in which G79 alone points out of the lumen (green band) and differ only at G77 (grey band, arrow): $\cos\theta_z$ $0.44$ against $0.27$ and $0.26$, every other residue agreeing to within $0.08$ (largest $0.074$, at T75). The G77 reorientation sets the V76--G77 and G77--Y78 backbone-plane spacings and the S2 coordination distance (Extended Data Table~\ref{tab:validation_summary}). \textbf{(g)}~Fraction of late-trajectory frames (window as in c,e) with on-axis occupancy $\geq$4 ions, $\leq$3 ions, and any water in the filter core; per-replica values in Extended Data Fig.~\ref{fig:kcsa_validation_si}.}\label{fig:transmembrane_channel}
\end{figure}

Outside the filter all three hold a fluid ($L_\alpha$) bilayer, and every replica's second-half backbone RMSD stays below $1.5$~\AA---largest $1.45$, on the one seed whose column later opens and whose trajectory peaks at $1.68$---so no model reaches the filter by unfolding its scaffold (Fig.~\ref{fig:transmembrane_channel}d). The deep sites agree: at S3 and S4, permanently occupied, all three give the same K$^+$--carbonyl-oxygen distance at close to eightfold coordination, contracted $\approx$$0.2$~\AA{} from the crystallographic model~\cite{zhou2001kcsa} as finite-temperature dynamics should be (Extended Data Table~\ref{tab:validation_summary}). The classical models therefore fail neither at ion--protein coordination---what their Lennard-Jones and Coulomb terms were fitted to---nor by deforming the filter.

The bilayer is where our potential is weakest, and releasing the barostat makes it worse. Under isotropic coupling it drifts where the classical one does not---thinning, tilting and taking up hydrocarbon-core water over a nanosecond. Three replicas run to $400$~ps with the membrane normal free thin \emph{further} on a matched window, $38.4$ against $40.2$~\AA{} isotropic and $42.0$ classical, while area per lipid---pinned by construction when the box is isotropic---expands to $53.0$ from $49.3$~\AA$^2$, neither converged. The constraint was therefore masking part of this deficit rather than creating it: the bilayer is genuinely thinner. Director-referenced chain order, which needs no box, settles below classical within $10$~ps under both couplings and further when the box is free ($0.237$ and $0.222$ against $0.257$, sn-1 chain); acyl covalent geometry is identical under both (Extended Data Table~\ref{tab:validation_summary}). At $n=3$ this is a direction, not a magnitude. The filter results are tested against this drift (Methods; Discussion).

What differs is the G77 carbonyl, which with Y78 cages site S2. Under UBio-MolFM it points differently, and that single reorientation widens the V76--G77 backbone-plane spacing and narrows G77--Y78, while spacings not involving G77 agree across all three potentials to within $0.10$~\AA. S2 follows the cage, staying coordinated as tightly as the deep sites in four of five quantum-derived replicas and emptying under both fixed-charge ones (Fig.~\ref{fig:transmembrane_channel}f; Extended Data Table~\ref{tab:validation_summary}).

The divergence propagates into the ion column. The start is deliberately ion-loaded---five K$^+$ across S1--S4 and the cavity---and the state it relaxes \emph{into} is what we characterize. Under all three the excess ion leaves within the first tens of picoseconds ($2$--$74$~ps across the fifteen replicas), always from S1, whose uncapped ``half-cage'' topology---not any one potential---selects it (Fig.~\ref{fig:transmembrane_channel}a,f), leaving a four-ion column spanning the central cavity and S2--S4---four ions on axis, three at filter sites. From there the potentials part. Under UBio-MolFM that column is dehydrated in every replica and contiguous in four of five throughout: all three upper spacings simultaneously below 4~\AA{} in $78$--$89\%$ of late frames, $28\%$ in the fifth, at least four ions on axis in $95\%$, and no water in the filter core in any frame (Fig.~\ref{fig:transmembrane_channel}c,e,g). Under neither classical treatment does the contact column form in any frame of any replica---yet both hold their \emph{closest} pair within half an \AA ngstr\"om of ours, so the pairwise minimum does not reveal the failure; the column does, breaking at S3--S2 immediately below the vacated S2 and at S$_\text{cav}$--S4, where water replaces ions. Three further replicas with the membrane normal free keep the column dry and compact---two contiguous on the same criterion, the third under $6$~\AA{} throughout without opening (Methods)---and restoring the truncated long-range electrostatics would tighten the contact pair rather than loosen it, so neither the fixed aspect ratio nor the $8$~\AA{} cutoff closes that pair (SI~S4.4).

One UBio-MolFM replica leaves this state before its trajectory ends, widening every quantum-derived number above. Seed 126 holds the contact column for $700$~ps then translocates outward, S3--S2 opening progressively to $13.7$~\AA{} over the remaining $300$~ps; on-axis occupancy falls below four $\approx$$90$~ps later still. The filter stays dry throughout, but the compact column does not re-form. We do not interpret the event: one replica cannot separate a feature of the potential from a rare fluctuation, and we cannot exclude a contribution from the membrane drift above (SI~S4.4). The other four occupy the contact state in $78$--$89\%$ of late frames with no drift resolvable at this length (Methods; Extended Data Fig.~\ref{fig:kcsa_validation_si}).

The three-way design answers a second question needing no structural ground truth---what the 12-6-4 model's induced-dipole term buys---and across ten classical replicas the answer is nothing: neither forms the contact column, both lose on-axis occupancy and admit filter water at rates that, pooled over five replicas each, are indistinguishable, and only the failure's \emph{shape} differs---within each family individual replicas differ widely. A pairwise, mean-field induced-dipole correction is no substitute for the many-body screening a quantum-derived potential carries self-consistently; the C4 term redistributes the failure without removing it (Extended Data Fig.~\ref{fig:kcsa_validation_si}).

How the conducting filter is organized remains contested, and the closest experimental probe to the state we simulate reads it as water-separated rather than contiguous (Discussion). What keeps the two from colliding head-on is occupancy: those spectra report a filter at its equilibrium load, two ions with water between them, where ours starts with five ions on axis and relaxes to a four-ion column spanning the cavity and S2--S4. Our measurement is therefore conditional---\emph{given} that column, it is in contact and dry---and a data point rather than a resolution, no experimental structure existing for this state. What it establishes is that at $10^5$ atoms the potential alone decides between two qualitatively different filter organizations, that the difference falls where added-range electronic structure should matter~\cite{wang2026scalable}, and that it is localized enough to name the atom responsible.

\clearpage

\section*{Discussion}

The accuracy--scale trade-off that has shaped biomolecular simulation for decades is, on this evidence, no longer one of principle but of cost. One potential, never refit, ties a purpose-built ion-specific force field at the RNA metal site where that model is the gold standard, then parts from fixed-charge models exactly where their physics is known to break: a classical potential flattens the Cyclosporine~A landscape until no conformer is preferred, and ten classical replicas never assemble the anhydrous contact column a quantum-derived one holds in four of five. No single result carries that claim; the pattern does. Matching a specialist on its own ground is evidence of accuracy, departing from it where it fails evidence of transferability. Two things make both possible: DFT-level force accuracy holding past a thousand atoms, where strictly-local potentials degrade~\cite{wang2026scalable}, and a hybrid receptive field that carries a $10^5$-atom solvated membrane onto one GPU.

The field itself does not agree on how the conducting filter is organized. The direct knock-on picture---adjacent sites held by ions in contact---rests on long simulations at physiological voltage~\cite{kopfer2014knockon,kopec2018} and on solid-state NMR finding the conduction pathway water free, albeit in a NaK2K construct rather than KcsA~\cite{oster2019}. The alternating ion--water picture is the reading of the crystallographic densities~\cite{zhou2001kcsa,moraiscabral2001}, the first conduction free-energy calculations~\cite{berneche2001}; two-dimensional infrared spectroscopy of site-labelled KcsA supports it directly, its spectra matching water-separated and not adjacent-site configurations~\cite{kratochvil2016,ryan2023}. That last work is the closest experimental probe to the state we simulate, and it does not agree with us. Two things keep the disagreement from being decisive, both expanded in SI~S4.4: the infrared assignment runs through spectra computed on classical trajectories, and configurations with three or more ions and water in S1 reportedly reproduce them too~\cite{kopec2018}; and which mechanism a simulation produces traces in part to the water model and ion parameters alone~\cite{bosio2026}. That cuts both ways---why comparing potentials is worth doing, and why no one potential's answer settles the question.

Real limitations remain, and the largest is cost. Per force evaluation UBio-MolFM is three to four orders of magnitude slower than a highly optimized fixed-charge engine, further behind once its shorter timestep is counted. That confines these trajectories to the nanosecond scale: the zero-voltage KcsA runs probe structural fidelity, not conduction, which needs an applied membrane potential and far longer sampling. Other gaps are narrower but real. Relative-energy tracking on nucleic acids still trails the strongest baseline, reflecting sparse DNA/RNA coverage; glycans, protein--ligand binding free energies, reactive chemistry and transition metals beyond Mg$^{2+}$ remain untested. The CsA surface is converged over the \emph{cis}-amide branch carrying the permeable conformer, not the full aqueous ensemble, amide isomerization being too slow to equilibrate at this level of theory. Electrolyte ion-pairing rests on few pairs. And the KcsA comparison fixes the start; occupancy statistics for a filter prepared at resting load need independent long trajectories. Those trajectories carry a protocol limitation, and releasing it does not help our case. Under an isotropic barostat the bilayer cannot relax area and thickness independently, so every collective membrane observable measures that constraint as well as the potential. Three replicas with the normal free split that confound, and the answer is unflattering: freed, the bilayer thins further and spreads rather than returning to the classical reference, so the constraint was masking part of the deficit rather than making it. The thinness is real, and director-referenced chain order---which needs no box, though it is measured on the saturated sn-1 chain alone---sits below classical under both couplings and further under the free one, plausibly because no extended bilayer enters training (Methods). At $n=3$ and $400$~ps, with both observables still moving, that is a direction and a lower bound rather than a magnitude, and we make no mesoscopic membrane claim. Untouched either way is what the filter result rests on: local lipid covalent geometry, the protein fold and every filter distance.

That classical force fields remain the right instrument wherever their accuracy suffices is not a concession but the point. The case for a quantum-derived potential is conditional rather than universal, and the conditional version is the more useful one: it is decisive precisely where fixed-charge physics fails. On the CsA surface the polarization-dependent latch gating permeability is flattened away, and no amount of extra classical sampling recovers it---the error lives in the potential, not the statistics. Elsewhere the accuracy is redundant. Which regime one is in is the practical question, and a transferable potential validated against experiment across four tiers makes it answerable rather than a matter of taste.

What this points toward is not a better benchmark but a routine instrument---one a structural biologist or medicinal chemist reaches for on an ordinary day. Both sides of the cost--accuracy trade-off are still moving, and the residual gap is an engineering target rather than a fundamental one. Quantization and distillation, better neighbour-list and kernel scheduling, multiple-timestep integration, and QM/MM partitioning that reserves the learned potential for the chemically decisive region should together recover further orders of magnitude, while broader top-down sampling of nucleic acids, lipids, glycans and protein--ligand complexes will widen what one untuned potential covers. The full stack---dataset, weights and inference engine---is released openly to that end (see Data and Code availability).

\newpage
\section{Methods}
\subsection{Data Construction} \label{sec:data_construction}
A foundation model for biological simulation needs quantum-mechanical data spanning the conformational complexity and system scales of macromolecular environments, and existing public datasets fall short in complementary ways. SPICE~\cite{spice} and its 2.0 revision~\cite{spice2} pair high-level labels with broad drug-like chemistry but remain too small for large-scale pretraining; OMol25~\cite{omol25} provides $\approx$140~million high-precision DFT calculations but caps system size at 350 atoms, far below a full protein and too small to cover the long-range interactions that stabilize macromolecules.

UBio-Mol26 closes that gap. It targets large biological macromolecules---drug-like molecules, protein segments, DNA/RNA fragments and lipid components, all in explicit solvent---together with cross-modal interactions such as drug--amino acid complexes and DNA--protein interfaces, and holds $\approx$19~million configurations reaching $1{,}370$ atoms at a mean of $\sim$414 (dominated by the \texttt{def2-SVP} subset, whose own mean is $449$), an order of magnitude above OMol25's mean of 50.

\subsubsection{Data Construction Strategy}
Two strategies are combined: ``bottom-up'' enumeration for unbiased coverage of chemical space, ``top-down'' sampling for biological relevance.

\paragraph{Bottom-up: unbiased enumeration.}
Biological systems are built from a finite set of units, so we enumerated those units combinatorially rather than sampling them: the $6{,}840$ tripeptides of three \emph{distinct} standard amino acids ($20\times19\times18$), with the analogous enumeration for DNA/RNA base pairs and lipid components. These are then combined with drugs and nucleic-acid fragments in explicit solvent, so the model learns the building blocks systematically rather than at whatever frequency a structural database happens to contain them. Cross-modal interfaces, DNA--protein included, come from this construction rather than from top-down sampling in the current release.

\textit{Where this construction stops: the bilayer.} Lipids are the one family it cannot carry to its own length scale. A phospholipid is long, and once solvated it consumes the atom budget a cluster calculation allows, so the enumerated lipid entries are single molecules or tail-joined pairs in explicit water rather than extended bilayer patches: no configuration in the corpus presents a membrane over the area on which its collective structure is defined. The periodic fidelity, the only one that constrains condensed-phase collective response, contains no lipid at all---its composition is peptide, nucleic acid, ions and neat water (\emph{Dataset Composition} below). Per molecule this costs nothing measurable---lipid is in fact our strongest held-out class, force error $11.9$ against the best baseline's $33.0$~meV/\AA, the largest of the four margins (Table~\ref{tab:rel_energy_force}). What is missing is the extended phase, and that is a plausible origin for the one bilayer deficiency that survives the barostat confound of \S\ref{subsec:transmembrane_channel}: director-referenced chain order settling $10\%$ below a fixed-charge reference within $10$~ps and holding there. The two are different failures---the barostat compromises what we can \emph{measure} of the bilayer, this gap would compromise the potential itself---and we name this one as a candidate rather than a demonstrated cause. It is the coverage gap the next release should close.

\paragraph{Top-down: biological relevance.}
To anchor that coverage in real structures we also sampled from biological assemblies---protein only at present, nucleic-acid assemblies planned. AlphaFold Protein Structure Database~\cite{afdb} structures are solvated in explicit water and spherical clusters excised around chosen residues, each retaining its local environment of neighbouring residues and waters; truncated boundaries are chemically capped rather than left as bare ions, following GEMS~\cite{gems}. The preparation and clustering pipeline is in SI~S1, and the two strategies are summarized in Fig.~\ref{fig:framework}a.

\subsubsection{Data Generation Pipeline and Computational Methodology}
High-fidelity density functional theory (DFT) on large biological systems is prohibitive, so labels are multi-fidelity. The $\omega$B97M-D3 functional~\cite{wb97md3} with a mixed basis~\cite{weigend2005def2} (def2-TZVP for H and metal ions, def2-TZVPD elsewhere) lifts SCF convergence on large dense systems from $<$$20\%$ to $>$$60\%$; a larger portion at the faster def2-SVP basis expands coverage tenfold for under $10\%$ of the compute. Both are finite spherical clusters, which cannot constrain the macroscopic response of the bulk condensed phase, so a third, \emph{periodic} fidelity ($63{,}603$ configurations at revPBE0-D3/\texttt{TZV2P-MOLOPT-PBE0} with \texttt{CP2K}~2026.1) is generated separately (SI~S1).

Generation has two stages: \textit{assembly and relaxation}, using Packmol and AmberTools~24.8 with OpenMM~8.5 to prepare solvated clusters free of steric clashes, and \textit{high-fidelity calculation}, using GPU4PySCF~1.7 and ASE~3.27 for geometry optimization and MD sampling. Computational parameters, basis sets and tools are given in SI~S3.

\subsubsection{Dataset Composition}
UBio-Mol26 comprises three subsets of complementary purpose and fidelity: (i) a scale-coverage \texttt{def2-SVP} subset ($16{,}252{,}134$ configurations, mean $449$ atoms; $55.2\%$ protein, $19.6\%$ lipid, $15.0\%$ DNA, $9.5\%$ RNA, $0.7\%$ drug-like) supplying structural and size coverage across large solvated systems; (ii) a high-fidelity \texttt{def2-TZVPD} subset ($2{,}531{,}865$ configurations, mean $193$ atoms; $68.5\%$ protein, $12.7\%$ ions, $8.1\%$ lipid, $5.7\%$ RNA, $4.1\%$ drug-like, $0.8\%$ DNA) anchoring local accuracy; and (iii) a \emph{periodic} density-correction subset ($63{,}603$ configurations, mean $314$ atoms; $88.5\%$ peptide/DNA/RNA, $11.5\%$ ions and neat water, the water boxes being counted inside that ionic fraction) at revPBE0-D3/\texttt{TZV2P-MOLOPT-PBE0} with \texttt{CP2K}~2026.1~\cite{cp2k} in the Gaussian and plane-waves formalism (SI~S1). The three sum to $18{,}847{,}602$ labelled configurations, the $\approx$19~million quoted throughout. Per-subset size distributions and category fractions are in Extended Data Fig.~\ref{fig:data_overview}a,b. The periodic subset's neat-water boxes anchor the equation of state underlying the thermodynamic-consistency results of \S\ref{subsec:hydration_solvation}. Together they give $\approx$19\,M labels, reaching $\approx$160\,M with OMol25's 140\,M.

Compositional analyses confirming the intended coverage---t-SNE phase-space overlap with OMol25, carbon chemical-environment frequencies (UBio-Mol26 enriched in methylene and amide groups where OMol25 is dominated by aromatics), pair-distance distributions that stay broad beyond the $5$--$6$\,\AA{} at which small-molecule datasets decay, and per-element counts including trace ions---are given in SI~S1 and Extended Data Figs.~\ref{fig:data_overview} and~\ref{fig:data_distributions}.

\subsection{Model Architecture}
\label{sec:model}

Three requirements bind the architecture at once: $\mathrm{SO}(3)$ equivariance, a receptive field reaching the non-covalent distances that stabilize macromolecular conformations, and the hardware efficiency to propagate $>$$10^5$ atoms on one GPU. We meet the latter two with an unmodified E2Former-V2 backbone~\cite{huang2026e2former,li2025e2former} and one application-specific choice on top of it, the layer stack.

The backbone is not restated here. Its node-centric Wigner-$6j$ factorization, which moves tensor-product cost from edges to nodes; \emph{Equivariant Axis-Aligned Sparsification}, which uses an $\mathrm{SO}(3)\!\rightarrow\!\mathrm{SO}(2)$ change of basis to replace dense contractions with sparse re-indexing ($\sim$6$\times$ at the convolution stage); and the fused on-the-fly attention kernel, which streams over neighbours without materializing edge tensors and so holds activation memory at $\mathcal{O}(|\mathcal{V}|)$ rather than $\mathcal{O}(|\mathcal{E}|)$ (up to $20\times$ higher TFLOPS)---all are derived, benchmarked and ablated in those two reports and summarized in SI~S2. We use them as published and claim no part of them; specific to this work are the corpus (\S\ref{sec:data_construction}), the curriculum (\S\ref{sec:training_strategy}) and the stack below.

\subsubsection{Hybrid-cutoff layer stack}

UBio-MolFM places a four-layer hybrid-cutoff stack on that backbone. The first three layers are short-range equivariant attention on a radius graph with $r_{\text{short}} = 5\,\text{\AA}$ over \emph{all} atoms, hydrogens included, carrying the local many-body interactions and the hydrogen-bond geometry that defines secondary structure. The fourth extends the cutoff to the full $0.1$--$8\,\text{\AA}$ sphere, with hydrogen excluded by the \emph{neighbour} mask alone---every atom, hydrogen included, still acts as a query centre and receives this layer's update, but only heavy atoms enter the neighbourhoods it is built from. That reaches the second hydration shell, salt bridges and short non-covalent contacts while suppressing the quadratic growth in H--H pairs that would otherwise dominate cost on a solvated system. Three short-range layers beneath one mid-range layer of heavy-atom neighbourhoods give an effective receptive field of $\sim$$23\,\text{\AA}$ at the near-linear scaling and constant-degree neighbour lists of a strictly cutoff-based model: no bipartite long-range graph, fragment coarse-graining or explicit Ewald summation enters the potential anywhere in this work. (Ewald summation does appear in this study, but only outside the model: in the classical reference simulations, which evaluate their own $1/r$ tail by particle-mesh Ewald, and in the post-hoc truncation control of \S\ref{sec:simulation_protocols}.)

\paragraph{What is deliberately absent: an explicit electrostatic term.}
No term in this stack is a Coulomb interaction. Beyond the $8\,\text{\AA}$ neighbour cutoff an atom's environment reaches it only through composition across layers, so the $1/r$ monopole tail is never evaluated analytically and the potential is finite-ranged by construction. Three things bound what that costs here, and one thing it rules out. First, screening. Every system simulated is net-neutral and condensed-phase, so no uncompensated monopole exists at any distance, and where free charges do exist they screen each other over the range the model already resolves---the Debye length is $\approx$$8\,\text{\AA}$ at the $0.15\,\text{mol/L}$ ionic strength of the electrolyte and KcsA runs, and of the same order in the RNA box, whose 25 neutralizing Na$^+$ sit at a comparable concentration. In the two systems with no free ions---bulk water and solvated Cyclosporine~A---the leading interaction beyond the cutoff is dipolar rather than monopolar and decays far faster than $1/r$. Second, the periodic \texttt{tzv2p} tier is labelled under full periodic electrostatics (\S\ref{sec:data_construction}), so the model is fitted in part against forces that already contain the long-range condensed-phase contribution; what it can learn from them is the portion predictable from a local environment, which in a screened liquid is most of it. Third, the direction of any residual error is not in our favour: a fixed-charge force field evaluates the tail explicitly by Ewald summation, so truncation handicaps our side of every comparison here, not the classical side. What it rules out is any observable carried by an unscreened or externally imposed field---electro-osmotic flow, conduction under an applied transmembrane potential, the long-wavelength dielectric response of the liquid. That is one reason the KcsA trajectories run at zero voltage and are scoped to structural fidelity (\S\ref{subsec:transmembrane_channel}); an applied field would need an explicit long-range term, not merely longer sampling.

Multi-scale information is combined by late fusion---the two scales' invariant ($\ell = 0$) channels are concatenated and projected back to the working width---which keeps the total energy rotationally invariant. Atomic forces are the analytical gradient $\mathbf{F} = -\nabla_{\mathbf{R}}E$, so energy and forces are consistent by construction rather than by penalty---the property the $NVE$ energy-conservation test of \S\ref{subsec:hydration_solvation} measures directly.

\subsection{Training}
\label{sec:training_strategy}
Training UBio-MolFM spans two very different data regimes: large-scale chemical diversity (OMol25, $\approx$140M frames) and highly heterogeneous biological systems (UBio-Mol26, $\sim$19M configurations at 15--1{,}370 atoms). Because naive graph-count batching load-imbalances badly across that size range, we use a three-stage curriculum---three sequential runs---together with an atom-balanced distributed data loader.

\subsubsection{Curriculum Learning and Infrastructure}
The checkpoint lineage is as follows. Stage~1 (S1) is randomly initialized and trained on OMol25 with an energy head and a \emph{separately parameterized} force head; its 1{,}000k-step checkpoint was selected from a run whose cosine schedule was configured for a 1{,}400k-step horizon. Stage~2 (S2) loads those weights model-only, non-strict, with the optimizer and scheduler reset, retires the parameterized force head, and runs its own 1{,}400k steps on the same corpus---not a continuation to a cumulative total---with forces obtained by automatic differentiation of the predicted energy, $\mathbf{F}=-\nabla_{\mathbf{R}}E$. The two OMol25 stages share the corpus, the objective, the loss weights and the architecture; \emph{only} how forces are produced separates them, and S2's autograd form is the self-consistency the reported model inherits. Stage~3 (S3) is the first and only exposure to UBio-Mol26: a single mixed-dataset run of 400k steps, which produces the reported model. Energy and force supervision are present in every run, so the lineage contains no energy-only initialization phase.

The mixed stage draws from \emph{four} data branches, one per reference source, balanced by the sampling ratios $0.15$ (\texttt{omol25}), $1$ (\texttt{tzvpd}), $2$ (\texttt{tzv2p}, at revPBE0-D3) and $0.2$ (\texttt{svp}). Their objectives differ because their labels do. The \texttt{omol25} branch keeps the default energy-plus-force objective, and its sampling is deliberately non-uniform: because the target is large biological assemblies, configurations above $200$ atoms ($\approx$2\,M labels) are drawn at ten times the rate of smaller ones, so the branch preserving broad chemistry still concentrates on the size range this model is built for. The \texttt{tzvpd} and \texttt{tzv2p} branches disable their energy terms and minimize the full-vector per-atom loss $\lVert\widehat{\mathbf F}_i-\mathbf F_i\rVert_2$ on retained atoms---a vector loss, not one on force magnitude alone---which also bypasses rather than reconciles the functional offsets between the three references. The \texttt{svp} branch, drawn from the largest subset, is supervised through the same energy head as the other branches, by a magnitude-gated directional hinge on that head's autograd forces
\begin{equation}
  \mathcal{L}_{\mathrm{dir}}=\operatorname{mean}_i\!\left[\operatorname{ReLU}\!\left(1-e^{-(\lVert\mathbf F_i\rVert_2-0.2)}\right)\operatorname{ReLU}\!\left(0.6-\cos(\widehat{\mathbf F}_i,\mathbf F_i)\right)\right].
  \label{eq:dirhinge}
\end{equation}
Because that hinge uses only the \emph{direction} of each \texttt{def2-SVP} force, the largest subset is used well below its size. To recover part of it, $0.05$ of the $0.2$ \texttt{svp} ratio is routed through a second, separately parameterized \texttt{svp} energy head on the default energy-plus-autograd-force objective, the remaining $0.15$ staying on the directional hinge. That low-weight branch is the only route by which a biomolecular \emph{energy} label reaches the model; its gradients do reach the shared backbone, but every result reported in this work is evaluated through the \texttt{omol25} head. The Stage-3 checkpoints therefore carry two energy heads, \texttt{omol25} and \texttt{svp}, each paired with a parameter-free gradient-force head. All three objectives above train the single \texttt{omol25} head---the directional hinge included, since the $\widehat{\mathbf F}_i$ it scores is that head's autograd force---and the auxiliary head carries the $0.05$ slice alone. Three choices in this section are in fact one: because the two other-functional branches supervise forces alone and the auxiliary \texttt{svp} head absorbs the only biomolecular energy labels, every reported energy is read from the \texttt{omol25} head at the single level of theory it is calibrated against, $\omega$B97M-V/\texttt{def2-TZVPD}---also the level the Cyclosporine~A surface is spot-checked at (\S\ref{sec:simulation_protocols})---so no offset between reference functionals is ever implicitly reconciled inside a predicted energy.

Filtering is also two-level. The static \texttt{def2-TZVPD} source is first sample-filtered to form \texttt{def2-tzvpd-uma-filtered}, removing $20.7\%$ of configurations; training then applies a \emph{per-atom} gate at $\tau=0.8$ (with the configured force-norm gate of 0.05). Together they suppress $\approx$$27\%$ of force labels---not a configuration-discard rate attributable to $\tau=0.8$ alone.

The atom-budgeted loader packs molecules up to a fixed $\texttt{bs\_atom}$ limit of $2{,}048$ atoms per batch, unchanged across all three stages, and an atom-centric neighbourhood representation with dense reductions avoids the memory conflicts of sparse edge operations. Curriculum stages, dataset balancing ratios and atom-centric message construction are described in SI~S2.

\subsubsection{Reproducibility}
\label{sec:reproducibility}

\paragraph{Compute.}
S1 was trained on 32 NVIDIA H20 GPUs (141~GB) and S2 on 64 of the same; only S3 ran on 64 NVIDIA A100 GPUs (40~GB). All three used PyTorch~2.7.0. All three used FP32, without mixed precision, loss scaling or stochastic rounding. The reported S3 checkpoint contains 23{,}505{,}963 parameters (23{,}503{,}618 trainable; $\approx$23.5M), and downstream MD experiments were run on a single NVIDIA H20 (141~GB) per trajectory. As world size times wall-clock, the useful lineage cost $\approx$102 / 403 / 494 GPU-days for S1 / S2 / S3, or $\approx$999 GPU-days in total. We report this figure because the cost that separates a quantum-derived potential from a classical one is incurred once, at training time, and is amortized over subsequent simulations.

\paragraph{Hyperparameters.}
Supplementary Table~\ref{tab:hparams} gives the lineage, one column per stage. Linear warm-up starts at $0.2$ times the peak learning rate rather than zero; cosine decay uses the current step divided by the full configured horizon, including warm-up, and has a floor of $0.01$ times the peak rate. The archived configuration files specify the remaining per-layer settings and are the authoritative record of the four branches, their sampling ratios and their objectives.

\paragraph{Random seeds.}
The reported checkpoints were trained with a single random seed (6666). Downstream MD uses independent seeds for initial velocities wherever a replica design calls for them, and that design differs by scenario: KcsA and 1L2X use \emph{five} replicas per potential, so their uncertainties are seed-to-seed standard deviations, with two-sided Welch $t$-tests between the two five-replica sets for 1L2X; water and the electrolytes are single trajectories with block-averaged uncertainties; and the Cyclosporine~A surface is a single 80-state umbrella campaign with window-level bootstrap, for which \emph{no} RNG seed was set in any window under either potential, so it is statistically but not bit-reproducible. Replica seed sets, and the per-window starting structures that fix the umbrella campaign instead, are listed in \texttt{REPRODUCIBILITY\_}\allowbreak\texttt{CHECKLIST.md} \S D of the reproducibility deposit.

\subsection{Benchmarks and Metrics}
\label{sec:eval_benchmarks}
The microscopic accuracy benchmark of Table~\ref{tab:rel_energy_force} is organized as a ladder of three tiers of increasing system size, constructed so that the first tier lies inside the baselines' own training distribution and the last lies far outside every model's training-size range.

\emph{OMol-Bio-10k} (309--350 atoms) is $10{,}000$ configurations from the largest biomolecular systems of the OMol25 validation split, keeping OMol25's released labels ($\omega$B97M-V/\texttt{def2-TZVPD}). Because this tier shares both the level of theory and the data distribution of the baselines, absolute per-atom energies are directly comparable, and we report energy and force MAE. The \emph{UBio-Mol26 TZVPD held-out} tier (390--909 atoms) is the largest structures of each biomolecular class at our own high-fidelity $\omega$B97M-D3/\texttt{def2-TZVPD} level (mixed basis, as above), excluded from every training stage; a constant offset between our reference functional and the baselines' makes absolute energies incomparable there while leaving forces unaffected, so we report force MAE alone. The \emph{TZVP extreme-size} tier (1{,}215--1{,}555 atoms) probes extrapolation beyond the training-size cap and is computed for evaluation only---no \texttt{def2-TZVP} data enter training at any stage. It comprises five categories at $\omega$B97M-D3/\texttt{def2-TZVP}---$43$ trajectories and $3{,}117$ labelled frames in total: geometry relaxations of solvated AFDB-derived protein ($10$ trajectories, $1{,}010$ frames), 10-mer DNA duplexes ($5$, $226$), 15-mer RNA ($5$, $505$) and lipid systems ($5$, $501$), plus finite-temperature protein MD ($18$ trajectories, $875$ frames). Per-category system sizes, frame counts and elemental coverage are given in Supplementary Table~\ref{tab:extreme_tier}. Being trajectories rather than independent configurations, they admit two offset-free energy measures alongside force MAE---relative energy referenced to each trajectory's first frame, and the frame-to-frame change $\Delta E = E_i - E_{i-1}$, which reports how faithfully a model tracks local PES topology. Metrics are defined below.

\subsubsection{Metrics and baseline settings}
On the held-out and extrapolation tiers above we report force MAE, relative-energy MAE and $\Delta E$ MAE, all normalized per atom for comparability across tiers, with predicted and reference series differenced independently wherever a difference is taken; absolute per-atom energy MAE is reported only on OMol-Bio-10k, where model and reference share a level of theory. Exact definitions are given in SI~S3 (\emph{Metric definitions}). All baselines are evaluated with released weights and their authors' recommended inference settings; MACE-OMol---the MACE architecture~\cite{mace} trained on OMol25~\cite{omol25}, not the organic-molecule MACE-OFF model---is run in float64 as its authors specify, all other models in float32. Evaluation scripts are included in the public code release.

\subsection{Downstream Simulation Protocols}

\label{sec:simulation_protocols}

All downstream molecular dynamics (MD) simulations use UBio-MolFM (Stage~3) as a custom calculator under the ASE~3.27.0/PyTorch~2.7.0 interface (entry point \texttt{interface/ase/cli.py}), with velocity-Verlet integration, conservative-force evaluation $\mathbf{F} = -\nabla_{\mathbf{R}}E$, and---for canonical or isothermal--isobaric ensembles---a Langevin thermostat; per-system timesteps, ensembles, and barostats are specified in each protocol below. Every scenario uses the identical Stage-3 checkpoint through its \texttt{omol25} energy head (\S\ref{sec:training_strategy}), so no per-system or per-fidelity switching occurs anywhere, and no system-specific parameters, restraints or reparameterizations are applied to the ML potential. Every atom---water and ions included---is propagated on that potential; classical force fields appear only in the explicitly labelled pre-equilibration steps and matched classical references. Trajectory analysis uses MDAnalysis~2.10.0~\cite{mdanalysis} with NumPy and SciPy, plus in-house tools. Per-system random-seed assignments are listed in \texttt{REPRODUCIBILITY\_}\allowbreak\texttt{CHECKLIST.md} \S D of the reproducibility deposit.

\paragraph{Analysis conventions shared by all scenarios.}
Every scalar compared between potentials is computed on the equilibrated second half of each production trajectory; distributions and radial functions use the full length. In the five-replica scenarios a quoted scalar is the mean over \emph{per-replica} values rather than a property of the pooled or mean curve, so a peak read off a plotted mean curve can differ from the quoted mean by one bin width. Per-scenario windows, bin widths and tests are given below.

\paragraph{Solvation and ionic solutions (Scenario 1).}
A cubic box of $\approx$25.0~\AA{} is packed with 512 water molecules using Packmol~\cite{packmol} (AmberTools~24.8~\cite{ambertools}) and briefly equilibrated under Amber-FF14SB~\cite{ff14sb}/TIP3P~\cite{tip3p} before switching to UBio-MolFM for production---no classical water model enters thereafter. Three 1-ns trajectories were run at a $0.5$\,fs timestep: $NVT$ (300\,K, Langevin thermostat), $NPT$ (300\,K, 1\,bar) using a Monte-Carlo barostat, and $NVE$ (no thermostat, started from the same classically equilibrated box as the other two rather than from an UBio-MolFM-equilibrated configuration). For the $NPT$ density benchmark the pretrained baselines (UMA-S-1p2, MACE-OMol, and DPA-4 through its \textbf{OMol head}) ran under their authors' recommended stress-based barostats; equilibrium density is a state property and does not depend on the barostat family. DPA-4 is multi-head, and the OMol head is used for it everywhere in this work: it is the head trained on OMol25, so all three baselines and our own Stages~1--2 share one training distribution and the density comparison turns on what a periodic condensed-phase tier adds rather than on which head was selected. Its other heads target materials or water-specific data; selecting the best-suited head per baseline would make the panel uninterpretable, and the price of not doing so is that these runs bound the OMol-trained head rather than DPA-4 as a whole (Extended Data Fig.~\ref{fig:density_baselines_si}). The 0.15\,mol/L NaCl and KCl solutions use a $\approx$50.0~\AA{} box (4033 water molecules and 11 cation/anion pairs each, matching physiological ionic strength) under the same 1-ns $NPT$ protocol. Radial distribution functions use a $0.06$~\AA{} bin width, with coordination integrated to the first RDF minimum. The $NVE$ energy-drift rate is the slope of a linear fit to the per-atom total energy. Because that run begins on a classically equilibrated box it relaxes onto our own surface, and it does so quickly rather than throughout: $3.9$~meV/atom moves out of the potential reservoir and into the kinetic one inside the first $25$~ps, carrying the box from the $299$~K of its starting frame to a stationary $328$~K; that transfer is the same read directly off the kinetic energy or inferred from $\tfrac{3}{2}k_\text{B}\Delta T$. Over the remaining $975$~ps the box is stationary---the temperature rises $0.8$~K, a seventh of its own instantaneous spread of $5.6$~K---while the potential and kinetic terms creep upward at $0.06$ and $0.12$~meV/atom per nanosecond. Their \emph{sum} is the drift, and it is recovered two ways that agree to $1\%$: $0.179$~meV/atom/ns on the stationary stretch alone against $0.181$ from a linear fit to the total over the whole run. The transient therefore does not manufacture it, and we quote the whole-run figure, $1.81\times10^{-4}$~eV/atom/ns. Separating the transient from the stationary stretch makes the conservation test more stringent rather than less---on the stationary stretch energy is still moving between the reservoirs while the total holds flat to $0.4\%$ of the kinetic energy per nanosecond---and the transient is why the kinetic trace of Fig.~\ref{fig:thermo}b rises early. Nothing reported from this trajectory other than the drift depends on its temperature; the density, structure and diffusion all come from the $NPT$ run.

\textit{Self-diffusion.} The water self-diffusion coefficient is taken from the pure-water $NPT$ trajectory (20{,}001 frames at $50$~fs, whose mean temperature is $298.6$~K) as the slope of the mean-square displacement of the $512$ molecular centres of mass against $6t$. Displacements are unwrapped in fractional coordinates against the \emph{instantaneous} cell, so box breathing does not enter them; whole-box centre-of-mass drift is removed and every time origin used. The fit window is $10$--$100$~ps, where the MSD log--log slope is $0.970$ and the motion is therefore diffusive; the value is stable to $\pm0.03\times10^{-5}$~cm$^2$\,s$^{-1}$ across windows from $5$--$50$ to $20$--$200$~ps, with a standard error of $\pm0.04$ over five independent $200$-ps blocks. Because a $24.94$~\AA{} box is small enough for the periodic hydrodynamic self-interaction to matter, we apply the Yeh--Hummer correction $D_\infty = D_\text{PBC} + \xi k_\text{B}T/(6\pi\eta L)$ with $\xi = 2.837297$ for a cubic cell~\cite{yeh2004}, evaluated with the experimental shear viscosity of water at $298$~K ($0.890$~mPa\,s) because the model's own viscosity was not computed. The correction adds $0.28\times10^{-5}$~cm$^2$\,s$^{-1}$ to an uncorrected $1.60$; both are reported, and since it far exceeds the statistical uncertainty the viscosity dominates the error budget. At this box size the increment is half what a 64-molecule \emph{ab initio} box requires ($+0.56$ at $L=12.42$~\AA), so the comparison against corrected literature values is not inflated by our own correction.

\paragraph{Mg$^{2+}$ coordination in the BWYV RNA pseudoknot 1L2X (Scenario 2).}
\textit{System.} We simulated the 27 standard nucleotides of the BWYV pseudoknot (PDB 1L2X~\cite{egli2002metal}, chain~A residues 2--28; \texttt{GCGCGGCACCGUCCGCGGAACAAACGG}; 582 RNA heavy atoms), renumbered 1--27 throughout, so a crystal residue index equals ours plus one. The 5$'$-terminal GTP (crystal residue~1) is omitted because Amber~OL3 templates no triphosphate cap. Of the six crystallographic Mg$^{2+}$ we retained only the inner-sphere one on the phosphate backbone (crystal residue~29, on the OP1 of crystal G2 $=$ our G1) with its five inner-shell crystallographic waters; the five surface-bound ions are not modelled. The system is neutralized with 25 Na$^+$ and solvated in a 12~\AA{} TIP3P box (7{,}300 waters; 22{,}804 atoms), initially $71.6\times63.9\times64.9$~\AA.

\textit{Shared pre-equilibration, and the two restraints it used.} All ten replicas start from a single configuration equilibrated in OpenMM~8.5 under the classical setup of (ii) below: energy minimization ($\leq$$5{,}000$ iterations), $100$~ps of $NVT$ at $300$~K, then $500$~ps of $NPT$ at $300$~K and $1$~bar, with a LangevinMiddle integrator at $1.0$~ps$^{-1}$ friction, PME at a $10$~\AA{} real-space cutoff, H-bond constraints and rigid water permitting $dt = 2$~fs, and the barostat applied every $25$ steps. The box contracts to $65.7\times58.6\times59.6$~\AA{} at $1.020\,\text{g\,cm}^{-3}$. Two restraints act during equilibration: a harmonic Mg$^{2+}$--OP1 bond ($k = 100$~kcal/mol/\AA$^2$ at $r_0 = 1.957$~\AA, taken from the input coordinates so it neither pushes nor pulls at $t=0$), because fixed-charge divalent parameters do not reliably hold an inner-sphere Mg--phosphate contact and the ion would relocate to an outer-sphere position before production began; and position restraints on the 582 RNA heavy atoms ($k = 10$~kcal/mol/\AA$^2$). Both are released in full for production under both potentials, so the inner-sphere coordination reported below is each potential's own unbiased result rather than an imposed constraint.

That single configuration is the \emph{common} start for both potentials, so the benchmark measures drift of the folded, Mg-bound state from an identical origin. Two properties of that origin follow from the restraints and apply equally to both sides. The RNA position restraint was held at full strength through the $NPT$ stage, so the starting fold sits close to the crystal geometry and has been freely relaxed by neither potential; part of the RMSD each accumulates is therefore relaxation away from a crystal-biased structure rather than degradation of the fold, which is why we report RMSD as drift from the common origin and claim nothing about proximity to the crystal. And while that fold is crystal-derived rather than classically relaxed, the surrounding solvent and the box dimensions are what the classical setup produced, so what residual preparation bias exists favours the classical baseline and not us.

\textit{Two potentials, only the PES differs.} (i)~UBio-MolFM (Stage~3), with all atoms---water included---on the ML potential. (ii)~A classical reference: Amber~OL3~\cite{zgarbova2011ol3} for RNA, TIP3P water, and the Li--Merz \emph{12-6-4} divalent-ion model~\cite{li2014ion} for Mg$^{2+}$ (the C4 polarization term active; a gold-standard, ion-specific Mg$^{2+}$--nucleic-acid parameterization), run in OpenMM~8.5. Both are unbiased (no restraints) $NPT$ at 300~K, 1~bar, $dt = 0.5$~fs. Each potential is run as \textbf{five independent replicas} of $1.0$~ns from that common configuration, differing only in the seed driving thermostat noise and initial velocities: $42$, $84$, $126$, $168$, $210$ for UBio-MolFM (the default and successive multiples of it) and $42$, $1$, $2$, $3$, $4$ classically. Seeds are recorded per replica in the released data.

\textit{Analysis.} Heavy-atom RMSD and radius of gyration are computed per frame after Kabsch superposition on the 582 RNA heavy atoms, referenced to the common start. Coordination number (O atoms within 2.8~\AA), the Mg$^{2+}$--nearest-phosphate-oxygen distance (inner-sphere threshold 2.6~\AA) and the Mg$^{2+}$--O$_\text{water}$ radial distribution function are computed over the trajectory. Coordination geometry is characterized by three probability densities over the coordinating oxygens (within 2.8~\AA): the O--Mg--O angle (octahedral \emph{cis}/\emph{trans} symmetry), the Mg$^{2+}$--O$_\text{P}$--P angle (monodentate binding pose) and the Mg$^{2+}$--O$_\text{P}$ distance, each normalized to unit area with peaks reported as histogram modes. Every comparison between potentials in this scenario is a two-sided Welch unequal-variance $t$-test on the per-replica values, $n=5$ per potential, with Cohen's $d$ tabulated beside it in the deposit; this is the test behind every $p$ quoted for this system, in the text and in Fig.~\ref{fig:rna_mg_analysis}. Canonical Watson--Crick occupancy counts the eight stem pairs with $\ge$2 base--base hydrogen bonds ($<3.4$~\AA) per frame; of the twelve base--base contacts detected, the other four are geometric near-contacts or the labile C7--G11 pair, open in every replica of both potentials, so the all-contact mean ($\approx$$0.76$) is not a measure of stem integrity and is not reported. Scalars are recomputed on the second half of each trajectory ($0.5$--$1.0$~ns), so the comparison is made after the shared starting structure has been forgotten, and are reported as five-replica means $\pm$ seed-to-seed standard deviation with a two-sided Welch $t$-test ($n=5$ versus $n=5$). Two statistics are deliberately \emph{not} treated as results. The Mg$^{2+}$--O$_\text{P}$--P mode does separate the two potentials at $1.0$~ns ($147.1\pm10.3^\circ$ against $165.5\pm2.5^\circ$, $p=0.015$), but it is not converged and we do not build on it: individual replicas move by $10$--$26^\circ$ when the window is extended from $0.8$ to $1.0$~ns (one shifts $164.5\to138.5^\circ$), the five-replica scatter is four times the classical, and the mean lies above the $120$--$140^\circ$ reported for crystallographic monodentate binding, so we quote the width and decline the geometric interpretation. Nor is any single replica quoted alone: seed~42, common to both sets, is the highest-RMSD of the five classical runs and among the highest of ours, so it represents neither.

\paragraph{Cyclosporine A free-energy landscape (Scenario 3).}
\textit{System.} The closed crystallographic conformer (CSD refcode DEKSAN~\cite{loosli1985csa}) is solvated in a $39.2$~\AA{} cubic box of $1{,}932$ water molecules (5{,}992 atoms total, net charge $0$). Every atom, water included, is propagated on the UBio-MolFM potential---no classical water model is used. Dynamics are $NVT$ at $300$~K (Torch Langevin/BAOAB, friction $10^{-3}$~fs$^{-1}$, $dt = 0.5$~fs, centre-of-mass fixed).

\textit{Collective variables.} Residues are numbered by the canonical cyclosporin~A sequence, MeBmt1--Abu2--Sar3--MeLeu4--Val5--MeLeu6--Ala7--\textsc{d}-Ala8--MeLeu9--MeLeu10--MeVal11; Abu2 and Val5 are the macrocycle's only $\alpha$-aminobutyryl and valyl residues and carry two of its four free backbone N--H donors. (In the deposited topology and analysis scripts these two residues appear as \texttt{ABA} and \texttt{VAL} with internal residue IDs 6 and 9, offset $+500$ in PDB 1IKF chain~C.) $\text{CV}_1 = d(\text{C{=}O}_{\text{Abu2}},\text{C{=}O}_{\text{Val5}})$ is the transannular backbone carbonyl--carbonyl distance (0-based atom indices $105,150$); the latch coordinate $\text{CV}_2 = \text{HB}_1 = d(\text{N}_{\text{Abu2}},\text{O}_{\text{Val5}})$ is the Abu2~N--H$\cdots$O=C~Val5 heavy-atom distance (indices $103,151$), and $\text{HB}_2$ is its reciprocal Val5~N--H$\cdots$O=C~Abu2. Harmonic biases use $k_1 = 1.0$~eV/\AA$^2$ ($23.06$~kcal/mol/\AA$^2$) on $\text{CV}_1$ and $k_2 = 0.45$~eV/\AA$^2$ on $\text{CV}_2$ (2D windows only; $\text{HB}_1$ recorded unbiased in the 1D windows).

\textit{Enhanced sampling and estimator.} The reaction coordinate is tiled by $24$ forward and $24$ reverse one-dimensional windows (centres $\text{CV}_1 = 4.7\to9.3$~\AA, spacing $0.2$~\AA) plus $32$ localized two-dimensional patch cells on $(\text{CV}_1,\text{HB}_1)$ over the latch region---$80$ biased states in total, seeded from steered closed$\leftrightarrow$open pulls and, for the patches, from the nearest $(\text{CV}_1,\text{HB}_1)$ frame mined across both 1D directions. Each potential is seeded from its own steered trajectory, never from the other's. The exact starting structure of every window is deposited for both potentials---$79$ distinct structures for the $80$ classical states, the two 2D cells at $\text{CV}_1 = 5.7$~\AA{} with $\text{HB}_1$ targets $3.8$ and $4.1$~\AA{} having been seeded from the same mined frame, which sat $0.30$~\AA{} from the former's target and exactly on the latter's; the ML side shares none. Their restraints differ, so MBAR treats them as distinct states, and the only cost is that those two cells' initial conditions are correlated where the other thirty are independent. \emph{No} RNG seed was set in any window, so the campaign is reproducible statistically but not bit-for-bit, and the deposited structures are what pin it down. Production is $400.8$~ps per state (median 1{,}600 frames at $0.25$~ps/frame, $0.5$~fs step) for 1D windows and 2D patches alike, four of the eighty running $1$--$2$~ps short of that; MBAR weights unequal per-state sample counts correctly, so no reported number depends on the difference. The first $20\%$ of each is discarded, and the convergence test of Fig.~\ref{fig:csa_pmf}g splits the retained $320.6$~ps in half by frame index. Overlap between states adjacent \emph{in the collective-variable lattice} is at worst $0.080$ within each 1D ladder and $0.116$ between adjacent patch cells; consecutive matrix indices that straddle two ensembles are $\text{CV}_1$-distant and their vanishing overlap is not a gap in the ladder (Extended Data Fig.~\ref{fig:csa_diagnostics_si}). A single global two-dimensional histogram-MBAR (\texttt{pymbar}~4.2.1~\cite{shirts2008mbar}) over all $80$ states yields $F(\text{CV}_1,\text{CV}_2)$; $F(\text{CV}_1)$ follows by marginalizing $\text{HB}_1$. The bands accompanying every free energy are the spread of $20$ window-level bootstrap resamples, and we quote them as a range rather than as a confidence interval because at $B=20$ they cannot be a calibrated one: the $2.5$th percentile interpolates between the two smallest replicates, so the band is in practice the min--max of the draws. Its coverage is unknown and its width is a statement about window-to-window spread. Two further limits: replicates whose own minimum lands on a column the $\min$-count mask removes are dropped silently, so $19$ of the $44$ $\text{CV}_1$ columns are built from fewer than $20$ values, three from fewer than ten and one from none---all of them at the sparsely sampled ends, none at a stationary point we quote---and no uncertainty anywhere in this campaign is derived from the effective sample count. What the comparison rests on is that the ranges do not overlap at either stationary point, which under this reading is the stronger statement: no single resample of either campaign reached the other. Sampling-connectivity diagnostics are in Extended Data Fig.~\ref{fig:csa_diagnostics_si}. Two properties of the analysis grid bound what may be read off it. First, the $\min$-count mask is applied to the one-dimensional marginal only, not to the two-dimensional histogram it sums over, so the surface itself rests in part on sparsely populated cells: $91\%$ of our finite two-dimensional bins and $57\%$ of the classical ones hold fewer samples than that threshold, some only one. Masking them would change $F(\text{CV}_1)$ rather than clean it, so we measured what they are worth instead---dropping every bin below five samples moves our curve by $0.004$~kcal/mol in shape and our span by $0.003$, and moves the classical \emph{span} by $0.009$, but moves the classical \emph{shape} by $1.2$ and its curve outside $4.7$--$9.0$~\AA{} by up to $1.8$. We therefore quote the classical curve only inside that window, which is where both of its extremes lie, and nothing about its shape (\S\ref{subsec:csa_thermodynamics}). Second, the grid stops at $\text{HB}_1 = 10$~\AA{} while the classical latch opens past it: $0.6\%$ of classical samples ($1{,}348$ of $224{,}480$, $1{,}209$ from one window) and $0.009\%$ of ours fall outside it. Those samples still enter the MBAR solve and the state free energies; they populate no histogram bin, so they are absent from the surface alone. The two-dimensional cells exist because $\text{CV}_1$ is degenerate against $\text{HB}_1$: one carbonyl--carbonyl distance admits both latched and unlatched microstates, so a one-dimensional ladder in either direction alone misestimates the depth of the closed basin and the forward and reverse estimates disagree systematically---a naive 1D PMF is not reproducible. Lifting that degeneracy reconciles the two directions to $0.83\,\text{kcal/mol}$ at worst, with the residual confined to the sparsely sampled basin and wall ends and contained by the bootstrap interval throughout (Fig.~\ref{fig:csa_pmf}e,g).

\textit{Classical-force-field baseline.} To isolate the role of the potential, the identical solvated box is \emph{reparametrized} (not re-solvated) and re-run classically. The baseline is deliberately an \emph{automatically parameterized} one: the 196-atom cyclic peptide---seven of its eleven residues $N$-methylated, and built from non-standard residues (MeBmt, Abu, Sar, \emph{N}-methyl-Leu/Val, \textsc{d}-Ala) that ff14SB does not template---is assigned \textbf{GAFF2} atom types with \textbf{AM1-BCC} charges (AmberTools~24.8 \texttt{antechamber}, single-point charges on the MD geometry to avoid gas-phase macrocycle collapse; \texttt{parmchk2} for missing terms) and \textbf{TIP3P} water. This is what a practitioner obtains for such a molecule without bespoke parameterization, and it is the condition our comparison is about.

Every parameter \texttt{parmchk2} supplied is torsional---the mass, bond, angle and van der Waals sections of its output are empty---and its ten proper dihedrals all sit on the four residues carrying a free backbone N--H, analogized at zero penalty from the tertiary-amide nitrogen type. Two of those four are the $\text{HB}_1$ and $\text{HB}_2$ donors, so the torsions about the latch coordinate are among the borrowed ones; we record this rather than leave it to be found. The full 30-line file and a per-term account are given in Supplementary Table~\ref{tab:csa_frcmod}. More consequentially, general small-molecule torsions carry no correction specific to $N$-methylated peptide backbones---a term documented to be what prevents classical models from localizing CsA between its open and closed geometries, and whose construction for this molecule required fitting $\varphi/\psi$ maps to MP2 dipeptide scans~\cite{yamane2022csaff}. That molecule-specific CHARMM force field is public, and we did not adopt it for reasons of scope rather than cost: hand-built for this one molecule and validated against its NMR spectra, it would answer how a bespoke, expert-parameterized potential performs---a different and equally legitimate question. We make no claim about what it would give, and the main-text conclusion is scoped to automatically parameterized models accordingly.

\textit{Classical protocol and controls.} We use OpenMM~8.5 (\texttt{LangevinMiddle\hspace{0pt}Integrator}, $300$~K, friction $1.0$~ps$^{-1}$, $dt = 2$~fs, $NVT$, PME, $1.0$~nm real-space cutoff, H-bond constraints). The same CV atoms, spring constants, window layout ($24+24$ 1D $+\,32$ 2D), production length ($400.8$~ps per state, $4{,}008$ recorded frames; the 2D patch alone contributes $89{,}792$ samples) and global 2D-MBAR analysis are applied, so the two differ \emph{only} in the potential-energy surface and in two protocol details that both favour the classical side---it discards the first $30\%$ of each window rather than $20\%$, and it is allowed the longer timestep and rigid X--H bonds controlled for below. Lattice-neighbour overlap is comparable ($0.109$ 1D, $0.124$ 2D). The effective sample count is not: the classical campaign reaches $N_\text{eff} = 14{,}244$ against our $7{,}105$, twice ours. We state that plainly rather than call the two comparable, because it means the reference is the better-sampled of the pair and nothing in this comparison is tilted our way by sampling density. $\text{HB}_2$, absent from the classical bias files, is recomputed from the saved trajectories at $1$~ps spacing; the measured statistical inefficiency is $\approx$$2.3$~ps so this costs nothing, and reprocessing $\text{HB}_1$ the same way reproduces its $0.1$-ps-sampled PMF to $0.039$~kcal/mol MAE.

\textit{What the classical surface does and does not support.} The classical estimate is the less converged of the two: its production halves differ by up to $1.93$~kcal/mol (at $\text{CV}_1 = 5.88$~\AA) and its forward and reverse 1D estimates by $1.81$, against $0.83$ for UBio-MolFM. Its \emph{flatness} is nonetheless robust to the split---each half independently spans at most $2.1$~kcal/mol over $4.7$--$9.1$~\AA{} against $5.51$ for the quantum surface---so the main-text comparison rests on the absence of a funnel, not the shape of the residual noise. The position of the classical minimum is correspondingly not a result: bin-to-bin scatter exceeds the depth of any candidate well, the bootstrap interval at the lowest bin excludes zero, and the 1D and 2D estimates place that bin $2.6$~\AA{} apart.

Because the classical runs use rigid X--H bonds at $dt = 2$~fs where ours are unconstrained at $0.5$~fs, the \emph{entire} 80-state campaign was additionally re-run at $0.5$~fs without X--H constraints. The two surfaces differ by less than either campaign's own half-to-half drift, so that difference decides nothing; the paired three-window test bounds the protocol effect instead, and both protocols agree that the classical landscape is $3$--$4\times$ flatter than ours. SI~S4.3 gives every number and the reasoning. We report the constrained $2$-fs campaign throughout because it is standard classical practice, \emph{not} because it is uniformly conservative. Neither protocol bears on the potential-energy-surface comparison below, computed on fixed geometries with no integrator, constraints or solvent at all. One difference is irreducible: TIP3P is rigid under both classical protocols because it is parameterized that way, whereas every water molecule in our simulations is flexible.

\textit{Quantum-chemical spot-check of the model's own surface.} One hundred conformers were drawn from the production windows at a spacing exceeding the measured correlation time, covering the closed basin, the open minimum, the region between and the steric wall, and spanning $101\,\text{kcal/mol}$. Each was recomputed as a gas-phase single point on the $196$-atom solute at three levels: $\omega$B97M-V/\texttt{def2-TZVPD} (the OMol25 level, on which the model's single shared energy head is calibrated); $\omega$B97M-D3(BJ) with the mixed basis, grid level 3 and $10^{-6}$ SCF threshold used to collect UBio-Mol26's own high-fidelity tier; and revDSD-PBEP86-D4/\texttt{def2-TZVPD} as an independent high-accuracy reference. Convergence was reached for $99$, $93$ and $83$ of the $100$ respectively; all comparisons are on relative energies after removing each series' mean, and confidence intervals are $4000$-sample bootstraps resampling the closed and open groups independently. The model reproduces its calibration level to $0.44\,\text{kcal/mol}$---closer than the three references stand to each other---and the $\approx$$3\,\text{kcal/mol}$ disagreement among those references falls cleanly on the nonlocal-versus-pairwise dispersion divide rather than on the model. Per-region errors and the interval arithmetic behind that attribution are in SI~S4.3 and Extended Data Table~\ref{tab:csa_qm}.

\textit{The classical potential-energy surface, tested directly against DFT.} The flatness of the classical free-energy profile could in principle come from sampling, solvation, the water model, the integrator or the constraints rather than the potential, so we evaluated the same classical Hamiltonian as single points on the 100 fixed conformers of the DFT spot-check, removing all of those at once. Both sides are gas-phase solute-only (196 atoms, no cutoff, no constraints), matching how the DFT references were computed; the solute is sliced from the production topology with ParmEd so its parameters are byte-identical to those the umbrella sampling used, and atom ordering was verified element-by-element against the frame files. Relative energies, shift-corrected as elsewhere, depart from the three DFT references by $9.63$--$9.90$~kcal/mol at Pearson $r=0.79$--$0.80$, against $0.44$--$1.45$ and $r\geq0.995$ for UBio-MolFM on the same geometries (Extended Data Table~\ref{tab:csa_qm}). The classical energy spread over the set, $107$~kcal/mol, \emph{exceeds} the DFT spread of $101$, so the error is misordering rather than compression. Two checks bound the objection that the AM1-BCC charges, fitted on the single closed crystal conformer, under-weight the transannular hydrogen bonds: the error is worst at the \emph{open} minimum rather than in the closed basin, and re-deriving the charges from an open conformer moves relative energies by at most $12\%$ of the discrepancy. Both are quantified in SI~S4.3, and a multi-conformer charge fit cannot close a $10$~kcal/mol gap.

\paragraph{KcsA transmembrane channel (Scenario 4).}
\textit{System construction.} The starting structure derives from PDB 1K4C~\cite{zhou2001kcsa} (KcsA tetramer), prepared in a deliberately ion-loaded state with five K$^+$ in the pore---four at the canonical filter sites S1--S4 and one in the central cavity (S$_\text{cav}$). Construction ran in two stages, and the bilayer analysed here is the second. First, the tetramer was placed in a coarse-grained membrane with \texttt{insane} under the MARTINI model and back-mapped to all atoms with \texttt{cg2at} (both taken from their GitHub \texttt{main} branches), which fixes the transmembrane pose; that stage used DPPC, so the protein coordinates entering the second stage were relaxed in a different lipid from the one simulated. Second, the POPE:POPG (3:1) bilayer reported here was built around that pose: protonation states were assigned with GROMACS~2025.4 \texttt{gmx pdb2gmx} under CHARMM36 ---protonation only, no CHARMM parameter enters any simulation---and lipid hydrogens with Open~Babel~3.1.1; \texttt{packmol-memgen} packed the bilayer around the preoriented channel, solvated it and added salt, with Packmol filling the solvent shell; and \texttt{tleap} applied the force fields (\texttt{leaprc.protein.ff14SB} $+$ \texttt{leaprc.lipid21} $+$ \texttt{leaprc.water.tip3p}), set the box, neutralized it and wrote the topology (all three tools from AmberTools~24.8). The result is brought to 0.15\,mol/L KCl, yielding a final system of $\approx 108{,}964$ atoms (352 lipids: 264 POPE and 88 POPG; 19{,}440 water molecules; 130 K$^+$ and 54 Cl$^-$, with the K$^+$ excess over Cl$^-$ neutralizing the anionic POPG headgroups and the protein).

\textit{Equilibration.} An initial 500\,ps equilibration under Amber-FF14SB/Lipid21~\cite{dickson2022lipid21}/TIP3P brings the bilayer to physiological area-per-lipid at 310\,K and 1\,atm under semi-isotropic $NPT$. This scenario runs at 310\,K rather than the 300\,K used elsewhere because the bilayer is POPE-rich: POPE's main gel-to-fluid transition lies near 298\,K, so at 300\,K the membrane would sit only marginally inside its fluid $L_\alpha$ phase and could retain gel-like domains over sub-nanosecond trajectories, whereas 310\,K places it $\approx$12\,K above. All three potentials use the same temperature, so their comparison is unaffected, and the measured chain order (Fig.~\ref{fig:transmembrane_channel}b) with the gauche fraction and covalent geometry below confirm a fluid $L_\alpha$ bilayer in every replica of all three. The final equilibrated configuration is then switched to UBio-MolFM~(S3) for a further 500 steps of energy minimization only before production.

\textit{Production.} From the equilibrated configuration we ran five independent UBio-MolFM~(S3) production replicas under $NPT$ at 310\,K and 1\,atm, using Velocity-Verlet integration ($dt = 0.5$\,fs) with a Langevin thermostat at a friction of 1\,ps$^{-1}$. Each runs $1.0$\,ns (seeds 42, 84, 126, 168 and 210), matching the classical references in length. The pressure is controlled by a Monte-Carlo barostat that scales all three box vectors by the same factor, so the box aspect ratio is fixed at its equilibrated value; this is \emph{isotropic}, not semi-isotropic, coupling. The replicas are seeded by independent initial velocities, and because the $500$-step minimization above was run per replica rather than once, their starting \emph{positions} also differ slightly---by up to $0.07$~\AA{} on the widest filter spacing, two orders of magnitude below the divergence the comparison reports. As matched classical references, the same system was run from the equilibrated configuration \emph{before} that minimization, since it is defined on our potential and not theirs; the offset this leaves at the first frame is $0.14$~\AA{} on the S$_\text{cav}$--S4 spacing, which is enough to matter only because the contiguity criterion falls inside it: the classical start sits at $4.08$~\AA{} and ours at $3.94$, on opposite sides of the $4$~\AA{} line. It therefore accounts for the classical column being absent \emph{at frame zero} and for nothing after: within $10$~ps every classical replica has opened S3--S2 to $5.6$--$6.2$~\AA, and over the second half the two classical families average $9.7$ and $9.2$~\AA{} (Fig.~\ref{fig:transmembrane_channel}e), one to two orders of magnitude beyond the offset. Final-frame values are not quoted: in one 12-6 replica the ion ordering reverses, so the signed spacing changes sign and no common range exists. The excess ion's departure time is the first drop below five on-axis ions that \emph{persists} for $10$~ps, not the first frame below five: several replicas dip below five a dozen or more times before the column settles, so a first-frame criterion reports where the sampling grid landed rather than an event. The choice changes four of the fifteen replicas and no others, by $22$ to $58$~ps, and moves the ensemble range from $2$--$68$ to $2$--$74$~ps. Individual departure times are not well conditioned beyond that: each replica's value is exported at hold thresholds from $0$ to $50$~ps, and on seven of the fifteen it moves further as the threshold lengthens, the ensemble range reaching $2$--$100$~ps at $20$~ps and $2$--$176$~ps at $50$~ps. The range quoted in the text is the $10$~ps one and no claim rests on any single replica's value; what survives every threshold is that all fifteen lose the excess ion early and from S1 under all three potentials. The comparison is run in OpenMM~8.5 (\texttt{LangevinMiddleIntegrator}, $dt = 0.5$\,fs) under two fixed-charge treatments---the induced-dipole-corrected Li--Merz 12-6-4 model~\cite{li2014ion} and the plain 12-6 Lennard-Jones model with Joung--Cheatham monovalent-ion parameters~\cite{joung2008jc}---five replicas of $1.0$\,ns each, under \emph{semi-isotropic} $NPT$ in which the membrane normal couples independently of the bilayer plane. All three potentials are therefore compared at equal $n=5$ and equal length, on $\approx$500 frames each; an earlier 12-6 run that carries no recorded RNG seed, and is therefore not bit-reproducible, is excluded from the quantitative statistics. One 12-6-4 replica was launched before the driver accepted an explicit seed and carries none, so that trajectory is not bit-reproducible from its inputs; it is retained as one of the five, and it is also the replica whose cavity ion descends into carbonyl contact, so the weakest claim below rests on the least reproducible run.

\textit{Stationarity at the final frame.} Whether either ensemble is still relaxing when its trajectory ends was tested on the closest K$^+$--K$^+$ distance, the observable contiguity is defined on, by least squares over $52$--$1000$~ps with the uncertainty from $20$ contiguous blocks. All five 12-6-4 replicas are still spreading, $+0.29$ to $+0.54$~\AA/ns, every one significant; of ours, the four that hold the column carry no resolvable trend ($+0.001$ to $+0.046$~\AA/ns, none significant) and the fifth is seed 126 ($+0.63\pm0.21$~\AA/ns), characterized in \S\ref{subsec:transmembrane_channel}. One nanosecond bounds the resolvable rate rather than excluding a slower one, and the bound is what the comparison needs: the pooled classical spacings describe a filter better than the one those trajectories end with, so the length match understates the separation rather than inflating it. Occupancy and filter water, by contrast, are no longer measurably changing at the final frame in either ensemble: the classical runs have settled into their degraded state rather than still descending into it, and the trend test on those two observables has no power to say otherwise, its blocked interval on a $0$--$1$ indicator admitting any rate.

\textit{The cavity site is a kinetic difference, not a structural one.} The cavity ion's distance to its nearest filter carbonyl oxygen is the one filter observable that is not stationary on the classical side, and it must be read with its window. Resolved in $250$~ps windows over $50$--$1000$~ps (two UBio-MolFM replicas, seeds 42 and 84, against all five 12-6-4; $95\%$ intervals from five block means per run per window), the reference-minus-model separation falls from $1.65\pm0.06$~\AA{} at $50$--$300$~ps to $1.08\pm0.25$~\AA{} at $800$--$1000$~ps, a first-versus-last contrast of $-0.57\pm0.26$~\AA. Essentially all of that movement is on the classical side---$4.46$ to $3.90$~\AA{} against our $2.81$ to $2.82$~\AA---so the drift cannot manufacture the difference; what it means is that the reference is still relaxing at $1$~ns while we are not. Per replica it means more. One 12-6-4 run descends from $4.52$ to $2.89$~\AA---within $0.06$~\AA{} of our seed 84 and $0.07$ of our two-replica mean: the fixed-charge reference \emph{can} reach the carbonyl-contacting cavity state, and in four of five replicas does not do so within a nanosecond. We therefore claim only that it mostly does not reach that state on this timescale, and make no claim that the state is classically inaccessible. For the same reason no cavity-site separation is quoted without its window, and none is presented as an equilibrium value: a linear extrapolation closes the late-window gap at $\approx$$2.6$~ns, which bounds how fast it could close rather than predicting that it does, and fixing the equilibrium separation would need a classical replica several times longer than these.

\textit{Released-box control.} Because the isotropic coupling is a confound, three further UBio-MolFM replicas (seeds 124, 168, 210) were run under the same \emph{semi-isotropic} barostat as the classical references, to their design length of $400$~ps each. Seeds 168 and 210 pair with their isotropic namesakes; 124 is an independent seed with no isotropic counterpart, so two of the three are matched and one is not. They reproduce the filter observables: at least four ions on axis in $100\%$ of frames, no filter water in any frame, closest K$^+$--K$^+$ $3.37\pm0.21$~\AA, and the contiguous contact column in $86\%$ of late frames in two of the three. The third (seed 124) registers $0\%$ on the $4$~\AA{} criterion without the column ever opening: from $\approx$$140$~ps its S4--S3 pair settles at $4.7$--$5.1$~\AA{} and stays there, so its widest spacing reaches only $5.77$~\AA{} (95th percentile $5.27$), \emph{every} late frame holds all three below $6$~\AA{} and $80\%$ below $5$~\AA, with four or five ions on axis and no filter water throughout. Seed 126---the isotropic replica that genuinely translocates---opens to $16.05$~\AA{} and stays under $5$~\AA{} in $39\%$ of frames and under $6$~\AA{} in $43\%$. What differs between those two is the criterion, not the column, and the filter conclusions therefore do not depend on the aspect-ratio constraint.

These runs are also long enough to report on the bilayer, and they do not favour our potential. The thickness, area-per-lipid and chain-order values they give on the matched $200$--$400$~ps window are reported in the text and in Extended Data Table~\ref{tab:validation_summary}; against the classical reference the chain-order deficit is $14\%$ released and $10\%$ isotropic on each set's own late half, or $8\%$ isotropic on the matched window. Neither thickness nor area has an onset or a plateau---thickness declines from the first frame under both couplings and both are still moving monotonically when the runs end---so with $n=3$ these runs establish a direction and a lower bound on the deficit, not a value, and the isotropic figures quoted elsewhere are the conservative ones.

\textit{Electrostatic-truncation control.} The hybrid cutoff reaches $8$~\AA{} and carries no Ewald sum, so a reader may ask whether truncated electrostatics is what brings our ions closer together. We test it directly, on the trajectories themselves: for each stored frame we evaluate the electrostatic force on the filter K$^+$ ions twice, once with particle-mesh Ewald and once with the same interaction summed under a hard $8$~\AA{} cutoff, and take the difference $\Delta\mathbf{F}_\text{LR}$. This quantity is a property of the configuration and the cutoff radius alone, not of the potential that generated the frame, so it is computed identically on the classical and UBio-MolFM trajectories---which is what makes the comparison symmetric. Only the \emph{adjacent-pair differential} $(\Delta\mathbf{F}_\text{LR}^{i+1}-\Delta\mathbf{F}_\text{LR}^{i})\cdot\hat{\mathbf z}$ is interpretable: a long-range field is nearly common-mode across a column of ions on the pore axis, and a common mode translates the column without changing any spacing. Positive means expansive---restoring the missing term would widen that pair, so truncation would have been compressing it. Fourteen of the eighteen trajectories tested are net \emph{compressive} at their stiffest pair, including every UBio-MolFM replica that holds its contact column under either barostat. Restoring the missing long-range term would therefore tighten the contact pair, not loosen it, so truncation cannot be why that pair sits closer. It bounds no other pair, and at the cavity pair the sign reverses. The test is one-directional: it excludes an explanation, and does not establish that the spacings themselves are correct, which would need an \emph{ab initio} or QM/MM reference. Per-run values, the runs too soft to bound, the pair whose sign runs the other way and the reasons the single-ion quantities are not interpretable are in SI~S4.4 and Supplementary Table~\ref{tab:lr_electrostatics}.

Three things differ between the UBio-MolFM and classical runs: the potential-energy surface, the barostat coupling, and the $500$-step minimization described above, which only the UBio-MolFM side received. The first is the comparison. The second is a confound and the next paragraph bounds what it touches. The third is bounded above by its own magnitude, quantified with the production protocol: it displaces the first frame by $0.14$~\AA{} on one spacing, putting the classical start at $4.08$~\AA{} and ours at $3.94$ on opposite sides of the $4$~\AA{} contiguity criterion, and is one to two orders of magnitude smaller than the divergence that follows, so it can account for the classical column being absent at frame zero and for nothing later. Between the \emph{two classical} treatments nothing differs but the ion-induced-dipole (C4) term, so that contrast is clean. Throughput on the full $108{,}964$-atom system is $\approx$$0.24$~steps/s on one H20 under the production settings above, which use activation recompute; this is the water sweep of Extended Data Fig.~\ref{fig:throughput} interpolated to this system size ($4.2$~s per force evaluation at a $79.8$~GB peak), not a separate wall-clock measurement of these trajectories. We quote a single figure for this system rather than tabulating baselines beside it, because MACE-OMol does not run at this size at all and the comparison against UMA-S-1p2 is made properly---swept over system size rather than at a single point---in that figure.

\textit{What the barostat difference does and does not affect.} Because the UBio-MolFM box cannot change shape, the bilayer cannot relax area and thickness independently: with the lipid count fixed, the Voronoi area per lipid is set by the cross-sectional box area and held near its input value rather than sampled. The signature is direct---area per lipid ranges over $0.25$~\AA$^2$ within an isotropic UBio-MolFM trajectory against $0.95$--$1.05$~\AA$^2$ classically, a fourfold \emph{narrower} range, and over $5.4$~\AA$^2$ when the same potential runs with the normal free. Every collective membrane observable is consequently confounded on the isotropic side---area per lipid, hydrophobic thickness, chain tilt and hydrophobic-core hydration all measure the constraint as well as the potential---so none is quoted from those runs alone; what bounds them is the released-box control below, not the isotropic trajectories (\S\ref{subsec:transmembrane_channel}). Observables independent of the box aspect ratio are unaffected, and these are what the results rest on: local lipid covalent geometry, director-referenced chain order, the protein fold, and every filter distance and occupancy. As a check on the last, restricting the analysis to the first $300$~ps---over which the collective drift has accumulated least, it being monotonic from the first frame---moves the deep-site K$^+$--carbonyl distances by $0.005$ and $0.007$~\AA{} at S3 and S4 and, across the four replicas that hold the column throughout, the contiguous-contact fraction from $82$ to $85\%$, in the direction that would strengthen the reported contrast. The membrane question itself is no longer open in sign: the released-box replicas below carry the free-normal bilayer to $400$~ps and it thins further rather than recovering, so the constraint bounds how much we can \emph{measure} of the bilayer without being the reason it differs.

\textit{Validation observables.} All metrics run through one analysis pipeline with identical definitions for all three potentials, so the comparison is fair by construction. Seven families are computed---backbone and filter RMSD; bilayer area-per-lipid and per-carbon order parameter $S_{CD}$; on-axis filter occupancy; the closest K$^+$--K$^+$ distance with the three adjacent axial spacings, from which a \emph{contiguous direct-contact column} is a frame with all three simultaneously below $4$\,\AA; filter-core hydration; and filter-carbonyl tilt $\cos\theta_z$ against the pore axis---with exact definitions and cut-offs in SI~S4. Occupancy, hydration and spacing statistics are pooled over the late half of each trajectory; backbone RMSD is reported for all fifteen replicas. The necessity of an extended-range mechanism beyond a single short cutoff for large-biomolecular-fragment MLFFs has been established on a $\sim$1{,}200-atom benchmark with the same architecture family~\cite{wang2026scalable}, so we do not repeat a KcsA-specific extended-range ablation here.

\paragraph{Throughput, latency and memory benchmarking.}
The hardware-efficiency measurements of Extended Data Fig.~\ref{fig:throughput} use cubic boxes of liquid water as a size-controlled, chemically homogeneous workload spanning $\approx$3{,}000 to 69{,}000 atoms at standard runtime and to $120{,}000$ under activation recompute, all on one NVIDIA H20 (141~GB). Each timed step is a complete inference step---forward pass plus force evaluation---rather than energy alone, because force evaluation is what production MD requires. On this workload MACE-OMol and DPA-4 exhaust the device budget beyond $15{,}000$ atoms, whereas UMA-S-1p2 scales further under activation recompute but holds only $\sim$3{,}000~atoms\,s$^{-1}$ ($\approx$$0.03$~steps/s at $10^5$ atoms), against UBio-MolFM's $8.3\times$ at equal size. Every model is profiled twice where supported---standard runtime and with activation recomputation, which trades throughput for a much lower memory ceiling---and baselines~\cite{wood2025family,mace,li2026dpa4} run from released weights under their authors' recommended settings and precision. Being single-GPU measurements on identical hardware and workloads, the differences reflect computational structure rather than implementation-specific I/O. Timing protocol, warm-up and averaging counts, the out-of-memory criterion, and the two largest points measured but omitted from the plot for legibility are given in SI~S5 (\emph{Single-GPU Performance}).

\section*{Data availability}
A curated protein-focused subset of UBio-Mol26, \textbf{UBio-Protein26 5M}, is publicly available on Hugging Face at \url{https://huggingface.co/datasets/IQuestLab/UBio-Protein26} under an MIT license. The release contains a 5M training split (4.5M at \texttt{def2-SVP} and 0.5M at \texttt{def2-TZVPD}) and a 0.2M held-out test split, preserving the statistical characteristics of the full UBio-Mol26 distribution and providing a standardized high-fidelity benchmark for training and evaluating macromolecular foundation models. Source structures are derived from the AlphaFold Protein Structure Database and the Protein Data Bank. 
A companion reproducibility deposit, \textbf{UBio-MolFM-MD}, ships under an MIT license in the \texttt{reproduce-data/} directory of the model release, \url{https://huggingface.co/IQuestLab/IQuest-UBio-MolFM-V1.5}. It provides starting structures, simulation and analysis scripts with random seeds, and final observable values for every molecular-dynamics scenario reported here, with stride-decimated trajectories carrying positions and forces for all of them but one: the Cyclosporine~A umbrella campaigns ship as per-frame collective-variable time series plus each window's starting structure, because MBAR consumes the collective-variable histories rather than the coordinates. The deposit is sufficient to reproduce every reported observable; a per-scenario inventory is given in SI~S6. Full femtosecond-resolution trajectories (in excess of 1\,TB) are available from the corresponding author upon reasonable request.

\section*{Code availability}
The implementation of the E2Former-V2 architecture, together with the training configuration files that specify every hyperparameter of the three curriculum stages, is available at \url{https://github.com/IQuestLab/UBio-MolFM} under an MIT license. Pretrained weights for UBio-MolFM (Stage~3, \texttt{molfm-v1p5-stage-3.pt} with its \texttt{config.yaml}) are available on Hugging Face at \url{https://huggingface.co/IQuestLab/IQuest-UBio-MolFM-V1.5}, also under an MIT license, alongside the reproducibility deposit of the Data availability statement. Because the configuration files are the authoritative record of the training setup rather than a convenience copy, they are versioned with the code rather than restated here, while the tool versions, the training hardware and the per-scenario random seeds that complete the picture are recorded in \texttt{REPRODUCIBILITY\_}\allowbreak\texttt{CHECKLIST.md} of the reproducibility deposit, so a reader can recover the weights, the settings and the starting conditions that produced a given number. Standardized inference scripts and integrated modules for running molecular dynamics simulations on commodity GPU hardware---built on PyTorch and Triton-optimized kernels and including the SMD, US-MBAR, and KcsA simulation protocols described in Methods---are released together with the model weights.

\section*{Acknowledgements}
The authors thank the IQuest Research team and collaborators for their support.

\section*{Author contributions}
Five initials are close enough to need spelling out---three sharing a surname stem, two differing only by a final letter: Jia.Z., Jia~Zhang; Ji.Z., Ji~Zhang; Ju.Z., Junping~Zhao; J.J., Jack~Jia; J.J.C., JiaJun~Cheng.
Jia.Z. conceived and led the project, assembled the quantum-chemical reference data, designed the molecular-dynamics experiments, wrote the molecular-dynamics engine, and ran and analysed the simulations. L.H., F.P. and Jia.Z. designed and trained the model architecture and prepared the code and model release. L.H., F.P., Ju.Z., Ji.Z., J.J.C., Z.W., H.Y. and H.L. developed the inference-acceleration stack. A.J. and J.J. maintained and coordinated compute resources, and A.J. additionally contributed to quantum-chemical data generation. All authors contributed to writing and revising the manuscript. Jia.Z. is the corresponding author.

\section*{Competing interests}
The authors declare no competing interests.

\bibliographystyle{unsrt}
\bibliography{ref}

\clearpage
\section*{Extended Data}

\setcounter{figure}{0}
\setcounter{table}{0}
\captionsetup[figure]{name={Extended Data Figure}}
\captionsetup[table]{name={Extended Data Table}}
\captionsetup{font={footnotesize,stretch=1.0}}

\begin{figure}[H]
  \centering
  \resizebox{0.7\textwidth}{!}{\definecolor{oursC}{RGB}{38,68,143}%
\definecolor{umaC}{RGB}{198,66,132}%
\definecolor{maceC}{RGB}{209,110,22}%
\definecolor{dpaC}{RGB}{74,148,64}%
\definecolor{axg}{RGB}{120,126,148}%

\begin{tikzpicture}[font=\sffamily]
\begin{axis}[
  width=9.2cm, height=6.4cm, axis lines=left,
  line width=0.5pt, tick align=outside, tick style={axg, thin},
  label style={font=\fontsize{9}{11}\selectfont},
  tick label style={font=\fontsize{8}{9}\selectfont, text=axg},
  every axis x label/.style={at={(ticklabel cs:0.5)}, anchor=north, font=\fontsize{9}{11}\selectfont},
  grid=major, major grid style={axg!25, line width=0.3pt},
  xlabel={time (ps)}, ylabel={$\rho$ (g\,cm$^{-3}$)},
  xmin=0, xmax=100, ymin=0.45, ymax=1.25, xtick={0,25,50,75,100}, ytick={0.6,0.8,1.0,1.2},
  legend style={font=\fontsize{7.6}{9}\selectfont, draw=none, fill=none,
    at={(0.025,0.02)}, anchor=south west, row sep=-1pt, legend columns=3,
    /tikz/every even column/.append style={column sep=6pt}},
  clip=true,
]
\draw[axg, densely dashed, line width=0.8pt] (axis cs:0,0.997)--(axis cs:100,0.997);
\node[font=\fontsize{7}{8}\selectfont, text=axg, anchor=south east] at (axis cs:99,1.0) {experiment 0.997};
\addplot[umaC, line width=1.1pt, mark=none] table[col sep=comma, x=t_ps, y=rho]{data_thermo/density_uma.csv}; \addlegendentry{UMA-S-1p2}
\addplot[maceC, line width=1.1pt, mark=none] table[col sep=comma, x=t_ps, y=rho]{data_thermo/density_mace.csv}; \addlegendentry{MACE-OMol}
\addplot[dpaC, line width=1.1pt, mark=none] table[col sep=comma, x=t_ps, y=rho]{data_thermo/density_dpa4.csv}; \addlegendentry{DPA-4 (OMol)}
\addplot[oursC!60, line width=1.2pt, densely dashed, mark=none] table[col sep=comma, x=t_ps, y=rho]{data_thermo/density_water_s2.csv}; \addlegendentry{UBio-MolFM (S2)}
\addplot[oursC, line width=1.7pt, mark=none] table[col sep=comma, x=t_ps, y=rho]{data_thermo/density_water_ours.csv}; \addlegendentry{UBio-MolFM (S3)}
\end{axis}
\end{tikzpicture}}
  \caption{\textbf{Liquid-water density under unbiased $NPT$ dynamics: pretrained-baseline pathologies.} Mass density of a 512-molecule water box at 300~K, 1~bar. DPA-4, run here and throughout this work through its \textbf{OMol head} (as in Table~\ref{tab:rel_energy_force}), collapses toward a gas-like $0.57$--$0.62\,\text{g\,cm}^{-3}$, and MACE-OMol and UMA-S-1p2 over-densify by $\sim$$9$ and $\sim$$11\%$, whereas UBio-MolFM (S3) relaxes to and holds the experimental density (dashed line). That head is the right comparator for this work: it is the one trained on OMol25, the molecular-chemistry corpus our own Stages~1--2 train on and the distribution the first tier of Table~\ref{tab:rel_energy_force} is drawn from, so all four pretrained traces sit on one training distribution and the panel separates what a periodic condensed-phase tier adds from which head was picked. DPA-4's remaining heads target materials or water-specific data; the corollary is that this panel bounds the OMol-trained head rather than DPA-4 as a whole, and a water-specialised head would answer a different question---how well a water model does water, not whether a molecular potential transfers to the condensed phase. The fifth trace is the internal control: \textbf{Stage~2}, our own checkpoint before UBio-Mol26's periodic tier enters, over-densifies to $1.118\,\text{g\,cm}^{-3}$---$12.1\%$ high, the same failure and nearly the same magnitude as the two finite-cluster baselines, from the same architecture and the same starting box as the Stage-3 trace beside it. Baselines use the authors' recommended stress-based barostat over their 100-ps runs; both UBio-MolFM traces use a Monte-Carlo barostat over the same window. See SI~S4.1 for what this does and does not establish.}
  \label{fig:density_baselines_si}
\end{figure}
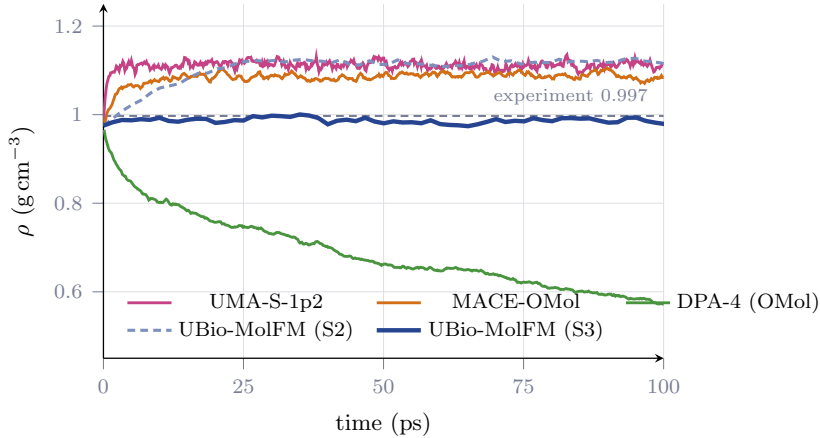

\clearpage

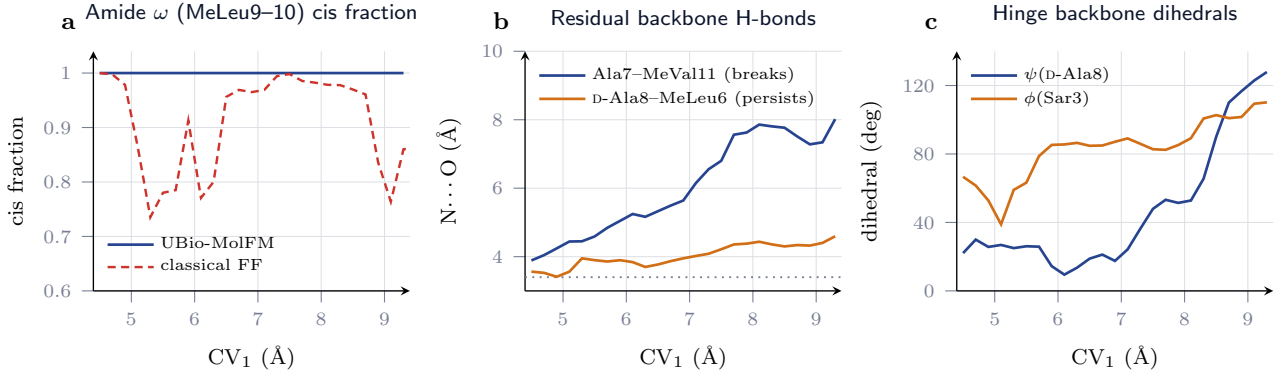
\begin{figure}[H]
  \centering
  \widefig{\definecolor{oursC}{RGB}{38,68,143}%
\definecolor{amberC}{RGB}{209,110,22}%
\definecolor{mmC}{RGB}{206,52,44}%
\definecolor{axg}{RGB}{120,126,148}%
\tikzset{panellabel/.style={font=\fontsize{9}{10}\selectfont\bfseries, text=black, anchor=south west}}%
\begin{tikzpicture}[font=\sffamily]
\pgfplotsset{
  every axis/.style={
    width=5.7cm, height=4.7cm, axis lines=left,
    line width=0.5pt, tick align=outside, tick style={axg, thin},
    label style={font=\fontsize{8}{9.5}\selectfont},
    tick label style={font=\fontsize{7}{8.5}\selectfont, text=axg},
    title style={font=\fontsize{8}{9.5}\selectfont, text=oursC!30!black},
    every axis x label/.style={at={(ticklabel cs:0.5)}, anchor=north, font=\fontsize{8}{9.5}\selectfont},
    grid=major, major grid style={axg!25, line width=0.3pt}, clip=true,
  },
  legstyle/.style={font=\fontsize{6.2}{7}\selectfont, draw=none, fill=none, row sep=-1.5pt, inner sep=1pt, legend cell align=left},
}

\begin{axis}[
  name=CIS, title={\textsf{Amide $\omega$ (MeLeu9--10) cis fraction}},
  xlabel={$\text{CV}_1$ (\AA)}, ylabel={cis fraction},
  xmin=4.4, xmax=9.4, ymin=0.6, ymax=1.04, xtick={5,6,7,8,9}, ytick={0.6,0.7,0.8,0.9,1.0},
  legend style={legstyle, at={(0.03,0.03)}, anchor=south west},
]
\addplot[oursC, line width=1.0pt, mark=none] table[col sep=comma, x=CV1_A, y=ML_cis_frac]{data_csa/si_cis_fraction_vs_cv1.csv}; \addlegendentry{UBio-MolFM}
\addplot[mmC, line width=0.9pt, densely dashed, mark=none] table[col sep=comma, x=CV1_A, y=MM_cis_frac]{data_csa/si_cis_fraction_vs_cv1.csv}; \addlegendentry{classical FF}
\end{axis}
\node[panellabel] at ($(CIS.north west)+(-0.55cm,0.16cm)$) {a};

\begin{axis}[
  name=IMHB, at={($(CIS.east)+(1.5cm,0)$)}, anchor=west,
  title={\textsf{Residual backbone H-bonds}},
  xlabel={$\text{CV}_1$ (\AA)}, ylabel={N$\cdots$O (\AA)},
  xmin=4.4, xmax=9.4, ymin=3, ymax=10, xtick={5,6,7,8,9}, ytick={4,6,8,10},
  legend style={legstyle, at={(0.03,0.97)}, anchor=north west},
]
\draw[axg, dotted, line width=0.7pt] (axis cs:4.4,3.4)--(axis cs:9.4,3.4);
\addplot[oursC, line width=1.0pt, mark=none] table[col sep=comma, x=CV1_A, y=ML_N_ALA11_O_MVA4]{data_csa/si_other_imhb_vs_cv1.csv}; \addlegendentry{Ala7--MeVal11 (breaks)}
\addplot[amberC, line width=1.0pt, mark=none] table[col sep=comma, x=CV1_A, y=ML_N_DAL1_O_MLE10]{data_csa/si_other_imhb_vs_cv1.csv}; \addlegendentry{\textsc{d}-Ala8--MeLeu6 (persists)}
\end{axis}
\node[panellabel] at ($(IMHB.north west)+(-0.55cm,0.16cm)$) {b};

\begin{axis}[
  name=PHI, at={($(IMHB.east)+(1.5cm,0)$)}, anchor=west,
  title={\textsf{Hinge backbone dihedrals}},
  xlabel={$\text{CV}_1$ (\AA)}, ylabel={dihedral (deg)},
  xmin=4.4, xmax=9.4, ymin=0, ymax=140, xtick={5,6,7,8,9}, ytick={0,40,80,120},
  legend style={legstyle, at={(0.03,0.97)}, anchor=north west},
]
\addplot[oursC, line width=1.0pt, mark=none] table[col sep=comma, x=CV1_A, y=ML_psi_MLE8]{data_csa/si_phipsi_vs_cv1.csv}; \addlegendentry{$\psi$(\textsc{d}-Ala8)}
\addplot[amberC, line width=1.0pt, mark=none] table[col sep=comma, x=CV1_A, y=ML_phi_SAR7]{data_csa/si_phipsi_vs_cv1.csv}; \addlegendentry{$\phi$(Sar3)}
\end{axis}
\node[panellabel] at ($(PHI.north west)+(-0.55cm,0.16cm)$) {c};

\end{tikzpicture}%
}
  \caption{\textbf{Conformer diagnostics for the Cyclosporine~A landscape.} \textbf{(a)}~\emph{cis} fraction of the MeLeu9--MeLeu10 amide ($\omega$) across all umbrella windows: UBio-MolFM stays essentially fully \emph{cis} ($\approx$1.00; mean $\omega=-6.5^\circ$), the classical force field slightly less so ($\approx$0.94; $-37^\circ$). The \emph{cis}$\to$\emph{trans} isomerization that would convert the closed (four intramolecular H-bonds) form into the all-\emph{trans} ``very open'' form of 1IKF~\cite{altschuh1992csa} and the cyclophilin complex is never sampled, so the entire surface is the \emph{cis} manifold. \textbf{(b)}~Two backbone hydrogen bonds outside the CV set: Ala7--MeVal11 (N$\cdots$O) ruptures as the ring opens (3.9$\to$7.3~\AA), whereas \textsc{d}-Ala8--MeLeu6 remains semi-static (3.5$\to$4.3~\AA), so even the open state keeps one transannular contact and is not hydrogen-bond-free (unlike 1IKF); dotted line, 3.4~\AA\ cut-off. \textbf{(c)}~Hinge dihedrals: beyond $\sim$8.5~\AA{} $\psi$(\textsc{d}-Ala8) swings by $\sim$90$^\circ$ (26$^\circ\to$90$^\circ\to$128$^\circ$) and $\phi$(Sar3) by $\sim$30$^\circ$, so further opening requires genuine backbone reorientation---consistent with the rising free energy of the wall (both residues sitting in flexible regions of the ring, this is reorientation evidence rather than proven strain).}\label{fig:csa_conformer_si}
\end{figure}

\clearpage

\begin{figure}[H]
  \centering
  \setlength{\abovecaptionskip}{4pt}
  \widefig{\definecolor{oursC}{RGB}{38,68,143}%
\definecolor{umaC}{RGB}{198,66,132}%
\definecolor{maceC}{RGB}{209,110,22}%
\definecolor{dpaC}{RGB}{74,148,64}%
\definecolor{axg}{RGB}{120,126,148}%
\tikzset{panellabel/.style={font=\fontsize{9}{10}\selectfont\bfseries, text=black, anchor=south west}}%

\begin{tikzpicture}[font=\sffamily]
\pgfplotsset{
  every axis/.style={
    width=5.6cm, height=5.1cm, axis lines=left,
    line width=0.5pt, tick align=outside, tick style={axg, thin},
    label style={font=\fontsize{8.1}{9.5}\selectfont}, tick label style={font=\fontsize{7.2}{8.5}\selectfont, text=axg},
    title style={font=\fontsize{8.1}{9.5}\selectfont, text=oursC!30!black},
    every axis x label/.style={at={(ticklabel cs:0.5)}, anchor=north, font=\fontsize{8.1}{9.5}\selectfont},
    xmin=0, xmax=74, xtick={0,20,40,60},
    grid=major, major grid style={axg!25, line width=0.3pt},
    clip=false,
  },
  ourS/.style={oursC, line width=1.7pt, mark=*, mark size=1.1pt, mark options={fill=oursC, draw=oursC, line width=0.2pt}},
  ourD/.style={oursC, line width=1.4pt, densely dashed, mark=none},
  umaS/.style={umaC, line width=1.0pt, mark=triangle*, mark size=1.7pt, mark options={fill=umaC!30}},
  umaD/.style={umaC, line width=1.0pt, densely dashed, mark=none},
  maceS/.style={maceC, line width=1.0pt, mark=square*, mark size=1.5pt, mark options={fill=maceC!30}},
  dpaS/.style={dpaC, line width=1.0pt, mark=diamond*, mark size=1.8pt, mark options={fill=dpaC!30}},
}

\begin{axis}[
  name=A, title={\textsf{Throughput}},
  ylabel={throughput (10$^3$ atoms\,s$^{-1}$)},
  ymin=0, ymax=42, ytick={0,10,20,30,40},
]
\addplot[umaS] coordinates {(3,6.3175)(6,6.2647)};
\addplot[maceS] coordinates {(3,7.3647)(6,7.3595)(9,7.4058)(12,7.2472)(15,5.4906)};
\addplot[dpaS] coordinates {(3,8.2523)(6,8.3004)(9,8.292)(12,8.2859)(15,8.2953)};
\addplot[umaD] coordinates {(3,3.0806)(12,3.1034)(24,3.0938)(36,3.0917)(48,3.0927)(60,3.0816)(69,3.0747)};
\addplot[ourD] coordinates {(3,22.6167)(12,26.0559)(24,26.6927)(36,26.2533)(48,26.5082)(60,26.0931)(69,26.1176)};
\addplot[ourS] coordinates {(3,31.3854)(6,34.8684)(9,35.4639)(12,36.6068)(15,36.9113)(18,37.2587)(21,36.9614)(24,37.171)(27,37.5283)(30,37.2843)(33,37.7065)(36,35.7111)(39,37.0547)(42,36.9032)(45,36.7575)(48,35.5479)(51,36.9373)(54,36.5388)(57,36.9899)(60,36.0811)(63,36.2252)(66,36.1739)(69,36.216)};
\node[font=\fontsize{7.2}{8.5}\selectfont, text=oursC!30!black, anchor=west] at (axis cs:19,14.5) {$4$--$5\times$ vs baselines};
\end{axis}
\node[panellabel] at ($(A.north west)+(-0.55cm,0.16cm)$) {a};

\begin{axis}[
  name=B, at={($(A.east)+(1.55cm,0)$)}, anchor=west,
  title={\textsf{Latency}},
  xlabel={system size (10$^3$ atoms)},
  ylabel={latency (ms\,/\,step)},
  ymode=log, ymin=80, ymax=40000, ytick={100,1000,10000},
  yticklabel style={/pgf/number format/1000 sep=},
]
\addplot[umaS] coordinates {(3,474.87)(6,957.75)};
\addplot[maceS] coordinates {(3,407.35)(6,815.28)(9,1215.26)(12,1655.81)(15,2731.93)};
\addplot[dpaS] coordinates {(3,363.54)(6,722.86)(9,1085.39)(12,1448.24)(15,1808.26)};
\addplot[umaD] coordinates {(3,973.85)(12,3866.71)(24,7757.43)(36,11644)(48,15520.5)(60,19470.4)(69,22441.2)};
\addplot[ourD] coordinates {(3,132.65)(12,460.55)(24,899.12)(36,1371.26)(48,1810.76)(60,2299.46)(69,2641.9)};
\addplot[ourS] coordinates {(3,95.59)(6,172.08)(9,253.78)(12,327.81)(15,406.38)(18,483.11)(21,568.16)(24,645.66)(27,719.46)(30,804.63)(33,875.18)(36,1008.09)(39,1052.5)(42,1138.11)(45,1224.24)(48,1350.29)(51,1380.72)(54,1477.88)(57,1540.96)(60,1662.92)(63,1739.12)(66,1824.52)(69,1905.23)};
\end{axis}
\node[panellabel] at ($(B.north west)+(-0.55cm,0.16cm)$) {b};

\begin{axis}[
  name=C, at={($(B.east)+(1.55cm,0)$)}, anchor=west,
  title={\textsf{Peak GPU memory}},
  ylabel={peak memory (GB)},
  ymin=0, ymax=155, ytick={0,50,100,150},
]
\draw[axg, dashed, line width=0.6pt] (axis cs:0,141)--(axis cs:74,141);
\node[font=\fontsize{5.6}{6.5}\selectfont, text=axg, anchor=south east] at (axis cs:73,142.5) {H20 $\cdot$ 141\,GB};
\addplot[umaS] coordinates {(3,52.435)(6,104.985)};
\addplot[maceS] coordinates {(3,28.207)(6,56.266)(9,84.286)(12,111.911)(15,140.054)};
\addplot[dpaS] coordinates {(3,26.33)(6,52.551)(9,78.748)(12,104.743)(15,131.004)};
\addplot[umaD] coordinates {(3,12.096)(12,18.783)(24,28.516)(36,39.107)(48,50.548)(60,62.95)(69,72.835)};
\addplot[ourD] coordinates {(3,2.535)(12,8.917)(24,17.659)(36,26.493)(48,35.168)(60,43.919)(69,50.583)};
\addplot[ourS] coordinates {(3,6.698)(6,12.485)(9,18.44)(12,24.367)(15,30.313)(18,36.239)(21,42.164)(24,48.05)(27,54)(30,59.878)(33,65.814)(36,71.745)(39,77.663)(42,83.534)(45,89.534)(48,95.266)(51,101.252)(54,107.024)(57,113.056)(60,118.769)(63,124.85)(66,130.72)(69,136.564)};
\end{axis}
\node[panellabel] at ($(C.north west)+(-0.55cm,0.16cm)$) {c};

\node[anchor=north, font=\fontsize{7.2}{8.5}\selectfont] at ($(A.south west)!0.5!(C.south east)+(0,-1.25cm)$) {%
  \begin{tikzpicture}[baseline]
    \draw[oursC, line width=1.7pt] (0,0)--(0.42,0) node[right, text=black, xshift=1pt]{UBio-MolFM (ours)};
    \draw[umaC, line width=1.0pt] (3.15,0)--(3.57,0) node[right, text=black, xshift=1pt]{UMA-S-1p2};
    \draw[maceC, line width=1.0pt] (5.55,0)--(5.97,0) node[right, text=black, xshift=1pt]{MACE-OMol};
    \draw[dpaC, line width=1.0pt] (8.05,0)--(8.47,0) node[right, text=black, xshift=1pt]{DPA-4};
    \draw[black, line width=1.1pt] (9.95,0)--(10.37,0) node[right, text=black, xshift=1pt]{standard};
    \draw[black, line width=1.1pt, densely dashed] (12.05,0)--(12.47,0) node[right, text=black, xshift=1pt]{activation recompute};
  \end{tikzpicture}};
\node[anchor=north, font=\fontsize{6.6}{7.8}\selectfont] at ($(A.south west)!0.5!(C.south east)+(0,-1.66cm)$)
  {\textcolor{axg}{peak memory at 120{,}000 atoms under activation recompute (off axis):}\ \
   \textcolor{oursC}{UBio-MolFM \textbf{87.8}\,GB} \textcolor{axg}{$\cdot$}\
   \textcolor{umaC}{UMA-S-1p2 \textbf{142.2}\,GB}};
\end{tikzpicture}}
  \captionsetup{font={footnotesize,stretch=1.0}}
  \caption{\textbf{Single-GPU inference throughput, latency and peak memory versus system size} (cubic water boxes, one H20 with 141\,GB; forward\,$+$\,force evaluation). Solid lines, standard runtime; dashed, activation recompute, which trades throughput for memory. All values are measured on one H20; no hardware scaling is applied. The workload is liquid water at its \emph{standard} density throughout, never a dilute or gas-phase box: at that density every atom carries its full complement of neighbours, which is what sets the cost of the solvated biomolecular systems this model is built for. Throughput published for these baselines elsewhere was obtained on other workloads and does not agree with what we measure here; those figures and these are not comparable, and the numbers in this panel should be read only against each other. In standard mode UBio-MolFM plateaus at $\approx$37{,}000~atoms\,s$^{-1}$ from 9{,}000 atoms upward---$5.1\times$ MACE-OMol and $4.4\times$ DPA-4 at 12{,}000 atoms---while using a quarter of their memory: at 15{,}000 atoms UBio-MolFM peaks at $30.3$~GB against $140.1$~GB (MACE-OMol) and $131.0$~GB (DPA-4), both of which then go out of memory at 18{,}000 atoms; UMA-S-1p2 reaches $105.0$~GB at 6{,}000 atoms and fails at 9{,}000. The same plateau carries UBio-MolFM to $69{,}000$ atoms, the largest box built, at $136.6$~GB. Activation recompute trades $28\%$ of that throughput as a size-matched mean over the sizes both sweeps reach ($\approx$26{,}000~atoms\,s$^{-1}$ against $\approx$37{,}000) for a $2.7\times$ lower memory slope---$0.73$ against $1.97$~GB per $1{,}000$ atoms---and holds an $8.3\times$ throughput advantage over UMA-S-1p2 out to the largest size run (120{,}000 atoms), where UBio-MolFM peaks at $87.8$~GB against UMA-S-1p2's $142.2$~GB. The dashed line marks the H20's \emph{nominal} $141$~GB rather than a measured ceiling: peaks are the allocator's own maxima, the $142.2$~GB point completed, and the next size up ($123{,}000$ atoms) is where UMA-S-1p2 runs out of memory. Both sweeps are measured with the same parameter freeze, so the solid--dashed gap is what recomputation costs and nothing else. UBio-MolFM and UMA-S-1p2 both additionally support TensorFloat-32, an opt-in worth roughly $30\%$ throughput on this hardware; it is off in every measurement plotted here and in every production trajectory, on both models alike. TF32 truncates the mantissa of the matrix products that yield the forces, and those forces are what the integrator accumulates over the $2\times10^6$ steps of a nanosecond at $0.5$~fs---the property the $NVE$ test of Fig.~\ref{fig:thermo}b certifies---so we neither trade it for speed in production nor claim it in a throughput figure. Both curves are conservative by about that margin, and symmetrically so.}\label{fig:throughput}
\end{figure}
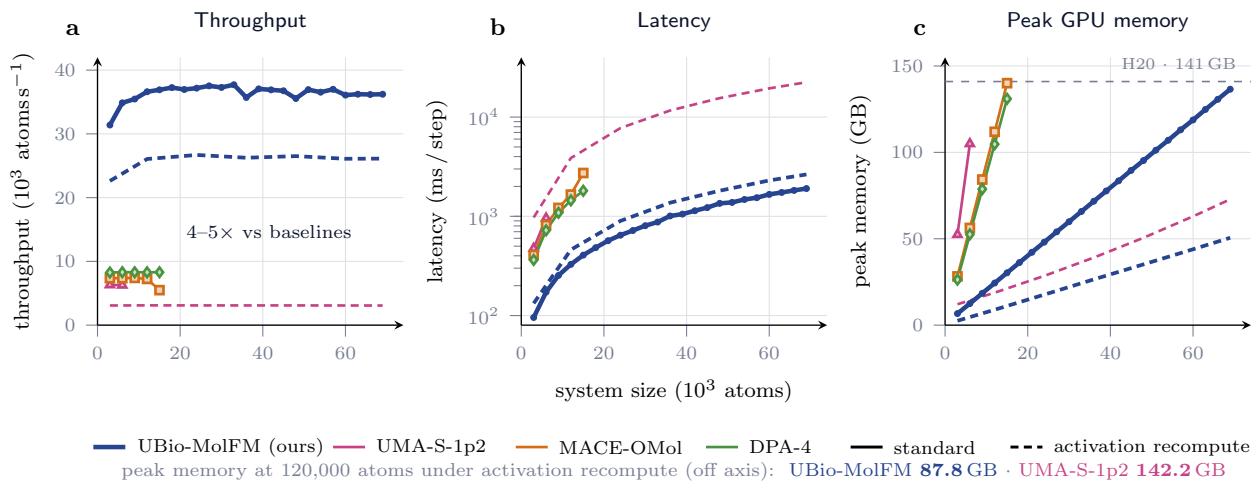

\clearpage

\begin{figure}[H]
  \centering
  \resizebox{0.95\linewidth}{!}{\definecolor{oursC}{RGB}{38,68,143}%
\definecolor{mmC}{RGB}{206,52,44}%
\definecolor{mm6C}{RGB}{181,141,54}%
\definecolor{contactC}{RGB}{74,159,128}%
\definecolor{axg}{RGB}{120,126,148}%
\tikzset{panellabel/.style={font=\fontsize{9}{10}\selectfont\bfseries, text=black, anchor=south west}}%
\begin{tikzpicture}[font=\sffamily]
\pgfplotsset{
  every axis/.style={
    width=7.2cm, height=5.0cm, axis lines=left,
    line width=0.5pt, tick align=outside, tick style={axg, thin},
    label style={font=\fontsize{8}{9.5}\selectfont},
    tick label style={font=\fontsize{7}{8.5}\selectfont, text=axg},
    title style={font=\fontsize{8.5}{10}\selectfont, text=oursC!30!black},
    every axis x label/.style={at={(ticklabel cs:0.5)}, anchor=north, font=\fontsize{8}{9.5}\selectfont},
    grid=major, major grid style={axg!25, line width=0.3pt}, clip=true,
  },
  csolid/.style={line width=1.0pt, mark=none},
  legstyle/.style={font=\fontsize{6.6}{7.6}\selectfont, draw=none, fill=none,
    row sep=-1.5pt, inner sep=1pt, legend cell align=left},
  seedbar/.style={ybar, bar width=8pt, enlarge x limits=0.08,
    every axis plot/.append style={bar shift=0pt}},
}

\begin{axis}[
  name=TS, title={\textsf{Closest-pair K$^+$--K$^+$ vs time}},
  xlabel={time (ps)}, ylabel={$d_\text{min}$(K$^+$,K$^+$) (\AA)},
  xmin=0, xmax=1000, ymin=2.8, ymax=5.0, xtick={0,300,600,900},
  ytick={3,3.5,4,4.5,5},
  legend style={legstyle, at={(0.97,0.97)}, anchor=north east},
]
\fill[contactC!14] (axis cs:0,3.0) rectangle (axis cs:1000,3.5);
\addplot[draw=none, name path=clo, forget plot] table[col sep=comma, x=time_ps, y=mm_lo]{data_kcsa/kk_timeseries.csv};
\addplot[draw=none, name path=chi, forget plot] table[col sep=comma, x=time_ps, y=mm_hi]{data_kcsa/kk_timeseries.csv};
\addplot[mmC!12, forget plot] fill between[of=clo and chi];
\addplot[draw=none, name path=glo, forget plot] table[col sep=comma, x=time_ps, y=mm6_lo]{data_kcsa/kk_timeseries.csv};
\addplot[draw=none, name path=ghi, forget plot] table[col sep=comma, x=time_ps, y=mm6_hi]{data_kcsa/kk_timeseries.csv};
\addplot[mm6C!14, forget plot] fill between[of=glo and ghi];
\addplot[draw=none, name path=mlo, forget plot] table[col sep=comma, x=time_ps, y=molfm_lo]{data_kcsa/kk_timeseries.csv};
\addplot[draw=none, name path=mhi, forget plot] table[col sep=comma, x=time_ps, y=molfm_hi]{data_kcsa/kk_timeseries.csv};
\addplot[oursC!14, forget plot] fill between[of=mlo and mhi];
\addplot[mm6C, csolid] table[col sep=comma, x=time_ps, y=mm6_mean]{data_kcsa/kk_timeseries.csv}; \addlegendentry{12-6}
\addplot[mmC, csolid] table[col sep=comma, x=time_ps, y=mm_mean]{data_kcsa/kk_timeseries.csv}; \addlegendentry{12-6-4}
\addplot[oursC, csolid] table[col sep=comma, x=time_ps, y=molfm_mean]{data_kcsa/kk_timeseries.csv}; \addlegendentry{UBio-MolFM}
\addplot[oursC!75, line width=0.55pt, dashed] table[col sep=comma, x=time_ps, y=minKK_A]{data_kcsa/kk_seed126.csv}; \addlegendentry{UBio-MolFM seed 126}
\node[font=\fontsize{6.2}{7}\selectfont, text=contactC!55!black, anchor=west] at (axis cs:120,3.06) {direct-contact band};
\end{axis}
\node[panellabel] at ($(TS.north west)+(-0.7cm,0.16cm)$) {a};

\begin{axis}[
  name=CT, at={($(TS.east)+(1.9cm,0)$)}, anchor=west, seedbar,
  title={\textsf{Contiguous contact column (per replica)}},
  ylabel={\% of late frames}, ymin=0, ymax=100, ytick={0,25,50,75,100},
  xmin=0.2, xmax=15.8,
  xtick={1,...,15},
  xticklabels={42,84,126,168,210,c4.1,c4.2,c4.3,c4.4,c4.5,c6.1,c6.2,c6.3,c6.4,c6.5},
  x tick label style={font=\fontsize{4.8}{5.6}\selectfont, text=axg, rotate=90, anchor=east},
  bar width=5pt, unbounded coords=discard,
]
\addplot[fill=oursC!85, draw=oursC!50!black, line width=0.3pt]
  table[col sep=comma, x=idx, y=molfm_contig]{data_kcsa/per_seed_wide.csv};
\addplot[fill=mmC!80, draw=mmC!50!black, line width=0.3pt]
  table[col sep=comma, x=idx, y=mm_contig]{data_kcsa/per_seed_wide.csv};
\addplot[fill=mm6C!85, draw=mm6C!50!black, line width=0.3pt]
  table[col sep=comma, x=idx, y=mm6_contig]{data_kcsa/per_seed_wide.csv};
\node[font=\fontsize{6.2}{7.2}\selectfont, text=oursC, anchor=north] at (axis cs:3,99) {UBio-MolFM};
\node[font=\fontsize{6}{7}\selectfont, text=mmC!70!black, anchor=north, align=center] at (axis cs:8,88) {12-6-4\\(0\% all)};
\node[font=\fontsize{6}{7}\selectfont, text=mm6C!60!black, anchor=north, align=center] at (axis cs:13,76) {12-6\\(0\% all)};
\end{axis}
\node[panellabel] at ($(CT.north west)+(-0.7cm,0.16cm)$) {b};

\begin{axis}[
  name=WT, at={($(TS.south west)+(0,-2.15cm)$)}, anchor=north west, seedbar,
  title={\textsf{Filter-core water (per replica)}},
  ylabel={\% of late frames wet}, ymin=0, ymax=100, ytick={0,25,50,75,100},
  xmin=0.2, xmax=15.8,
  xtick={1,...,15},
  xticklabels={42,84,126,168,210,c4.1,c4.2,c4.3,c4.4,c4.5,c6.1,c6.2,c6.3,c6.4,c6.5},
  x tick label style={font=\fontsize{4.8}{5.6}\selectfont, text=axg, rotate=90, anchor=east},
  bar width=5pt, unbounded coords=discard,
]
\addplot[fill=oursC!85, draw=oursC!50!black, line width=0.3pt]
  table[col sep=comma, x=idx, y=molfm_wet]{data_kcsa/per_seed_wide.csv};
\addplot[fill=mmC!80, draw=mmC!50!black, line width=0.3pt]
  table[col sep=comma, x=idx, y=mm_wet]{data_kcsa/per_seed_wide.csv};
\addplot[fill=mm6C!85, draw=mm6C!50!black, line width=0.3pt]
  table[col sep=comma, x=idx, y=mm6_wet]{data_kcsa/per_seed_wide.csv};
\node[font=\fontsize{6.2}{7.2}\selectfont, text=oursC, anchor=north, align=center] at (axis cs:3,92) {UBio-MolFM\\(0\% all)};
\node[font=\fontsize{6}{7}\selectfont, text=mmC!70!black, anchor=north east] at (axis cs:10,58) {12-6-4};
\node[font=\fontsize{6}{7}\selectfont, text=mm6C!60!black, anchor=north] at (axis cs:12.2,62) {12-6};
\end{axis}
\node[panellabel] at ($(WT.north west)+(-0.7cm,0.16cm)$) {c};

\begin{axis}[
  name=OC, at={($(WT.east)+(1.9cm,0)$)}, anchor=west,
  title={\textsf{On-axis filter occupancy}},
  ylabel={fraction of frames}, ymin=0, ymax=1.0, ytick={0,0.25,0.5,0.75,1.0},
  ybar=1.4pt, bar width=6.5pt, enlarge x limits=0.28,
  xtick={0,1,2}, xticklabels={5 ions,4 ions,$\leq$3 ions},
  x tick label style={font=\fontsize{7}{8}\selectfont, text=axg},
  legend style={legstyle, at={(0.97,0.97)}, anchor=north east},
]
\addplot[fill=oursC!85, draw=oursC!50!black, line width=0.3pt]
  table[col sep=comma, x expr=\coordindex, y=molfm_frac]{data_kcsa/occupancy.csv};
\addplot[fill=mmC!80, draw=mmC!50!black, line width=0.3pt]
  table[col sep=comma, x expr=\coordindex, y=mm_frac]{data_kcsa/occupancy.csv};
\addplot[fill=mm6C!85, draw=mm6C!50!black, line width=0.3pt]
  table[col sep=comma, x expr=\coordindex, y=mm6_frac]{data_kcsa/occupancy.csv};
\legend{UBio-MolFM,12-6-4,12-6}
\end{axis}
\node[panellabel] at ($(OC.north west)+(-0.7cm,0.16cm)$) {d};
\end{tikzpicture}%
}
  \caption{\textbf{Per-replica validation of the 108{,}964-atom KcsA simulation, all three potentials.} Five UBio-MolFM replicas (blue; seeds 42--210) versus five matched Li--Merz 12-6-4 replicas (crimson; c4.1--c4.5) and five matched plain 12-6 replicas (goldenrod; c6.1--c6.5), from the same equilibrated system under one $NPT$ protocol; the UBio-MolFM replicas additionally receive a $500$-step minimization on their own potential, a $0.14$~\AA{} first-frame offset that straddles the $4$~\AA{} contiguity criterion, bounded in Methods. \textbf{(a)}~Closest K$^+$--K$^+$ distance in the filter versus time (mean; shaded min--max envelope over replicas): UBio-MolFM sits in the direct-contact band throughout, both classical models just above it and close to each other. Envelopes span the full $1.0$~ns now common to all fifteen replicas; the single late departure of seed 126 is discussed in the main text. \textbf{(b)}~Fraction of late-trajectory frames forming a contiguous direct-contact column (all three upper axial spacings $<4$\,\AA{} simultaneously): $78$--$89\%$ for four UBio-MolFM seeds and $28\%$ for seed 126, whose late departure is described in the main text, against $0\%$ for every classical replica of either model. In those replicas the column is absent from the analysed half and from the first frame alike, though at the first frame only marginally: its widest spacing sits $0.08$~\AA{} above the criterion, a consequence of the $0.14$~\AA{} starting offset of Methods. By $10$~ps every classical replica has opened S3--S2 to between $5.6$ and $6.2$~\AA, so none offers an analogue of seed 126's late departure. \textbf{(c)}~Fraction of late-trajectory frames with water in the filter core: absent in all five UBio-MolFM seeds and strongly seed-dependent under both classical treatments---$0$--$86\%$ across the 12-6-4 replicas and $0$--$100\%$ across the 12-6 ones, the latter close to bimodal (two essentially dry, one at $18\%$, two continuously wet)---at pooled rates that do not separate the two models ($55$ against $44\%$, $p=0.70$). \textbf{(d)}~On-axis occupancy distribution (five, four, or $\leq$3 ions within 3\,\AA{} of the pore axis), on the same late-half window as (b) and (c): UBio-MolFM holds at least four ions in $95\%$ of frames (exactly four in $90\%$), whereas both classical treatments fall to $\leq$3 ions in $46\%$ and $37\%$ of frames respectively ($p=0.75$ between them). Occupancy itself is no longer measurably changing at the final frame in either ensemble, but the classical columns are still spreading---the closest K$^+$--K$^+$ distance drifts outward significantly in all five 12-6-4 replicas and in none of the four of ours that hold the state (Methods)---so the spacings in (b) describe a classical filter better than the one those trajectories end with. Three further UBio-MolFM replicas run under the classical references' semi-isotropic barostat to their $400$-ps design length reproduce (a), (c) and (d), and two of the three reproduce (b) at $86\%$; the third registers $0\%$ on the $4$~\AA{} criterion while holding every spacing below $6$~\AA{} throughout, a threshold effect rather than a lost column (Methods). No panel here therefore depends on the aspect-ratio constraint. Those replicas do change the \emph{membrane} picture, and not in this potential's favour (Methods). The residual five-ion fraction is the tail of the deliberately ion-loaded start and is present under all three potentials. Backbone RMSD and membrane order parameters are in main-text Fig.~\ref{fig:transmembrane_channel}b,d; the scaffold and per-site coordination anchors are in Extended Data Table~\ref{tab:validation_summary}.}\label{fig:kcsa_validation_si}
\end{figure}
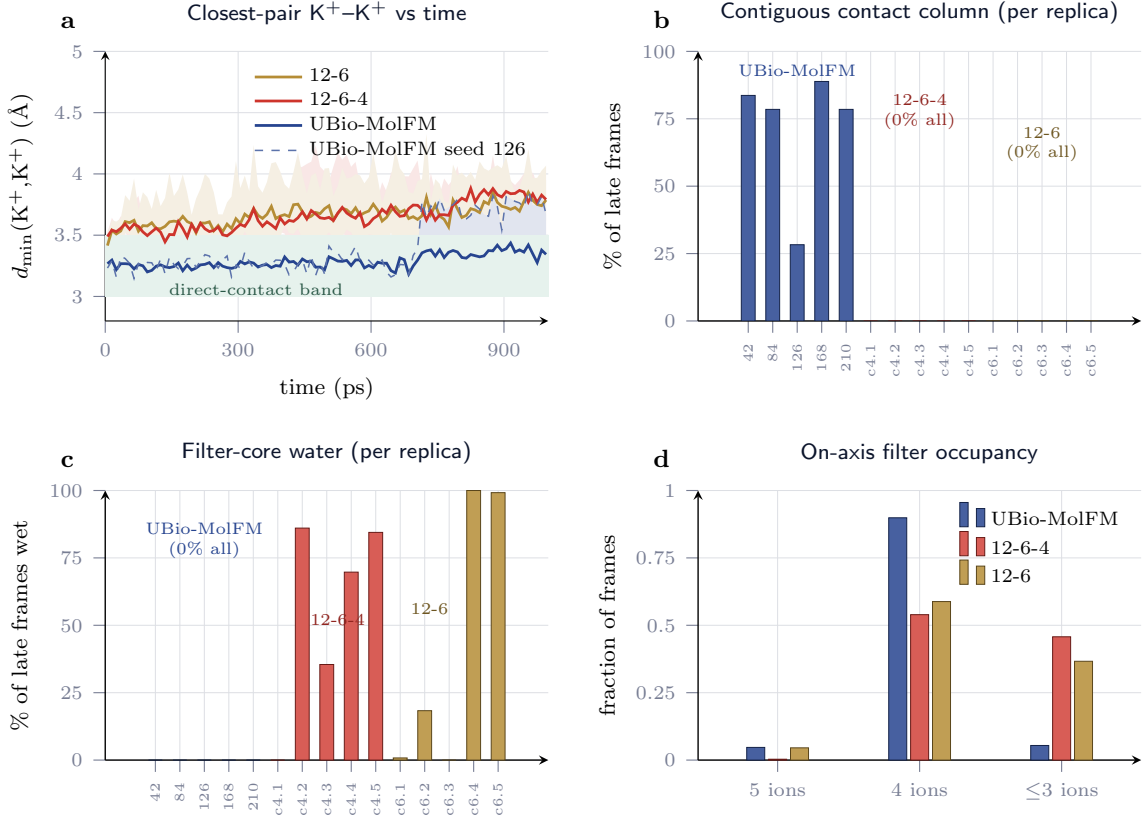

\clearpage

\begin{figure}[H]
  \centering
  \begin{subfigure}[b]{0.57\linewidth}
    \includegraphics[width=\linewidth]{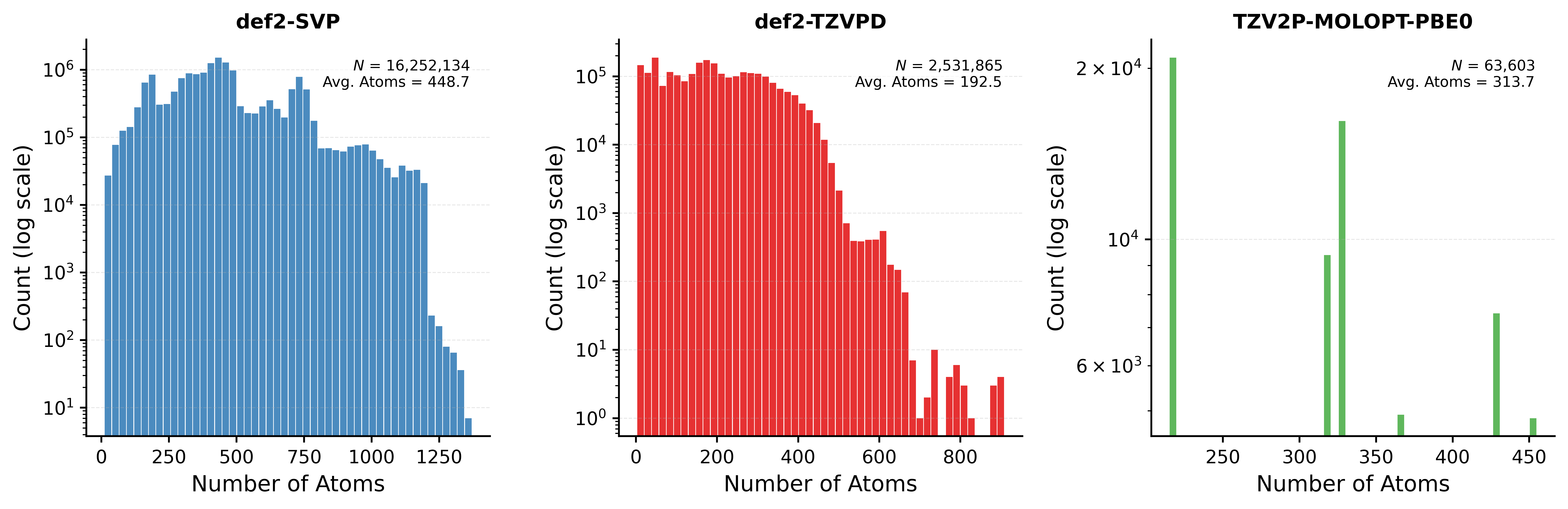}
    \caption{System-size distribution by fidelity}
    \label{fig:atom_counts}
  \end{subfigure}

  \vspace{0.20cm}
  \begin{subfigure}[b]{0.57\linewidth}
    \includegraphics[width=\linewidth]{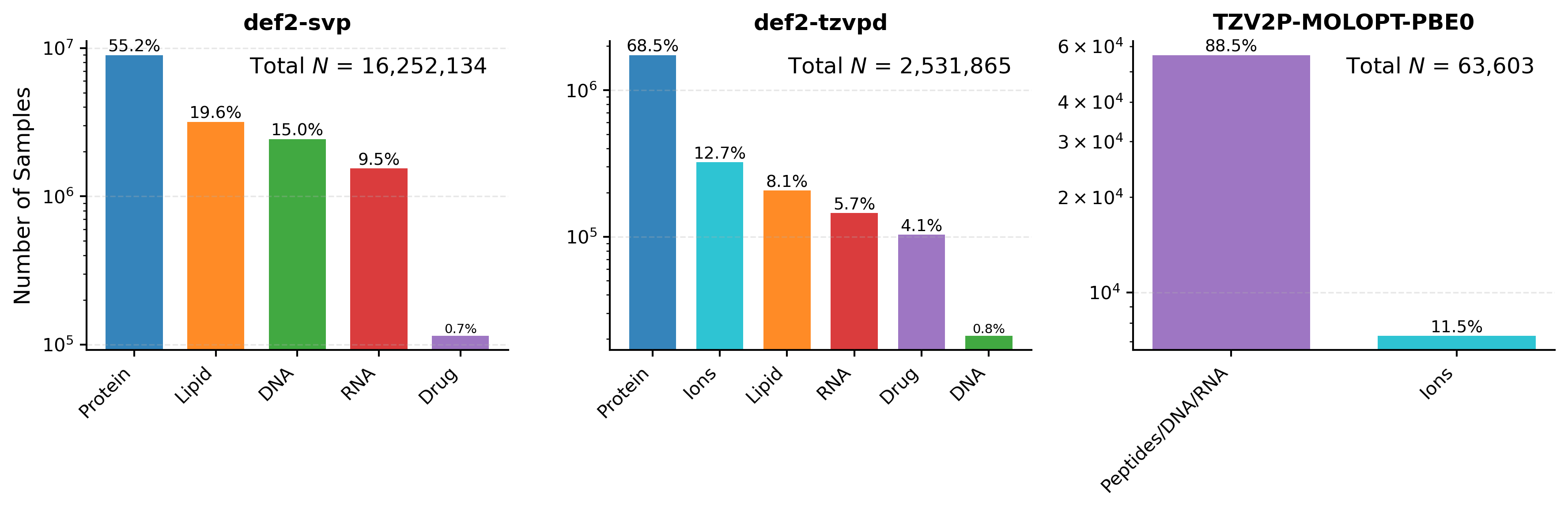}
    \caption{Composition by category, same three fidelities}
    \label{fig:proportion_all}
  \end{subfigure}

  \vspace{0.20cm}
  \begin{subfigure}[b]{0.58\linewidth}
    \includegraphics[width=\linewidth]{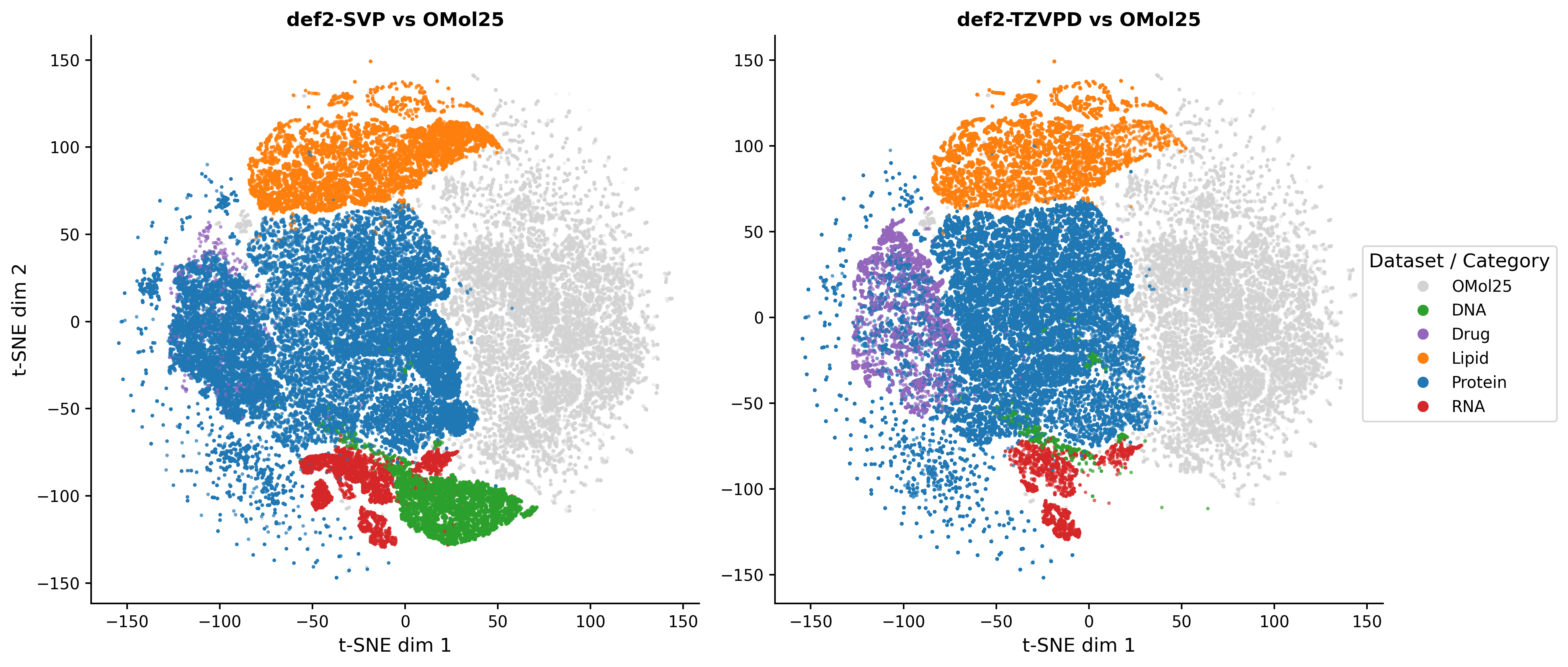}
    \caption{t-SNE feature space, UBio-Mol26 versus OMol25}
    \label{fig:tsne_comp}
  \end{subfigure}

  \vspace{0.20cm}
  \begin{subfigure}[b]{0.66\linewidth}
    \includegraphics[width=\linewidth]{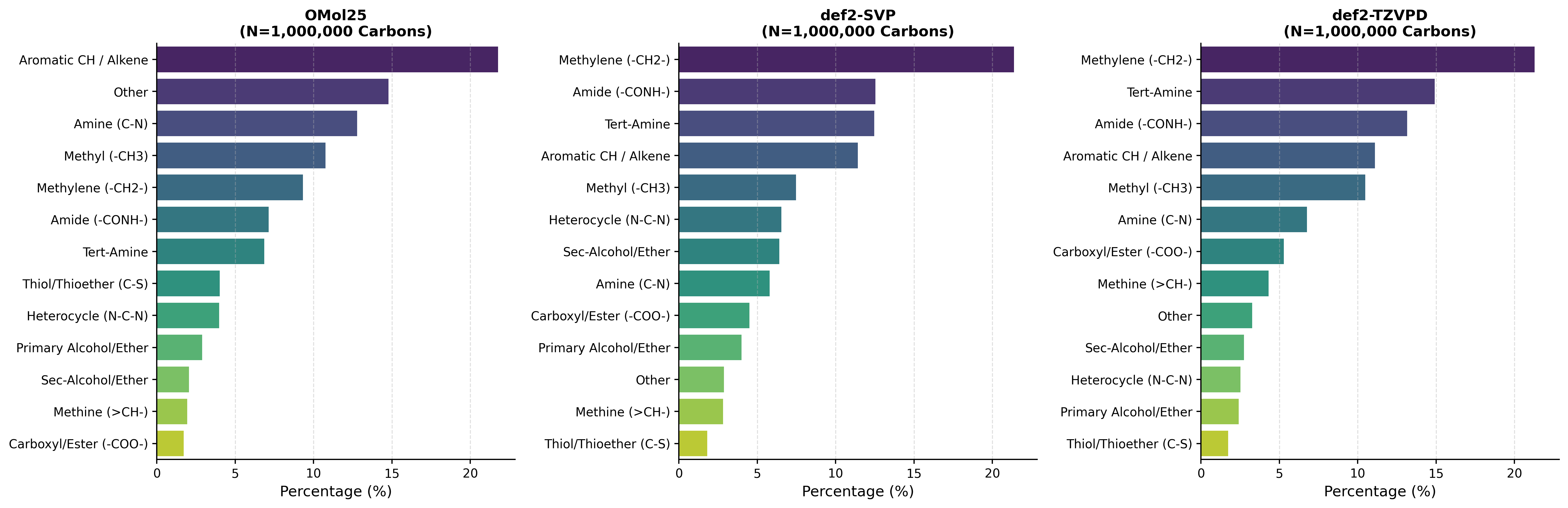}
    \caption{Carbon chemical-environment frequencies}
    \label{fig:organic_dist}
  \end{subfigure}
  \caption{\textbf{Scale, composition and chemical-space coverage of UBio-Mol26.}
    \textbf{(a)}~Atom-count histograms (log counts) and \textbf{(b)}~composition by biological category, for the three DFT fidelities in the same left-to-right order in both rows: the two finite-cluster subsets \texttt{def2-SVP} and \texttt{def2-TZVPD}, then the \emph{periodic} subset computed in CP2K~\cite{cp2k} at revPBE0-D3/\texttt{TZV2P-MOLOPT-PBE0}. Sample counts, mean system sizes and per-category fractions are printed on the panels and given in \S\ref{sec:data_construction}; the emphasis shifts from bulk protein/lipid/DNA coverage at \texttt{def2-SVP} to protein and ions at \texttt{def2-TZVPD}, while the periodic subset is peptide/DNA/RNA plus an ionic fraction that also contains the neat-water boxes. Only the periodic subset constrains the macroscopic pressure and density response of the condensed phase, which finite clusters leave unconstrained however large they are.
    \textbf{(c)}~One million sampled configurations in t-SNE-reduced feature space, by biological domain and basis set (\texttt{def2-SVP} versus \texttt{def2-TZVPD}); grey background, OMol25. UBio-Mol26 occupies the complementary macromolecular regime rather than densifying the same region. \textbf{(d)}~Functional-group frequencies for one million carbon atoms sampled from OMol25 (left) and the two finite-cluster subsets (\texttt{def2-SVP} centre, \texttt{def2-TZVPD} right): UBio-Mol26 is markedly richer in methylene and amide environments, consistent with its emphasis on proteins and lipids, where OMol25 is enriched in the aromatic groups typical of drug-like molecules.}
  \label{fig:data_overview}
\end{figure}

\clearpage

\begin{figure}[H]
  \centering
  \begin{subfigure}[b]{0.66\linewidth}
    \includegraphics[width=\linewidth]{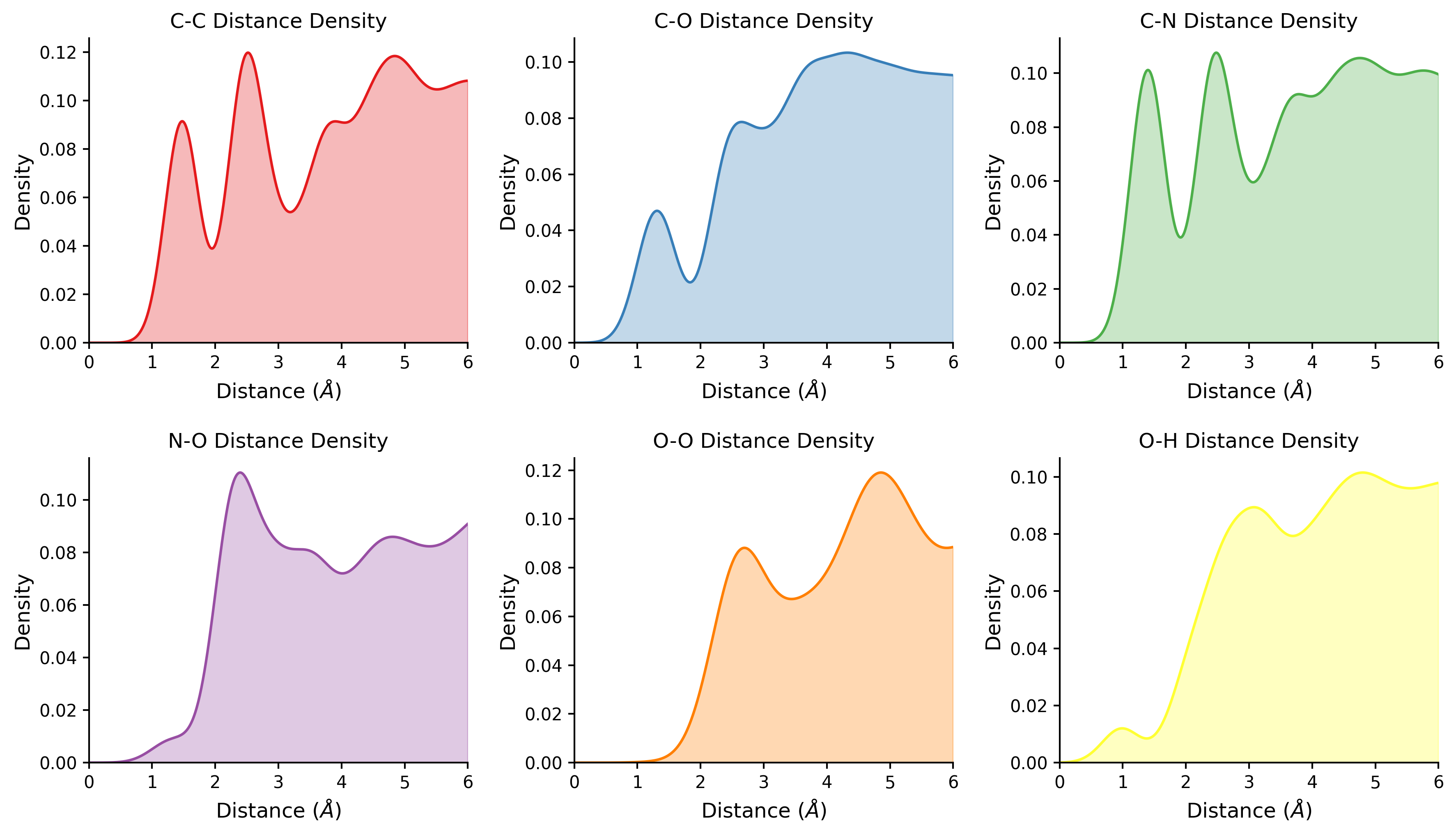}
    \caption{Pair distances, OMol25}
    \label{fig:pair_dist_omol}
  \end{subfigure}
  \vspace{0.25cm}
  \begin{subfigure}[b]{0.66\linewidth}
    \includegraphics[width=\linewidth]{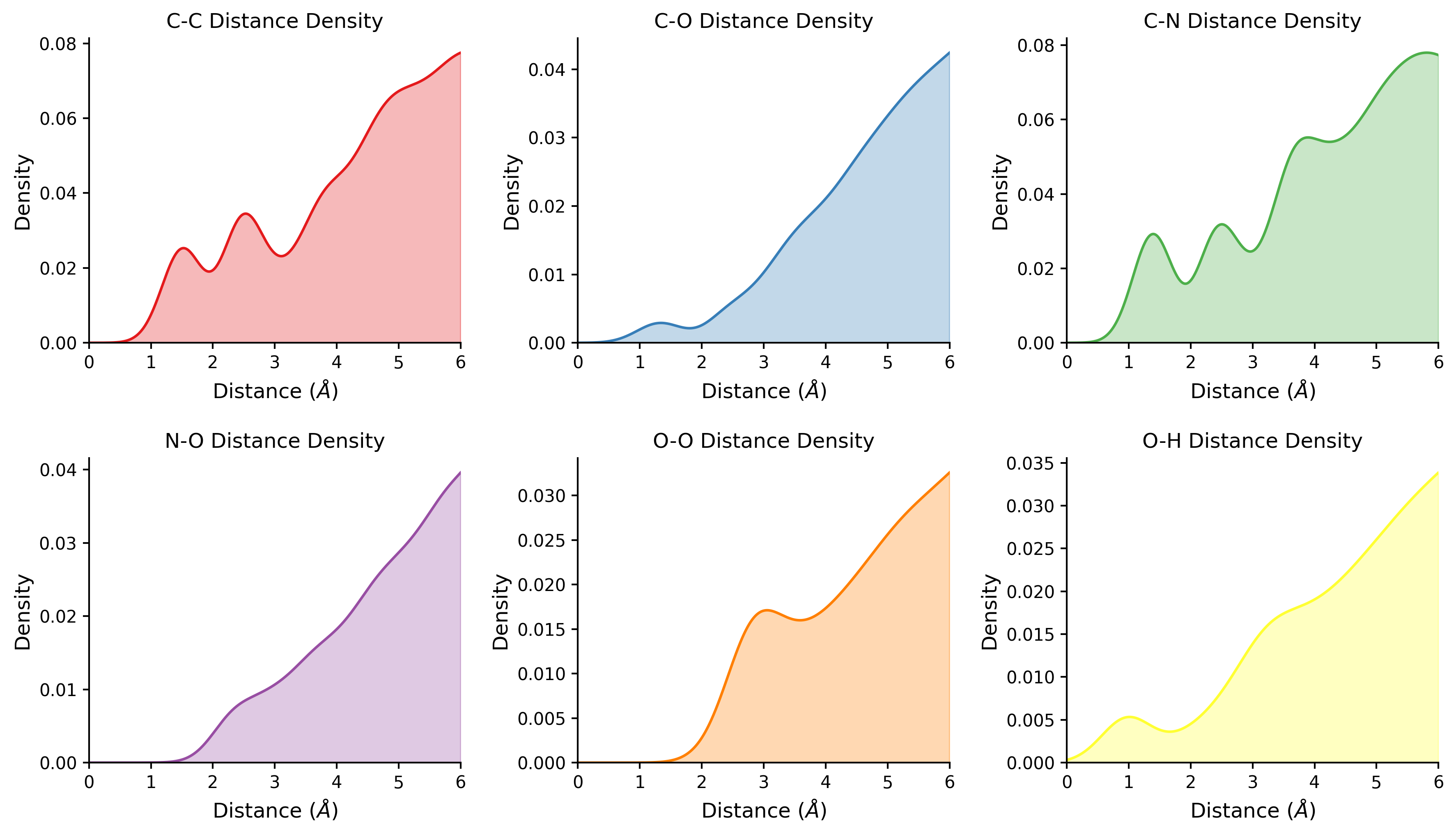}
    \caption{Pair distances, UBio-Mol26 (\texttt{def2-SVP})}
    \label{fig:pair_dist_svp}
  \end{subfigure}
  \vspace{0.25cm}
  \begin{subfigure}[b]{0.66\linewidth}
    \includegraphics[width=\linewidth]{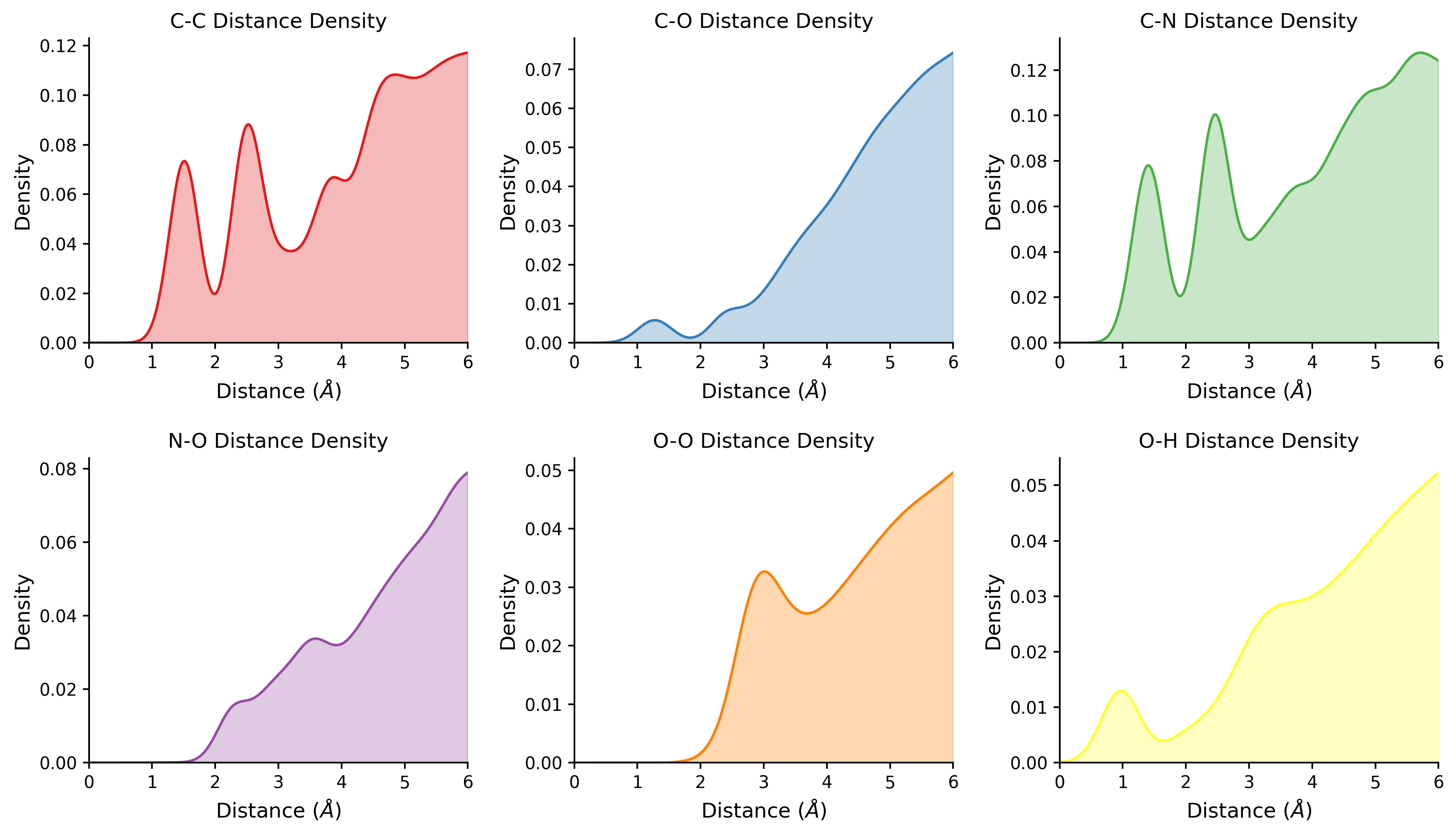}
    \caption{Pair distances, UBio-Mol26 (\texttt{def2-TZVPD})}
    \label{fig:pair_dist_tzvpd}
  \end{subfigure}
  \captionsetup{font={footnotesize,stretch=1.0}}
  \caption{\textbf{Interatomic-distance and elemental distributions.} \textbf{(a--c)}~C, N, O and H pair-distance distributions for OMol25 and the two UBio-Mol26 subsets. Small-molecule datasets lose structural correlation beyond $5$--$6$\,\AA; UBio-Mol26 retains it, which is what a receptive field reaching non-covalent distances has to be trained against. Elemental coverage is in \textbf{(d,e)} on the following page.}
  \label{fig:data_distributions}
\end{figure}
\begin{figure}[H]\ContinuedFloat
  \centering
  \setcounter{subfigure}{3}
  \begin{subfigure}[b]{0.92\linewidth}
    \includegraphics[width=\linewidth]{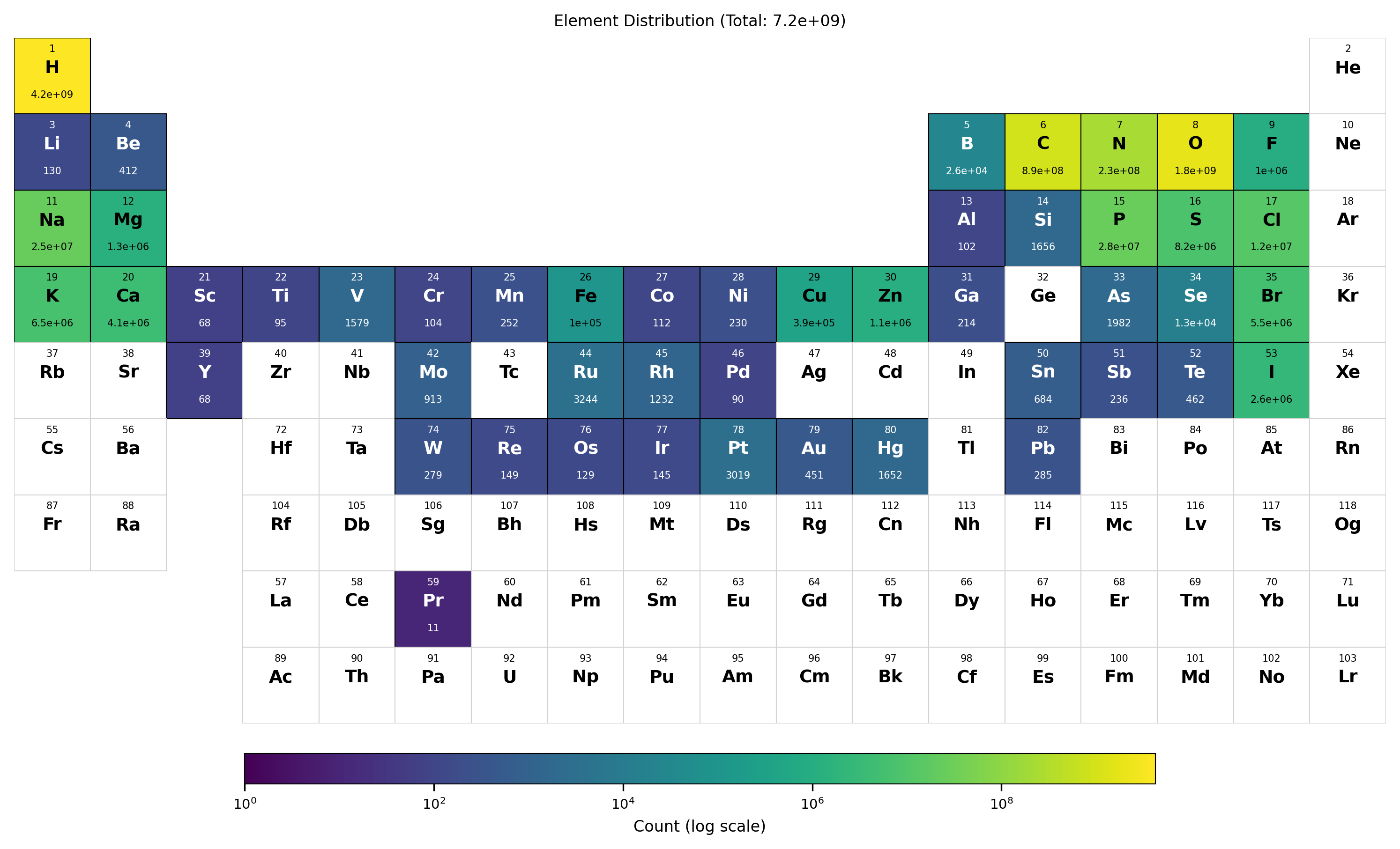}
    \caption{Elemental coverage, \texttt{def2-SVP}}
    \label{fig:atom_dist_svp}
  \end{subfigure}
  \vspace{0.3cm}
  \begin{subfigure}[b]{0.92\linewidth}
    \includegraphics[width=\linewidth]{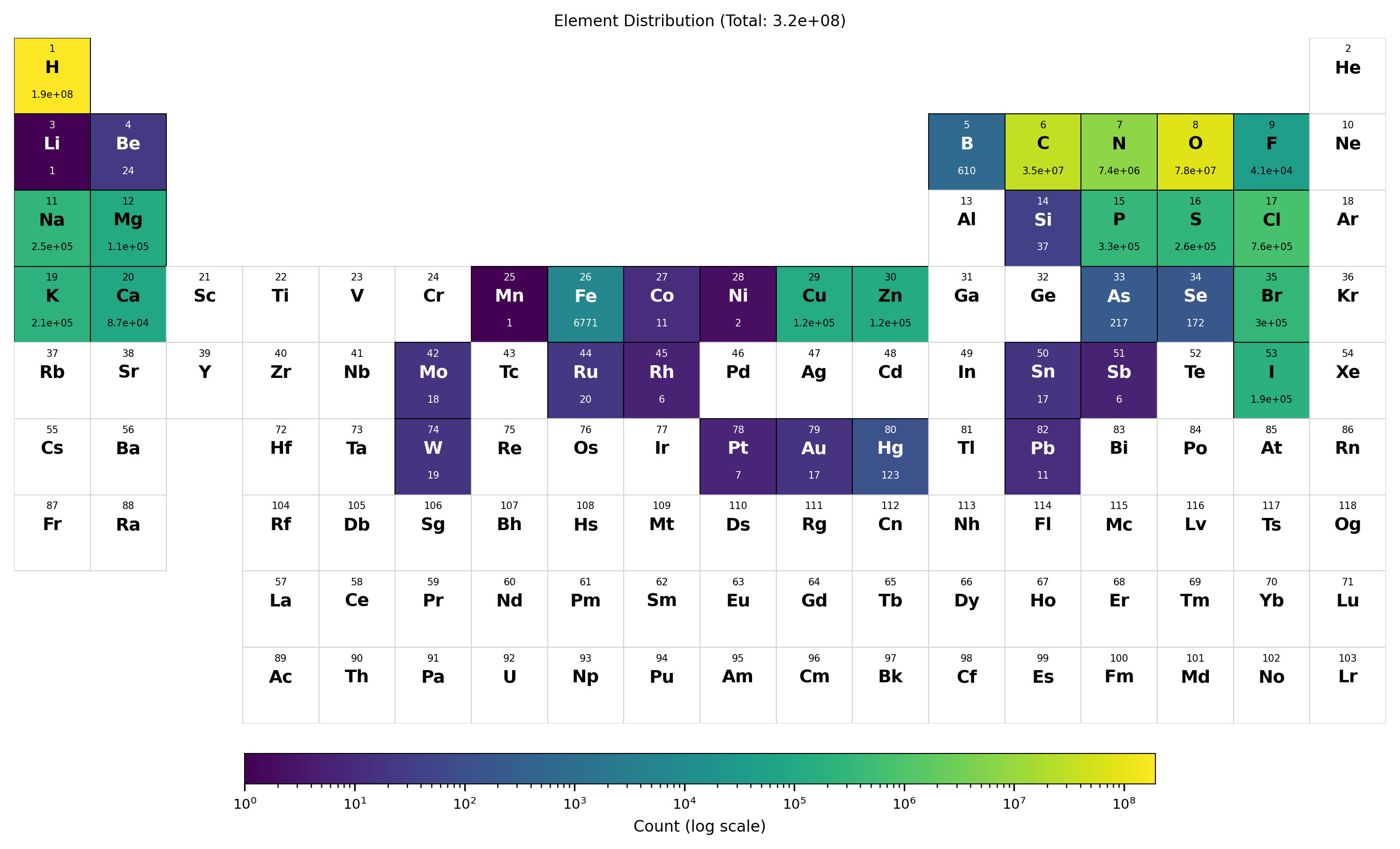}
    \caption{Elemental coverage, \texttt{def2-TZVPD}}
    \label{fig:atom_dist_tzvpd}
  \end{subfigure}
  \captionsetup{font={footnotesize,stretch=1.0}}
  \caption{\emph{(continued)} \textbf{(d,e)}~Element frequencies across the periodic table for each subset, showing coverage of the biologically relevant elements including the trace ions (Mg$^{2+}$, Zn$^{2+}$ and others) that the metal-site results depend on.}
\end{figure}

\clearpage

\begin{figure}[H]
  \centering
  \makebox[\linewidth][c]{%
    \begin{subfigure}[b]{0.4693\figwidth}
      \includegraphics[width=\linewidth]{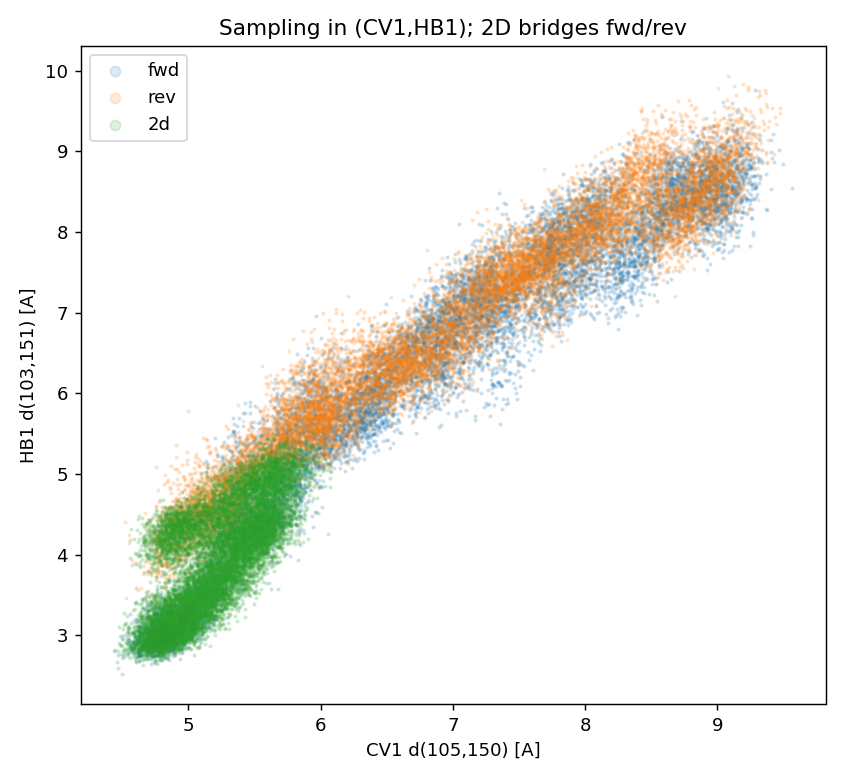}
      \caption{Sampling coverage in $(\text{CV}_1, \text{CV}_2)$}
      \label{fig:csa_sampling}
    \end{subfigure}%
    \hspace{12pt}%
    \begin{subfigure}[b]{0.5057\figwidth}
      \includegraphics[width=\linewidth]{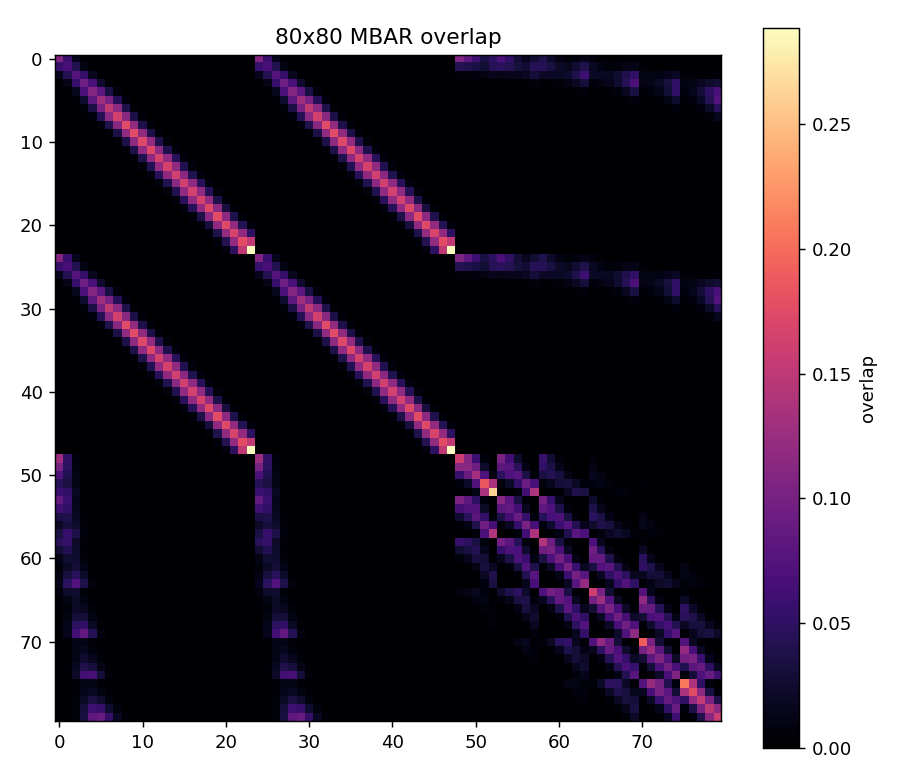}
      \caption{80$\times$80 MBAR state-overlap matrix}
      \label{fig:csa_overlap}
    \end{subfigure}%
  }
  \caption{\textbf{Phase-space connectivity of the 80-state Cyclosporine~A 2D-MBAR.} (a)~Every production frame in the $(\text{CV}_1, \text{CV}_2)$ plane, coloured by ensemble (forward 1D, reverse 1D, 2D patches); the localized patches bridge the two one-dimensional manifolds that a single reaction coordinate leaves disconnected. (b)~MBAR state-overlap matrix over all 80 states (24 forward + 24 reverse 1D windows + 32 patch cells). Overlap is quantified between states that neighbour one another \emph{in the collective-variable lattice}, not in matrix index order: the worst such pair is $0.080$ within each 1D ladder (23 pairs, median $0.235$) and $0.116$ between adjacent patch cells (50 pairs, median $0.164$); the classical baseline is comparable at $0.109$ and $0.124$. No state is isolated, so the combined ensemble is connected and the global estimate well-conditioned---underpinning the PMF of Fig.~\ref{fig:csa_pmf}. The off-diagonal zeros visible where the matrix crosses from one ensemble to the next are index artefacts, not gaps: state 48 begins the patch block at $\text{CV}_1=4.7\,\text{\AA}$ while state 47 is the reverse window at $9.3\,\text{\AA}$, so the two are $4.6\,\text{\AA}$ apart and are not expected to overlap.}\label{fig:csa_diagnostics_si}
\end{figure}

\clearpage

\begin{table}[htbp]
  \centering
  \captionsetup{singlelinecheck=false}
  \caption{\textbf{Quantitative validation summary across all condensed-phase tiers.} Each row is one observable with an independent reference; dashes mark comparisons not applicable to that potential (no classical run performed, or a model-specific quantity). Classical columns are the matched references of each Results subsection: Li--Merz 12-6-4 for Mg$^{2+}$ and KcsA, plain 12-6 Lennard-Jones for KcsA only. KcsA values pool late-trajectory frames of all replicas; all three potentials are run at $n=5$. RNA Mg$^{2+}$ values are five-replica means per potential, each trajectory $1.0$~ns. The $g_\text{OO}$ first-minimum row carries no reference value: the benchmark X-ray study~\cite{skinner2013benchmark} quotes uncertainties for the maximum only, and ours records the mild over-structuring expected of DFT-quality water with classical nuclei. $^{\P}$Self-diffusion is the only row here on which the model is clearly worse than experiment, and is listed for that reason. It carries the Yeh--Hummer finite-size correction evaluated with the \emph{experimental} viscosity; the uncorrected value is $1.60$, and $18\%$ is a lower bound on the shortfall (Results; Methods). The four backbone-plane rows are listed individually because their four-pair mean (2.94/2.99/3.00~\AA) conceals the divergence: the two spacings involving the G77 carbonyl differ by up to 0.38~\AA{} between potentials while the other two agree to 0.10~\AA. Being coupled to S2 occupancy, these spacings are not independent controls; the occupancy-independent statements are the global backbone RMSD and the deep-site coordination. $^\dagger$The UBio-MolFM S2 value pools all five replicas, including seed 126; over the other four it is $2.60\pm0.08$~\AA, the figure quoted in the main text. $^\S$The UBio-MolFM replicas were run under an isotropic barostat and the classical ones under a semi-isotropic barostat, so on the UBio-MolFM side the box aspect ratio is fixed and the bilayer cannot relax its area and thickness independently (Methods). Area per lipid, hydrophobic thickness and hydrophobic-core hydration are therefore omitted from this table entirely: measured on those runs they report that constraint rather than the potential, and area per lipid in particular is pinned by construction---it ranges over $0.25$~\AA$^2$ within an isotropic trajectory against $5.4$ when the same potential runs with the normal free---so its numerical agreement across potentials is not evidence of anything. What those observables do under a released box is reported in Results and Methods, and is unfavourable to this potential. Chain tilt is flagged for the same reason; the four rows above it are box-independent and are the membrane statements the paper relies on. $^{\#}$Pooled, like every KcsA row, over each trajectory's own late half---$500$--$1000$~ps isotropic, $200$--$400$~ps released. The main text instead compares the two on the single matched $200$--$400$~ps window, where the isotropic value is $0.237$. $^{\parallel}$Every lipid structural quantity here is computed on the saturated palmitoyl \emph{sn}-1 chain alone---$352$ chains, $15$ resolved carbon positions per chain---and not on the unsaturated oleoyl \emph{sn}-2 chain. That avoids the double bond, as is conventional for an order parameter, but the chain a packing defect would show up in first is therefore unmeasured. $^\ddagger$Two spreads exist for this quantity and are not interchangeable: the width of the pooled late-half frame distribution ($0.20$, $0.25$, $0.27$~\AA), which is what the histogram of Fig.~\ref{fig:transmembrane_channel}c displays, and the replica-to-replica standard deviation ($0.13$, $0.10$, $0.18$~\AA), which is the convention Methods states for five-replica scalars and therefore the one quoted after a $\pm$ in the main text.}
  \label{tab:validation_summary}
  \footnotesize\setlength{\tabcolsep}{4pt}\renewcommand{\arraystretch}{1.15}
  \resizebox{\linewidth}{!}{%
  \begin{tabular}{@{}l l r r r l@{}}
    \toprule
    \textbf{Tier} & \textbf{Observable} & \textbf{UBio-MolFM} & \textbf{12-6-4} & \textbf{12-6} & \textbf{Reference} \\
    \midrule
    \multirow{6}{*}{Water}
      & Mass density (g\,cm$^{-3}$)              & 0.987 & --   & --   & 0.997 (std.\ tabulated) \\
      & Self-diffusion $D$ ($10^{-5}$\,cm$^2$\,s$^{-1}$)$^{\P}$ & 1.88 & -- & -- & 2.299 (PFG NMR)~\cite{holz2000} \\
      & $g_\text{OO}$ first peak (\AA)           & 2.79  & --   & --   & 2.80(1) (X-ray)~\cite{skinner2013benchmark} \\
      & $g_\text{OO}$ first-peak height          & 2.58  & --   & --   & 2.57(5) (X-ray)~\cite{skinner2013benchmark} \\
      & $g_\text{OO}$ first minimum              & 0.79  & --   & --   & --- (over-structured; see note) \\
      & O--O coordination number                 & 4.3   & --   & --   & 4.3(2) at 3.30(5)\,\AA~\cite{skinner2014structure} \\
    \addlinespace
    \multirow{3}{*}{Electrolyte}
      & Na$^+$--O first peak (\AA)               & 2.37  & --   & --   & 2.38 (exp.)~\cite{galib2017revisiting} \\
      & K$^+$--O first peak (\AA)                & 2.73  & --   & --   & 2.6--2.8 (exp.)~\cite{ohtaki1993hydration} \\
      & Cl$^-$--O first peak (\AA)               & 3.2   & --   & --   & 3.1--3.2 (exp.)~\cite{ohtaki1993hydration} \\
    \addlinespace
    \multirow{5}{*}{RNA Mg$^{2+}$}
      & Mg$^{2+}$--O$_\text{water}$ peak (\AA)   & 2.08  & 2.03 & --   & $2.09\pm0.04$ (XRD)~\cite{ohtaki1993hydration} \\
      & O--Mg--O \emph{cis} width $\sigma$ ($^\circ$) & 6.5 & 4.8 & -- & $6.3$ (AIMD)~\cite{yu2022mgtfsi} \\
      & O--Mg--O \emph{trans} width $\sigma$ ($^\circ$) & 5.1 & 3.7 & -- & $5.0$ (AIMD)~\cite{yu2022mgtfsi} \\
      & Mg$^{2+}$--O$_\text{P}$ width $\sigma$ (\AA)  & 0.076 & 0.042 & -- & --- (model prediction) \\
      & Inner-sphere coordination number         & 6.0   & 6.0  & --   & 6, octahedral~\cite{ohtaki1993hydration} \\
    \addlinespace
    \multirow{13}{*}{KcsA}
      & K$^+$--carbonyl O, sites S3/S4 (\AA)     & 2.56  & 2.61 & 2.58 & 2.7--2.85 (crystal)~\cite{zhou2001kcsa} \\
      & K$^+$--carbonyl O coordination, S3/S4   & 7.97  & 7.58 & 7.61 & --- (model prediction) \\
      & K$^+$--carbonyl O, site S2 (\AA)$^\dagger$ & 3.15  & 5.61 & 8.51 & --- (site vacated classically) \\
      & Backbone planes T75--V76 (\AA)           & 3.05  & 3.00 & 2.99 & --- (G77-independent) \\
      & Backbone planes V76--G77 (\AA)           & 3.00  & 2.83 & 2.85 & --- (set by G77 orientation) \\
      & Backbone planes G77--Y78 (\AA)           & 3.06  & 3.44 & 3.44 & --- (G77 orientation; S2 cage) \\
      & Backbone planes Y78--G79 (\AA)           & 2.66  & 2.69 & 2.72 & --- (G77-independent) \\
      & Closest K$^+$--K$^+$ (\AA)$^\ddagger$    & 3.32  & 3.74 & 3.71 & --- (model prediction) \\
      & Contiguous contact column (\% frames)    & 28--89 & 0   & 0    & --- (model prediction) \\
      & Acyl C--C bond (\AA)$^{\parallel}$       & 1.530 & 1.542 & 1.542 & --- (box-independent) \\
      & Acyl C--C--C angle ($^\circ$)$^{\parallel}$ & 113.4 & 113.3 & 113.2 & --- (box-independent) \\
      & Gauche fraction$^{\parallel}$            & 0.28  & 0.34 & 0.33 & --- \emph{(barostat-affected: $0.30$ released)} \\
      & Chain order, director-referenced$^{\parallel}$ & 0.232 & 0.257 & 0.257 & --- (box-independent; $0.222$ released, gap wider$^{\#}$) \\
      & Chain tilt ($^\circ$)$^{\S\parallel}$    & 33.2  & 25.0 & 25.2 & --- \emph{(barostat-confounded)} \\
    \bottomrule
  \end{tabular}}%

\end{table}

\clearpage

\begin{table}[htbp]
  \centering
  \caption{\textbf{Hydration properties and ion pairing for 0.15~mol/L NaCl.} UBio-MolFM values are averaged over the equilibrated second half of a 1-ns $NPT$ trajectory (bin width $0.06$~\AA; coordination numbers integrated to the first RDF minimum); DFT values are from Ref.~\cite{galib2017revisiting}. Contact ion pairs are rarely sampled at this concentration (11 pairs), so the Na--Cl RDF is dominated by the solvent-separated pair (SSIP); see Fig.~\ref{fig:thermo}e.}
  \label{tab:nacl_comparison}
  \begin{tabular}{llccc}
    \toprule
    \textbf{Interaction}    & \textbf{Method/Exp.}     & \textbf{First Peak (\AA)} & \textbf{CN}                     & \textbf{Ref.}              \\
    \midrule
    \multirow{5}{*}{Na--O}  & \textbf{UBio-MolFM}      & \textbf{2.37}             & \textbf{5.83}                   & \textbf{This Work}         \\
                            & Exp. (XRD)               & $2.384 \pm 0.003$         & $5.5 \pm 0.3$                   & \cite{galib2017revisiting} \\
                            & DFT (revPBE)             & 2.45                      & 5.7                             & \cite{galib2017revisiting} \\
                            & DFT (BLYP)               & 2.40                      & 4.9                             & \cite{galib2017revisiting} \\
    \midrule
    \multirow{2}{*}{Cl--O}  & \textbf{UBio-MolFM}      & \textbf{3.21}             & \textbf{7.45 (cutoff 3.87 \AA)} & \textbf{This Work}         \\
                            & Exp. (XRD)               & $\sim 3.2$                & --                              & \cite{hwang2021hydration}  \\
    \midrule
    Na--Cl                  & \textbf{UBio-MolFM}      & \textbf{$\sim$4.6 (SSIP)} & --                              & \textbf{This Work}         \\
    \bottomrule
  \end{tabular}
\end{table}

\clearpage

\begin{table}[H]
  \centering
  \caption{\textbf{Gas-phase potential-energy-surface spot-check for Cyclosporine~A.} Single-point energies (and forces where available) on 100 decorrelated 196-atom solute conformers drawn from the umbrella windows of Fig.~\ref{fig:csa_pmf}, stratified over five $\text{CV}_1$ regions. Energies are shift-corrected on each pair's own overlap, because the reference conventions differ by thousands of kcal/mol and only within-column differences are meaningful. \textbf{$\omega$B97M-V} is OMol25's level, on which the model's energy head is calibrated; \textbf{$\omega$B97M-D3(BJ), mixed basis} is exactly the protocol at which UBio-Mol26's high-fidelity tier was collected (def2-TZVP on H and metals, grid~3, SCF $10^{-6}$; \S\ref{sec:data_construction}); \textbf{revDSD-PBEP86-D4} is an independent high-accuracy reference. $\Delta\Delta E$ is the closed-minus-open systematic offset, the component that bears on the $3.54\,\text{kcal/mol}$ free energy. Three conclusions. The model's deviation from its calibration level ($0.44$) is smaller than any disagreement among the references themselves ($\geq$$1.18$). The $\approx$$3\,\text{kcal/mol}$ spread in $\Delta\Delta E$ falls on the nonlocal-versus-pairwise dispersion divide, not on the model: the two pairwise-dispersion references agree with each other to $+0.28$ while both sit $\approx$$3\,\text{kcal/mol}$ from nonlocal $\omega$B97M-V, and revDSD-PBEP86-D4 shares $\omega$B97M-V's uniform def2-TZVPD basis, so the split is not a basis-set artefact. And the classical force field, evaluated as single points on the \emph{same} geometries with no sampling, solvent, integrator or constraints, is $22\times$ further from $\omega$B97M-V than the model is, at a correlation of $0.80$ rather than $0.9996$; its energy spread over the set ($107\,\text{kcal/mol}$) exceeds DFT's ($101$), so its failure is misordering, not compression. $\Delta\Delta E$ is not tabulated for the classical rows because a $\pm$$10\,\text{kcal/mol}$ scatter at $r=0.80$ makes a two-region mean difference uninformative; the per-region breakdown is in Methods. Two reading notes. $n$ falls below the $100$ conformers attempted because each level of theory converges on a different subset: $99$ at $\omega$B97M-V, $93$ at $\omega$B97M-D3(BJ) and $83$ at revDSD-PBEP86-D4, the ref-vs-ref rows using each pair's own overlap ($92$, $82$, $78$). The failures are not uniform over the five sampled $\text{CV}_1$ regions: they fall preferentially in the compact \emph{closed} region ($1$ of $1$, $5$ of $7$ and $7$ of $17$ at the three levels, against that region's $20$ of $100$ share), as steric strain makes SCF hardest there. What bounds the resulting bias is that the model's own error is flat across all five regions ($0.29$--$0.51\,\text{kcal/mol}$; SI~S4.3), so the comparison is not being scored on the easy geometries. Consequently $\Delta\Delta E$ is a per-pair quantity and is \emph{not additive across rows}: $+0.52$ and $+3.33$ do not compose to the $-2.51$ of the model-vs-revDSD row, because each is shift-corrected on a different overlap subset.}
  \label{tab:csa_qm}
  \footnotesize
  \setlength{\tabcolsep}{5pt}
  \resizebox{\textwidth}{!}{%
  \begin{tabular}{llcccccc}
    \toprule
    \textbf{Comparison} & \textbf{Reference} & \textbf{$n$} & \textbf{$E$ MAE} & \textbf{$E$ RMSE} & \textbf{$F$ MAE} & \textbf{$r$} & \textbf{$\Delta\Delta E$} \\
    & & & (kcal/mol) & (kcal/mol) & (meV/\AA) & & (kcal/mol) \\
    \midrule
    \multirow{3}{*}{UBio-MolFM vs}
      & $\omega$B97M-V/def2-TZVPD (calibration level) & 99 & \textbf{0.44} & 0.58 & \textbf{7.2} & 0.9996 & $+0.52$ \\
      & $\omega$B97M-D3(BJ), mixed basis (collection level) & 93 & 1.09 & 1.33 & 31.5 & 0.9978 & $-2.21$ \\
      & revDSD-PBEP86-D4/def2-TZVPD (high-accuracy reference) & 83 & 1.45 & 1.80 & --- & 0.9950 & $-2.51$ \\
    \addlinespace
    \addlinespace
    \multirow{3}{*}{\begin{tabular}[t]{@{}l@{}}classical FF vs\\{\scriptsize(GAFF2/AM1-BCC,}\\{\scriptsize same geometries)}\end{tabular}}
      & $\omega$B97M-V/def2-TZVPD & 99 & \textbf{9.90} & 12.46 & --- & \textbf{0.801} & --- \\
      & $\omega$B97M-D3(BJ), mixed basis & 93 & 9.71 & 12.19 & --- & 0.790 & --- \\
      & revDSD-PBEP86-D4/def2-TZVPD & 83 & 9.63 & 12.18 & --- & 0.799 & --- \\
    \addlinespace
    \multirow{3}{*}{reference vs reference}
      & $\omega$B97M-D3(BJ) vs $\omega$B97M-V & 92 & 1.18 & 1.41 & --- & 0.9980 & $+2.73$ \\
      & revDSD-PBEP86-D4 vs $\omega$B97M-V & 82 & 1.50 & 1.86 & --- & 0.9951 & $+3.33$ \\
      & revDSD-PBEP86-D4 vs $\omega$B97M-D3(BJ) & 78 & 1.58 & 1.96 & --- & 0.9934 & $\mathbf{+0.28}$ \\
    \bottomrule
  \end{tabular}%
  }
\end{table}

\clearpage

\clearpage
\section*{Supplementary Information (SI)}

\setcounter{figure}{0}
\setcounter{table}{0}
\captionsetup[figure]{name={Supplementary Figure}}
\captionsetup[table]{name={Supplementary Table}}
\captionsetup{font={footnotesize,stretch=1.0}}

This Supplementary Information follows the order of the paper. S1 documents the training data, S2 the model and the training curriculum, S3 the evaluation protocol together with the cross-tier validation summary, and S4 the per-system supporting analyses in the order the Results present them---bulk water and electrolytes, the RNA Mg$^{2+}$ site, Cyclosporine~A, and the KcsA channel. S5 gives the single-GPU performance measurements, S6 the reproducibility inventory and S7 the supplementary tables. Floats carrying evidence for main-text claims are collected in the preceding \emph{Extended Data} section and are cited from here by number; the \emph{Supplementary} tables of S7 carry reproducibility reference material and are numbered independently from~1.

\subsection*{S1. Data}

\subsubsection*{S1.1 Generation and multi-fidelity protocol}

Applying def2-TZVPD directly to large biological systems is limited as much by SCF convergence as by cost: near 600 atoms the convergence rate falls from $>$90\% (def2-TZVP) to $<$20\% (def2-TZVPD). We therefore used $\omega$B97M-D3~\cite{wb97md3} (D3 dispersion in place of VV10) with a mixed basis---def2-TZVP for hydrogen and metal ions, def2-TZVPD elsewhere---which restored convergence to $>$60\%, and generated a larger def2-SVP portion (50--100$\times$ cheaper) to expand coverage tenfold.

Configurations were prepared by packing solutes and solvent with Packmol~\cite{packmol}, assigning initial parameters with AmberTools~\cite{ambertools}, minimizing and equilibrating in OpenMM~\cite{openmm}, then carving spherical clusters with MDAnalysis~\cite{mdanalysis}; single points were computed with GPU4PySCF~\cite{gpu4pyscf} through ASE~\cite{ase} at a $10^{-6}$~Hartree threshold, mixing optimization trajectories and MD-sampled structures 4:1. Because finite clusters cannot constrain bulk pressure and density responses, we added $63{,}603$ periodic condensed-phase configurations (mean $314$ atoms; $88.5\%$ peptide/DNA/RNA, $11.5\%$ ions and neat water, the latter counted inside the ionic fraction) at revPBE0-D3/\texttt{TZV2P-MOLOPT-PBE0} in CP2K~\cite{cp2k}; these anchor the bulk equation of state and ground the thermodynamic-consistency results reported in the main text.

A fourth set at $\omega$B97M-D3/\texttt{def2-TZVP} is generated for evaluation only and enters no training stage. It is not a pool of independent single points but a set of \emph{trajectories}---geometry relaxations and finite-temperature MD of one system each, every frame of a given shard sharing that system's composition and atom count. That is what makes the two offset-free energy measures well defined on this tier: relative energy referenced to each trajectory's first frame, and the frame-to-frame change $\Delta E = E_i - E_{i-1}$, neither of which requires the absolute per-atom energies to be comparable across functionals. Per-category trajectory and frame counts, sizes and elemental coverage are in Methods~\S\ref{sec:eval_benchmarks} and Supplementary Table~\ref{tab:extreme_tier}.

\subsubsection*{S1.2 Composition and coverage}

UBio-Mol26 spans drug-like molecules, proteins, DNA, RNA, lipids, water and biological ions across three DFT fidelities. Extended Data Fig.~\ref{fig:data_overview} covers system sizes, category composition, chemical-space position relative to OMol25 and carbon chemical-environment frequencies; Extended Data Fig.~\ref{fig:data_distributions} covers pair-distance and per-element distributions.

\clearpage
\subsection*{S2. Model and Training}

\subsubsection*{S2.1 Backbone}

The E2Former-V2 backbone---its node-centric Wigner-$6j$ factorization, Equivariant Axis-Aligned Sparsification, and fused on-the-fly attention kernel---is defined, derived, benchmarked and ablated in its own report~\cite{huang2026e2former} and that of its predecessor~\cite{li2025e2former}. We neither restate nor re-ablate it, and the architectural choice that is ours rather than inherited, the four-layer hybrid-cutoff stack, is given in Methods~\S\ref{sec:model}. What the backbone buys in throughput and memory on this work's systems is measured end-to-end in S5.

\subsubsection*{S2.2 Curriculum, branch objectives and force-label filtering}

The three logical stages are realized as three sequential runs: the two OMol25 runs S1 and S2, which differ in how forces are produced rather than in what they are trained on, and the single mixed-dataset run that is Stage~3. Methods~\S\ref{sec:training_strategy} gives the checkpoint lineage, per-run learning rates, step counts and device allocations; this section documents the mixed-stage data plumbing and the atom-balanced loader, which are the two parts a reimplementation is most likely to get wrong.

Energy and force supervision are present in every run, so there is no energy-only initialization phase at any point. Forces are obtained by automatic differentiation of the predicted energy from S2 onward; S1 alone carries a separately parameterized force head, which S2 retires. The Stage-3 checkpoints carry \emph{two} energy heads, \texttt{omol25} and \texttt{svp}, each paired with a parameter-free gradient-force head. Three of the four data branches share the \texttt{omol25} head; only the $0.05$ slice of the \texttt{svp} ratio described below instantiates the second. Biomolecular energy supervision is therefore present in the objective rather than absent from it, but it is marginal: it enters through that one slice, at a low sampling ratio, and on the head that is \emph{not} the one used at inference. Its gradients nonetheless reach the shared backbone, so we do not claim the model never sees a biomolecular energy label---only that biomolecular supervision is overwhelmingly force-based. The rest of the \texttt{def2-SVP} data supervises force \emph{directions} on the \texttt{omol25} head through the hinge below. All results reported in this work are evaluated through the \texttt{omol25} head.

Heterogeneity is handled by per-branch objectives rather than by per-fidelity output heads. The mixed loader carries \emph{four} branches, one per reference source, balanced by the sampling ratios $0.15$ (\texttt{omol25}), $1$ (\texttt{tzvpd}), $2$ (\texttt{tzv2p}, computed at revPBE0-D3) and $0.2$ (\texttt{svp}). Three objectives cover them. The \texttt{omol25} branch is the only one carrying a full energy-plus-force loss on the inference head. Its own sampling is size-weighted rather than uniform: OMol25 configurations above $200$ atoms, $\approx$2\,M labels, are drawn ten times as often as the rest, so the branch that exists to retain OMol25's chemical breadth is nonetheless concentrated at the upper end of OMol25's size range, closest to the systems this model targets. The two force-filtered branches, \texttt{tzvpd} and \texttt{tzv2p}, disable their energy terms and minimize the full per-atom force-vector residual $\lVert\widehat{\mathbf F}_i - \mathbf F_i\rVert_2$ on retained atoms; this is a vector loss, not a loss on force magnitude alone. The \texttt{svp} branch uses the magnitude-gated directional hinge of Methods Eq.~\eqref{eq:dirhinge}, which penalizes direction only once the reference force exceeds a magnitude threshold, so near-equilibrium atoms with ill-conditioned force directions do not dominate. Because the force-filtered branches carry no energy term, the $\omega$B97M-D3 versus $\omega$B97M-V offset between UBio-Mol26 and OMol25---and the revPBE0-D3 offset on \texttt{tzv2p}---is bypassed on those branches rather than reconciled. The hinge, however, throws away every SVP force magnitude, which under-uses the largest subset in the corpus; $0.05$ of the $0.2$ \texttt{svp} ratio is therefore split off onto the second \texttt{svp} energy head under the default energy-plus-autograd-force objective, with $0.15$ left on the directional hinge. The archived configuration files are the authoritative record of the branch-to-source mapping.

Force-label filtering is two-level and the two levels are often conflated. The static \texttt{def2-TZVPD} source is first \emph{sample}-filtered offline to form \texttt{def2-tzvpd-uma-filtered}, removing $20.7\%$ of configurations. Training then applies a \emph{per-atom} cosine gate at $\tau = 0.8$ to the prefiltered data, together with a force-norm gate at $0.05$. The two levels together suppress $\approx$$27\%$ of force labels. This is not a configuration-discard rate, and it is not attributable to $\tau = 0.8$ alone. The end-to-end lineage is summarized in Algorithm~\ref{alg:curriculum}.

\begin{algorithm}[H]
  \caption{Checkpoint lineage for UBio-MolFM: three logical stages as four runs}
  \label{alg:curriculum}
  \begin{algorithmic}[1]
    \Require OMol25 corpus $\mathcal{D}_O$ ($\approx$140M frames); UBio-Mol26 branch sources $\mathcal{D}_{\text{TZVPD}}, \mathcal{D}_{\text{TZV2P}}, \mathcal{D}_{\text{SVP}}$; branch sampling ratios $\mathbf{w}=(0.15,\,1,\,2,\,0.2)$ over $(\texttt{omol25},\texttt{tzvpd},\texttt{tzv2p},\texttt{svp})$
    \Ensure Stage-3 checkpoint $f_{\theta_{S3}}$, evaluated through the \texttt{omol25} head
    \Statex
    \Statex \textit{Every run below supervises energy and force jointly; no run has an energy-only objective. S1 predicts forces with a separately parameterized head; from S2 onward $\hat{\mathbf F} \gets -\nabla_{\mathbf R}\hat E$ by autograd.}
    \Statex
    \State \textbf{S1: pretraining} --- $\theta \gets$ random; train on $\mathcal{D}_O$ with $(\lambda_E,\lambda_F)=(4,10)$
    \State \hskip1em energy head \texttt{omol25} \emph{plus} a separately parameterized force head
    \State \hskip1em peak LR $4\!\times\!10^{-4}$, cosine horizon $1{,}400$k steps; take $\theta_{S1}$ at $1{,}000$k steps
    \Statex
    \State \textbf{S2: autograd forces} --- $\theta \gets \theta_{S1}$ \Comment{model-only load: optimizer and scheduler reset}
    \State \hskip1em retire the parameterized force head; $\hat{\mathbf F} \gets -\nabla_{\mathbf R}\hat E$ from here on
    \State \hskip1em train on the same $\mathcal{D}_O$, peak LR $2\!\times\!10^{-4}$, to $1{,}400$k cumulative steps ($400$k further) $\to \theta_{S2}$
    \Statex
    \State \textbf{S3: thermodynamic fusion} --- $\theta \gets \theta_{S2}$ (model-only); attach the second energy head \texttt{svp}
    \State \hskip1em \textsc{MixedRun}$(\theta,\ \text{LR}=5\!\times\!10^{-6},\ 400\text{k steps}) \to \theta_{S3}$ \Comment{first and only exposure to UBio-Mol26}
    \State \textbf{return} $\theta_{S3}$
    \Statex
    \Statex \textbf{procedure} \textsc{MixedRun}$(\theta, \text{LR}, N)$ \Comment{four branches, three objectives}
    \For{step $= 1 \dots N$}
      \State draw branch $b \sim \mathbf{w}$; draw an atom-balanced batch $B$ from source$(b)$ via \texttt{bs\_atom}
      \State $\hat E \gets f_\theta(B; \text{head}(b))$; $\hat{\mathbf F} \gets -\nabla_{\mathbf R}\hat E$
      \If{$b \in \{\texttt{tzvpd},\ \texttt{tzv2p}\}$} \Comment{force-filtered}
        \State retain atoms passing the per-atom gate $\cos(\hat{\mathbf F}_i,\mathbf F_i)\ge\tau$, $\tau=0.8$, and $\lVert\mathbf F_i\rVert_2 \ge 0.05$
        \State $\mathcal{L} \gets \operatorname{mean}_{i \in \text{retained}} \lVert\hat{\mathbf F}_i - \mathbf F_i\rVert_2$ \Comment{vector residual; energy term disabled}
      \ElsIf{$b = \texttt{svp}$ and on the \texttt{omol25} head ($0.15$ of its $0.2$)}
        \State $\mathcal{L} \gets \mathcal{L}_{\mathrm{dir}}$ of Eq.~\eqref{eq:dirhinge} \Comment{magnitude-gated directional hinge}
      \Else \Comment{\texttt{omol25}, and the $0.05$ \texttt{svp} slice on its own head}
        \State $\mathcal{L} \gets \lambda_E\lVert\hat E - E_{\text{ref}}\rVert_1 + \lambda_F\lVert\hat{\mathbf F} - \mathbf F_{\text{ref}}\rVert_1$
      \EndIf
      \State update $\theta$ by AdamW on $\mathcal{L}$, FP32 throughout
    \EndFor
    \State \textbf{return} $\theta$
  \end{algorithmic}
\end{algorithm}

\subsubsection*{S2.3 Atom-balanced data loading}

Configurations in the combined corpus span $15$--$1{,}370$ atoms, so batching by graph count leaves devices badly load-imbalanced: a batch of large solvated clusters and a batch of dipeptides differ by more than an order of magnitude in work. We therefore pack greedily by \emph{atom} budget, adding molecules to a batch until the \texttt{bs\_atom} budget of $2{,}048$ atoms is reached, which is why batch size appears as an atom count rather than a sample count throughout (Supplementary Table~\ref{tab:optim_schedule}). The node-centric neighbour representation that makes this packing efficient on-device---a dense $N \times k$ index tensor in place of sparse edge-centric gather/scatter, which removes concurrent-write (AtomicAdd) conflicts---is a property of the backbone rather than of this work, and is described in Refs.~\cite{huang2026e2former,li2025e2former}.

\clearpage
\subsection*{S3. Evaluation Protocol}

\subsubsection*{S3.1 Metric definitions}
\label{si:metric_definitions}
The three microscopic metrics of Table~\ref{tab:rel_energy_force} are all normalized per atom, so that values remain comparable across benchmark tiers whose system sizes differ by a factor of five. \emph{Force MAE} is the mean absolute error per atom, in meV/\AA, over every atom of every frame in a category. \emph{Relative-energy MAE} (relE) first sets the energy of each trajectory's initial frame to zero---independently for the predicted and the reference series, so that any constant cross-functional offset cancels---and then averages the per-frame absolute error in meV/atom. \emph{$\Delta E$ MAE} is the absolute error of the consecutive-frame energy difference $E_i - E_{i-1}$, again differenced independently for prediction and reference, also in meV/atom; because it discards everything but the local slope, it probes how faithfully a model tracks the topology of the potential-energy surface along a trajectory rather than its absolute placement. Absolute per-atom energy MAE is meaningful only where model and reference share a level of theory, which among our tiers is OMol-Bio-10k alone, and is reported there only.

\subsubsection*{S3.2 Quantitative validation summary}

Extended Data Table~\ref{tab:validation_summary} collects, in one place, every condensed-phase observable for which an experimental or first-principles reference exists, together with the matched classical value where one was computed. The main text discusses these results qualitatively and cites this table rather than repeating the numbers.

\clearpage
\subsection*{S4. Per-System Supporting Analyses}

The four subsections below follow the order of the Results: bulk water and physiological electrolytes, the structural Mg$^{2+}$ site in folded RNA, the Cyclosporine~A free-energy landscape, and the KcsA transmembrane channel.

\subsubsection*{S4.1 Bulk water and physiological electrolytes}

Benchmark-trained MLFFs can reproduce static energies and forces accurately yet drift in bulk density under unbiased $NPT$ dynamics; eSEN~\cite{esen}, for instance, systematically over-predicts liquid densities~\cite{qu2026allscaip}. We ran unbiased $NPT$ simulations of a 512-molecule water box (300~K, 1~bar) and measured the equilibrium mass density, comparing the pretrained potentials UMA-S-1p2~\cite{wood2025family}, MACE-OMol~\cite{mace,omol25}, and DPA-4~\cite{li2026dpa4} against UBio-MolFM (Supplementary Table~\ref{tab:density_benchmark}, Extended Data Fig.~\ref{fig:density_baselines_si}). The baselines diverge within the first tens of picoseconds---DPA-4 collapses toward a gas-like density while MACE-OMol and UMA-S-1p2 over-densify by $9$--$12\%$---whereas UBio-MolFM holds within $1\%$ of experiment for the full nanosecond.

\paragraph{What the periodic tier does: the Stage-2 control.}
That UBio-Mol26's periodic subset is what calibrates the condensed-phase equation of state (\S\ref{sec:data_construction}) is testable on our own lineage, because Stage~2 has seen only OMol25---finite clusters computed without periodic electrostatics---and Stage~3 is the first exposure to the periodic tier. Run on the identical 512-water box under the identical $NPT$ protocol and at a matched temperature (measured means $298.1$ and $298.6$~K against a $300$~K setpoint), Stage~2 leaves the packed starting density and plateaus at $1.118\pm0.006\,\text{g\,cm}^{-3}$, $12.1\%$ above experiment, where Stage~3 gives $0.987$, $1.0\%$ below. The Stage-2 failure is not merely large, it is \emph{the same failure the finite-cluster baselines make}: MACE-OMol and UMA-S-1p2 over-densify by $\sim$$9$ and $\sim$$11\%$, Stage~2 by $12\%$. Three independently trained models and three architectures share one property of their training data---no periodic condensed-phase labels---and one pathology.

Two limits bound the inference. The Stage-2 run is $100$~ps against Stage~3's $1$~ns; its density plateaus by $\approx$$20$~ps and is flat over the remaining $80$ (means by fifths $1.049$, $1.119$, $1.117$, $1.117$, $1.121$), so the plateau is established, but the two runs are not length-matched. And Stage~3 adds \emph{all} of UBio-Mol26 rather than the periodic tier alone, so what this experiment shows is that the biomolecular corpus fixes the density; attributing that to the periodic tier rests on the argument that the other two tiers are finite clusters and cannot constrain a macroscopic equation of state however large they are. A run with the \texttt{tzv2p} branch ablated would test the attribution directly, and we have not done it.

\subsubsection*{S4.2 Mg\texorpdfstring{$^{2+}$}{2+} site in the BWYV RNA pseudoknot}

Both potentials were run as five independent replicas from a single common equilibrated configuration, differing only in thermostat seed (Methods): five replicas of $1.0$~ns per potential, all ten complete and equal in length. Radial, angular and pose distributions are accumulated over all frames of all replicas of each potential, and the widths reported in the main text and in Extended Data Table~\ref{tab:validation_summary} are resolved per replica before they are compared, so that seed-to-seed scatter is separated from the thermal width being measured.

One quantity is deliberately reported as a width and not as a position. The Mg$^{2+}$--O$_\text{P}$--P pose-angle distribution is the broader one under UBio-MolFM, and at the full $1.0$~ns its \emph{mode} is also the lower one by a margin that is statistically significant ($147.1\pm10.3^\circ$ against $165.5\pm2.5^\circ$, $p=0.015$). We still decline to build on the mode, because it is not converged at this trajectory length: the per-replica modes span $26^\circ$ against the classical $7^\circ$, one replica moves by $26^\circ$ when the window is extended from $0.8$ to $1.0$~ns, four of the five sit at $138$--$148^\circ$ while the fifth is indistinguishable from classical, and the five-replica mean still lies above the $120$--$140^\circ$ reported for crystallographic monodentate binding. Only the width enters the claims of \S\ref{subsec:rna_metal_coordination} (Methods).

\paragraph{The external references are aqueous, not RNA-bound.}
No measurement of the thermal width of an inner-sphere metal coordination shell inside a folded nucleic acid exists, so the widths of \S\ref{subsec:rna_metal_coordination} are compared against \emph{ab initio} dynamics and scattering of \emph{aqueous} Mg$^{2+}$~\cite{yu2022mgtfsi,ohtaki1993hydration}. The transfer rests on two grounds. The coordination shell is chemically the same in both settings---five waters and one oxygen at octahedral angles---and the librational stiffness of that shell is set by the Mg$^{2+}$ ligand field, which is far stronger than any perturbation the fold can transmit through the single phosphate-anchored ligand. Neither ground is a bound. Tethering one of the six ligands to a rigid backbone could narrow the true width relative to the aqueous reference, and that would move the reference \emph{toward} the fitted 12-6-4 potential rather than toward ours; nothing in these trajectories excludes it. We therefore read the width agreement as consistency with the best available reference and not as a measurement of this site. The first-shell Mg--O distance is exempt, being a position rather than a width and correspondingly less sensitive to the surroundings.

Per-replica values for every observable in this scenario, together with the trajectories they are computed from, are in the UBio-MolFM-MD release (S6).

\subsubsection*{S4.3 Cyclosporine~A}

\paragraph{Choice of the two latch coordinates.}
The two latch coordinates are defined structurally rather than empirically: of the four free backbone N--H donors in the canonical sequence, Abu2 and Val5 form the reciprocal transannular pair, so $\text{HB}_1$ and $\text{HB}_2$ follow from the covalent topology and not from a prior occupancy screen (Methods).

\paragraph{2D-MBAR sampling diagnostics.}
Phase-space connectivity of the 80-state campaign---the state-overlap matrix, its
lattice-neighbour statistics and the per-window sample counts---is given in Extended Data
Fig.~\ref{fig:csa_diagnostics_si}, and the thresholds read off it in
Methods~\S\ref{sec:simulation_protocols}.

\paragraph{Conformer diagnostics: the \emph{cis}-manifold caveat.}
The main-text landscape (Fig.~\ref{fig:csa_pmf}) is sampled entirely on the \emph{cis}-amide manifold; the following diagnostics establish this and add mechanistic depth to the open state. All quantities are plotted against $\text{CV}_1$ and compared, where relevant, with the classical-force-field baseline.

\paragraph{Spot-check results in detail.}
Three findings support \S\ref{subsec:csa_thermodynamics}. The model's error is flat across all five sampled $\text{CV}_1$ regions ($0.29$--$0.51\,\text{kcal/mol}$), so no basin is preferentially misdescribed. The disagreement between the reference levels partitions cleanly along the dispersion treatment: the high-accuracy reference and the mixed-basis D3(BJ) reference, both using pairwise dispersion, agree with each other on the closed-versus-open energy to $0.28$~kcal/mol (CI $[-0.77,+1.37]$) while each differs from the nonlocal-VV10 reference by $+2.73$ (CI $[+2.32,+3.15]$) and $+3.33$ (CI $[+2.54,+4.17]$); those intervals are disjoint, and the VV10 boundary is also the only one across which the discrepancy varies systematically with $\text{CV}_1$. The model tracks its own calibration level to $0.52$~kcal/mol on the same quantity, which is why the offset is attributed to the dispersion model rather than to the potential.

\paragraph{The classical surface: per-region error and the charge-refit control.}
Two checks bound the objection that the AM1-BCC charges, fitted on the single closed crystal conformer with all four transannular hydrogen bonds formed, under-weight those hydrogen bonds. First, the classical single-point error is not concentrated in the closed basin: per-region MAE runs $7.8$--$11.4$~kcal/mol and is worst at the \emph{open} minimum (ratio $1.46$). Second, re-deriving the charges from an open conformer under the identical protocol shifts them most on precisely the carbonyl oxygens at issue ($+0.067$ to $+0.084$~e; overall $0.012$~e MAE) yet moves relative energies by only $1.17$~kcal/mol (max $4.85$, $r=0.998$)---at most $12\%$ of the discrepancy. A multi-conformer charge fit therefore cannot close a $10$~kcal/mol gap.

\paragraph{Classical-baseline control: the paired three-window test.}
Because the classical runs use rigid X--H bonds at $dt = 2$~fs where ours are unconstrained at $0.5$~fs, three closed-basin windows ($\text{CV}_1 = 4.7$, $4.9$, $5.1$~\AA) were first re-run unconstrained at $0.5$~fs with seeds, spring constants, thermostat, $100$~ps equilibration and $400$~ps production all held fixed. The biased $\text{CV}_1$ distributions were indistinguishable (histogram overlap $0.82$--$0.95$; Kolmogorov--Smirnov statistic $0.03$--$0.19$), and over the bins holding at least $1\%$ of each dataset the local PMF agreed to $0.105$~kcal/mol in depth and $0.081$~kcal/mol MAE in shape. We note a metric change: the pre-specified form of this test used \emph{all} finite bins and returned $1.10$~kcal/mol, but its outermost bins hold $6$--$24$ of $\sim$$7{,}000$ samples and imply a $5.1$~kcal/mol rise across $0.96$~\AA, which the $1.5$~kcal/mol global span contradicts. We report both values and treat the restriction to well-populated bins as a specification fix rather than a reinterpretation of the result.

\paragraph{Classical-baseline control: full 80-state campaign.}
The entire 80-state campaign was then re-run at $0.5$~fs without X--H constraints, and $F(\text{CV}_1)$ differs in shape from the $2$-fs surface by $0.386$~kcal/mol MAE over 43 bins, maximum $2.774$---above the $0.30$ threshold this test was pre-specified against, and $4.8\times$ the paired bound above. We report that number and then decline to use it as evidence, because it cannot carry any: each campaign's own half-to-half convergence drift is $1.925$ and $2.055$~kcal/mol, both \emph{larger than the span each is measuring} ($1.506$ and $1.224$) and roughly five times the difference in question, and subtracting two independently converged surfaces adds their uncertainties where the paired test does not. The aligned run is also the less converged of the two: its cumulative drift does not fall ($2.807 \to 2.657 \to 2.631$ over $140/210/281$~ps) where the $2$-fs run's does ($1.591 \to 1.380 \to 0.401$). The $0.10/0.30$ thresholds were themselves calibrated against the $\approx$$2.5$~kcal/mol potential-to-potential gap, which presumes each surface is determined to $\approx$$0.1$; neither is. So the trustworthy bound on the protocol effect remains the paired three-window value, $0.081$~kcal/mol in shape, and we do not claim that the two protocols sample different ensembles---only that the classical surface is not determined finely enough for the question to be settled this way. Deciding it would need both campaigns run until their half-to-half drift falls below $\approx$$0.3$~kcal/mol, for a quantity no conclusion here depends on. What the two protocols do agree on is the claim that matters: spans of $1.506$ and $1.224$ against UBio-MolFM's $4.973$ on the same window, so the classical landscape is $3$--$4\times$ flatter under either. The switch moves four quantities, and we state all four because they do not all point the same way. Three make the classical surface look \emph{worse}, so reporting the $2$-fs numbers is the conservative choice on each: the span narrows from $1.51$ to $1.22$~kcal/mol (both far below the quantum surface's $5.51$, which is the same quantity as the $4.973$ above measured over UBio-MolFM's own slightly wider $4.7$--$9.1$~\AA{} window rather than the $4.7$--$9.0$~\AA{} range the classical curve is restricted to); the lowest bin moves from $5.64$ to $9.42$~\AA, a further $3.8$~\AA{} on a surface whose 1D and 2D estimates already disagreed by $2.6$~\AA, confirming that the minimum position carries no information; and the residual $\text{HB}_1$ asymmetry vanishes entirely, unlatch/relatch becoming $0.52/0.50$ (ratio $1.0$) against $0.70/1.20$ (ratio $1.7$) at $2$~fs and $0.43/3.73$ (ratio $8.6$) under UBio-MolFM. The fourth goes the other way: the closed basin deepens from $1.04$ to $1.32$~kcal/mol, narrowing its gap to our $3.54$, so on that one number the $2$-fs protocol is the more favourable of the two rather than the less. We report the constrained $2$-fs campaign throughout because it is standard classical practice, not because it is uniformly conservative.

\paragraph{What neither control affects.}
Neither protocol enters the potential-energy-surface comparison of Extended Data Table~\ref{tab:csa_qm}, which is evaluated as single points on fixed geometries with no integrator, no constraints and no solvent. One difference between the two potentials is irreducible and is not a protocol choice: TIP3P is rigid under both classical protocols because it is parameterized that way, whereas every water molecule in the UBio-MolFM simulations is flexible.

\subsubsection*{S4.4 KcsA transmembrane channel}
\label{si:kcsa_observables}

\paragraph{Where this result sits in the knock-on debate.}
The Discussion sets this result against the two readings of the conducting filter, and states that the closest experimental probe to the state simulated here---two-dimensional infrared spectroscopy of site-labelled KcsA---does not agree with us~\cite{kratochvil2016,ryan2023}. Two qualifications given there in one clause each are worth expanding. First, the assignment is indirect: measured spectra are matched against spectra computed on classical molecular-dynamics configurations, and the direct-knock-on authors have shown that configurations carrying three or more ions with water in S1 reproduce them as well~\cite{kopec2018}, so the experiment discriminates only among the candidate configurations that were simulated for it. Second, which mechanism a simulation produces has since been traced in part to the force field rather than to the physics: under one choice of water model and ion parameters hard and soft knock-on coexist and interconvert reversibly, whereas TIP3P yields no soft-knock-on events at all~\cite{bosio2026}. That is at once the reason a comparison across potentials is informative and the reason no single potential's answer can be treated as an arbiter.

\paragraph{Seed 126: what bounds the membrane drift as its cause, and what does not.}
One UBio-MolFM replica leaves the contiguous-contact state at $700$~ps (\S\ref{subsec:transmembrane_channel}). The barostat-driven membrane drift is the tempting explanation. We can bound it in one direction and cannot exclude it in the other, and we state both.

\emph{What bounds it.} When the column opened at $700$~ps, seed 126 carried $\approx$$17$ hydrocarbon-core waters. Seed 42 sustains that level for $206$~ps and reaches the same maximum seed 126 ever does, $37$ waters, while holding its contact column in $83.5\%$ of those frames. The hydration seed 126 carried when it failed is therefore not \emph{sufficient} to break a column: a column survives that level, and a higher one, for $200$~ps. It is not a \emph{necessary} condition either, but neither is it excluded as a contributing one, and the timing does not help us here---seed 126 first reached $17$ waters at $684$~ps, essentially when its column opened, whereas seed 42 first reached it at $784$~ps, so the two replicas cannot be ordered by exposure before the event. Relatedly, seed 126's own maximum of $37$ waters is reached at $990$~ps, $290$~ps \emph{after} its column had already gone, so a correlation computed on end-of-trajectory hydration (final-frame core water, $r=-0.71$) reverses the causal order and should not be used.

\emph{What does not.} Over the matched $0$--$700$~ps window seed 126 was nonetheless the most hydrated of the five (peak $20$ waters against $16$, $12$, $11$ and $11$), and across the five replicas that pre-event maximum correlates negatively with the late contact fraction (Pearson $r=-0.89$, $n=5$, two-sided $p=0.046$). We report that correlation with its weakness attached, because the weakness is the point: it is carried by seed 126 itself. Dropping that replica takes $r$ to $-0.51$ ($p=0.49$), dropping any other leaves it between $-0.86$ and $-0.98$, and the rank-based Spearman coefficient over all five does not reach significance ($\rho=-0.71$, $p=0.18$). Using it to explain seed 126 would therefore be circular. A \emph{contributing} role for the coupling is not excluded---only a sufficient one. Nor is it the only candidate: this replica's backbone RMSD is also the highest of the five, and it crosses $1.5$~\AA{} at $640$~ps, some $60$~ps before the column opens at $696$~ps, reaching $1.68$~\AA{} by $962$~ps. The other four never cross that line. We can no more order scaffold drift against the event than membrane drift---one replica cannot---but it means two collective coordinates, not one, are moving ahead of the departure, and neither is excluded. Five replicas cannot separate a contributing artefact from a rare fluctuation, and this is a second reason, beyond rarity, why we do not interpret the event. Settling it needs the semi-isotropic replicas carried to $\approx$$700$~ps, which we have not done. The classical sets offer no comparison either way, because the contact column is absent from the first frame of all ten and never appears in $1.0$~ns, so no classical replica has an analogous state to lose.

\paragraph{Electrostatic truncation, and the three ways to misread this test.}
The control defined in Methods asks whether the $8$~\AA{} cutoff is what closes our contacts. It does not:
fourteen of eighteen trajectories are net compressive at their stiffest pair, twelve of them inside the
linear-response range, and among those every UBio-MolFM replica that holds its contact column under either
barostat (Supplementary Table~\ref{tab:lr_electrostatics}). Three
guardrails matter more than the numbers.

First, the \emph{single-ion} quantities carry no information about spacing and must not be quoted. The
mean missing long-range force per filter ion is $0.44$~eV/\AA, which sounds large, but a long-range field
is nearly common-mode across ions on the pore axis: it translates the column bodily and leaves every
spacing unchanged. Only the adjacent-pair differential survives that cancellation, which is why the
verdict is read from it alone.

Second, the ratio of measured to required differential is built from absolute values and is therefore
\emph{sign-blind}. A compressive pair---one for which the objection already fails---can still report a
ratio near $3$. The 12-6 seed-2 run does exactly that: ratio $2.96$ with a signed mean of $-0.043$~eV/\AA.
The ratio must always be read beside the sign, never instead of it.

Third, linear response has a range. Where the implied spacing shift exceeds $0.25$~\AA{} the target pair is
too soft for the harmonic estimate and no quantitative statement is made; six runs fall there. One is
UBio-MolFM seed 126, whose target pair is eight to ten times softer than the others precisely because the
vacancy it develops after $700$~ps inflates the spacing variance. Its nominal $+5.03$~\AA{} is the formula
used outside its domain, not a physical prediction, and is quoted nowhere.

The test is also one-directional by construction. A compressive result excludes truncated electrostatics
as the cause of the tight column; nothing here can establish that the spacings are right, which requires an
\emph{ab initio} or QM/MM reference. Nor can a K$^+$--K$^+$ distance scan substitute: $3.25$ and
$6.5$~\AA{} both lie inside $8$~\AA, and an isolated dimer lacks the environment the objection is
about---that scan answers a different question, pair-potential accuracy against DFT.

The verdict is also pair-specific, and one pair runs against us. It is the contact pair S4--S3---the pair
that sets the closest K$^+$--K$^+$ distance---that is compressive and that the control therefore protects.
At the cavity pair S$_\text{cav}$--S4 the differential has the opposite sign, strongly and consistently:
$+0.68$ to $+0.93$~eV/\AA{} in all five replicas, $9\sigma$ to $21\sigma$ from zero and expansive in
$90$--$100\%$ of frames. Truncation therefore plausibly does hold that pair tighter than
full electrostatics would---the cavity ion is the one most exposed to the far field. No magnitude follows:
the implied shifts, $+1.3$ and $+1.6$~\AA, are far outside the harmonic range, which is why we make no
quantitative claim about this spacing and quote none. What follows is a caveat with a definite sign, and it
reaches one published statistic. Because a contiguous column requires S$_\text{cav}$--S4 below $4$~\AA{} and
ours sits at $3.55$--$3.61$~\AA{} in the four replicas that hold it, only $0.4$~\AA{} inside that threshold, the contiguity \emph{fraction} is the one
filter number this control does not defend; the dehydration, the occupancy and the closest-contact distance
are unaffected. The third pair, S3--S2, is equally unboundable ($-1.0$ to $-1.6$~\AA{} implied) and, across the
five isotropic replicas, compressive; it is expansive in one released-box replica. Nothing is
rested on it either way.

\paragraph{Observable definitions.}
Every metric below is computed by one pipeline with identical definitions and cut-offs for all three potentials. \emph{(i)~Structural integrity}: protein backbone RMSD over N, C$\alpha$, C and O; C$\alpha$-only RMSD; and filter-backbone RMSD, each against the equilibrated starting structure after Kabsch superposition. \emph{(ii)~Bilayer area-per-lipid} by Voronoi tessellation of the leaflet phosphorus atoms. \emph{(iii)~Deuterium order parameter} $S_{CD}$ per carbon along the sn-1 chain. \emph{(iv)~On-axis filter occupancy}: the number of the five pore K$^+$ lying within $r_{xy}<3$\,\AA{} of the pore axis. \emph{(v)~Ion-column geometry}: the closest K$^+$--K$^+$ distance and the three adjacent axial spacings (S$_\text{cav}$--S4, S4--S3, S3--S2); a \emph{contiguous direct-contact column} is a frame in which all three spacings fall below $4$\,\AA{} \emph{simultaneously}, so the statistic reports a connected column rather than any single close pair. \emph{(vi)~Filter-core hydration}: the count of water oxygens lying within the ion column. \emph{(vii)~Filter-carbonyl tilt} $\cos\theta_z$: the cosine between each backbone C=O of Thr75--Gly79 and the pore axis; the ``half-cage'' signature, in which Gly79 alone points out of the lumen, reports the absence of a top cap on site S1.

The identity of the S2 cage is read off the measured contacts of the cutaway of Fig.~\ref{fig:transmembrane_channel}a: Y78 at $2.71$/$2.72$~\AA{} and G77 at $2.83$/$2.92$~\AA, with the next nearest backbone oxygen (V76) $5.14$~\AA{} away. The cage is therefore unambiguously Y78$+$G77; the full tetramer contributes four of each and four are in view in that panel.

The K$^+$--carbonyl coordination number counts filter carbonyl oxygens within $3.5$~\AA{} of the site's ion; its deep-site value, the one behind ``close to eightfold'' in the main text, is tabulated in Extended Data Table~\ref{tab:validation_summary}, and is $7.97$ at both S4 and S3 under UBio-MolFM against $7.90$/$7.27$ and $7.93$/$7.30$ under the two classical models. Occupancy, hydration and spacing statistics are pooled over the late half of each trajectory. Backbone RMSD is reported for every replica of all three potentials (UBio-MolFM, 5; 12-6-4, 5; plain 12-6, 5).

\clearpage
\subsection*{S5. Single-GPU Performance}
\label{si:throughput_protocol}

Hardware efficiency is what makes the mesoscale simulations of the main text possible on a single GPU; we characterize it here rather than in the main text. What the backbone buys \emph{on this work's systems} is a $4$--$5\times$ throughput gain over dense equivariant baselines at standard runtime, and that is what makes the $10^5$-atom trajectories reported here feasible on one GPU. Two ratios appear in this paper and they are not the same comparison. Activation recomputation \emph{costs} throughput, as it must---it lowers ours from a $36$--$37.7$ to a $25$--$26.9\times10^3$~atoms\,s$^{-1}$ plateau, $28\%$---but it is the setting under which the largest systems fit at all, and compared like for like there ($25.1$ against UMA-S-1p2's $3.0$ at $120{,}000$ atoms) the margin is $8.3\times$. The $4$--$5\times$ annotation in Extended Data Fig.~\ref{fig:throughput}a is the standard-runtime comparison against the strongest baseline at the same setting---$4.4\times$ DPA-4 and $5.1\times$ MACE-OMol at $12{,}000$ atoms, rising to $4.5\times$ and $6.7\times$ at $15{,}000$; the $8.3\times$ quoted in Methods is the recompute-versus-recompute comparison at $10^5$ atoms. Both are end-to-end measurements on our own workloads rather than component-level figures carried over from the backbone papers.

\paragraph{Timing protocol.}
Box sizes advance in 3{,}000-atom increments. Throughput is reported as atoms\,s$^{-1}$, latency as ms/step, and peak memory as the maximum allocated device memory over the measured steps. Each plotted point is the median over $6$ consecutive timed evaluations taken after $4$ warm-up evaluations that are discarded (they absorb compilation, CUDA-graph capture and kernel autotuning), i.e.\ $10$ evaluations per system size, with CUDA synchronization immediately before and after every timed step so that asynchronous kernel launches cannot be counted as completed work. A size is recorded as out-of-memory when the run exceeds the $141$~GB device budget; no gradient checkpointing beyond the model's own activation-recompute path is applied. Our model is profiled with its weights frozen (\texttt{requires\_grad=False}), so the autograd graph that yields $\mathbf{F}=-\nabla_{\mathbf{R}}E$ is built with respect to positions alone; this is the setting production MD runs in, and both sweeps---standard runtime and activation recompute---are measured under it, so the two are directly comparable. The baselines are profiled from released weights under their authors' recommended inference settings, which already do the equivalent, and their timings are unchanged within run-to-run noise. For the record, the freeze is worth $\approx$$25\%$ of throughput and roughly a third of peak memory to our model, and nothing measurable to the baselines.

\paragraph{Points measured but not plotted.}
The axis of Extended Data Fig.~\ref{fig:throughput} is truncated at $72{,}000$ atoms for legibility. Under activation recompute both UBio-MolFM and UMA-S-1p2 were profiled on the same GPU out to $120{,}000$ atoms, the largest size run: UBio-MolFM sustains $25{,}100$~atoms\,s$^{-1}$ at an $87.8$~GB peak, UMA-S-1p2 $3{,}010$~atoms\,s$^{-1}$ at $142.2$~GB---an $8.3\times$ throughput ratio at equal size, with UMA-S-1p2 already at the memory ceiling while UBio-MolFM still has a third of the device free. These points are reported rather than plotted because including them would compress the region where the baselines actually diverge. In standard mode the sweep ends at $69{,}000$ atoms because that is the largest box built, not because of a failure: throughput there is still $36{,}200$~atoms\,s$^{-1}$ at a $136.6$~GB peak, with no sign of the throughput collapse that marks memory oversubscription. The standard-mode memory slope of $1.97$~GB per $1{,}000$ atoms puts the extrapolated ceiling at $\approx$$71{,}000$ atoms, so $69{,}000$ is within one size step of the device limit in any case; past that point activation recompute, at $0.73$~GB per $1{,}000$ atoms, is the mode that reaches $10^5$---on this slope the $141$~GB budget would not bind until $\approx$$1.9\times10^5$ atoms---and it is the mode the KcsA production runs of \S\ref{subsec:transmembrane_channel} use. Interpolating the recompute sweep to that system's $108{,}964$ atoms gives a $79.8$~GB peak at $4.2$~s per force evaluation.

\clearpage
\subsection*{S6. Reproducibility Inventory}

The \textbf{UBio-MolFM-MD} release (see Data availability) covers every molecular-dynamics scenario reported in this work: liquid water, NaCl and KCl solvation, Cyclosporine~A hydrogen bonding in water and in vacuum, the CsA steered-MD and umbrella-sampling/MBAR free-energy calculations, the RNA 1L2X Mg$^{2+}$ system, and the 108{,}964-atom KcsA membrane channel together with its matched 12-6-4 and 12-6 classical references. For each scenario it contains (i) starting structures and equilibration inputs; (ii) ASE and OpenMM simulation scripts with the random seeds actually used; (iii) stride-decimated trajectories carrying positions and forces---$1$\,ps for the $1$-ns water and electrolyte boxes, with the $NPT$ water run additionally deposited at its full $50$\,fs sampling because the self-diffusion coefficient is read off that resolution, and $10$\,ps for the RNA and KcsA systems---sufficient to recompute every observable reported from a trajectory. The Cyclosporine~A umbrella campaigns are the one exception, and deliberately so: they ship as per-frame collective-variable time series together with the starting structure of each of the $80$ windows, because MBAR consumes the CV histories rather than the coordinates, so a decimated trajectory would be simultaneously larger and insufficient; (iv) the analysis scripts themselves (radial distribution functions and coordination numbers, self-diffusion from the mean-square displacement with the Yeh--Hummer finite-size correction, hydrogen-bond occupancies, MBAR potentials of mean force, backbone RMSD, sugar pucker, selectivity-filter occupancy, K$^+$--carbonyl coordination and carbonyl-plane geometry); and (v) the final observable values as machine-readable CSV, which are the same files the figures in this paper are plotted from. Per-system random-seed assignments are additionally tabulated in \texttt{REPRODUCIBILITY\_}\allowbreak\texttt{CHECKLIST.md}~\S D of the reproducibility deposit.

\clearpage
\subsection*{S7. Supplementary Tables}

\begin{table}[H]
  \centering
  \caption{\textbf{Checkpoint lineage and core training settings.} One column per curriculum stage, matching Fig.~\ref{fig:framework}b. S1 and S2 are both OMol25 runs and differ in how forces are produced: S1 carries a separately parameterized force head, which S2 retires in favour of automatic differentiation of the predicted energy. S3 is the only stage that sees UBio-Mol26. Fig.~\ref{fig:framework}b accordingly labels both OMol25 stages \emph{OMol25, 140M labels}. Values in the loss-weight row are the default energy/force coefficients, which the \texttt{omol25} branch and the auxiliary \texttt{svp} head use directly; the force-filtered \texttt{tzvpd} and \texttt{tzv2p} branches disable the energy term, and the $0.15$ of the \texttt{svp} branch that stays on the \texttt{omol25} head replaces both with a directional hinge, as described in the text. The four numbers in the S3 data row are the branch sampling ratios. ``Model-only'' denotes non-strict weight loading with a reset optimizer and scheduler.}
  \label{tab:hparams}
  \footnotesize
  \setlength{\tabcolsep}{3pt}
  \begin{tabular}{@{}>{\raggedright\arraybackslash}p{0.15\linewidth}>{\centering\arraybackslash}p{0.25\linewidth}>{\centering\arraybackslash}p{0.25\linewidth}>{\centering\arraybackslash}p{0.25\linewidth}@{}}
    \toprule
    \textbf{Setting} & \textbf{S1: pretraining} & \textbf{S2: autograd forces} & \textbf{S3: thermodynamic fusion} \\
    \midrule
    Training data & OMol25, $\approx$140M frames & OMol25, same corpus & Four-branch mixture: \texttt{omol25} $0.15$, \texttt{tzvpd} $1$, \texttt{tzv2p} $2$, \texttt{svp} $0.2$ \\
    Initialization & Random & S1 at 1{,}000k steps, model-only & S2 at 1{,}400k steps, model-only \\
    Energy heads & \texttt{omol25} & \texttt{omol25} & \texttt{omol25} + \texttt{svp} \\
    Force prediction & \textbf{Separate force head} & \textbf{Autograd}, $\mathbf{F}=-\nabla_{\mathbf{R}}E$ & Autograd, $\mathbf{F}=-\nabla_{\mathbf{R}}E$ \\
    Default $(\lambda_E,\lambda_F)$ & $(4,10)$ & $(4,10)$ & $(4,10)$ \\
    Peak LR & $4\!\times\!10^{-4}$ & $2\!\times\!10^{-4}$ & $5\!\times\!10^{-6}$ \\
    Total steps & 1{,}000k (lineage checkpoint of a 1{,}400k cosine horizon) & 1{,}400k (its own run, not cumulative) & 400k \\
    Warm-up & Configured linear warm-up & 8k steps & Configured linear warm-up \\
    Optimizer / precision & AdamW / FP32 & AdamW / FP32 & AdamW / FP32 \\
    Devices / useful cost & 32$\times$H20-141GB / $\sim$102 GPU-days & 64$\times$H20-141GB / $\sim$403 GPU-days & 64$\times$A100-40GB / $\sim$494 GPU-days \\
    \midrule
    \multicolumn{4}{@{}p{\linewidth}@{}}{\textbf{Architecture (shared across runs):} 23{,}505{,}963 total parameters (23{,}503{,}618 trainable), $L_{\max}=2$; 3 short-range equivariant layers ($r=5$~\AA, all atoms) plus 1 layer over the full $0.1$--$8$~\AA{} sphere whose neighbour mask excludes hydrogen while query/centre atoms still include all atoms; node clustering (\texttt{with\_cluster=node}) with zero-order attention (\texttt{attn\_type=zero-order}); on-the-fly equivariant attention via a fused Triton kernel.} \\
    \bottomrule
  \end{tabular}
\end{table}

\begin{table}[H]
  \centering
  \caption{\textbf{Optimization schedule per curriculum stage.} The three quantities most often needed to judge or reproduce a training run, kept apart from Supplementary Table~\ref{tab:hparams} so that that table carries only what defines the objective and the model class. Batch size is expressed as an atom budget rather than a graph count because system sizes span $15$--$1{,}370$ atoms, so graph-count batching would leave the devices badly load-imbalanced (SI~S2). The learning-rate shape is linear warm-up followed by cosine decay in every stage; warm-up length, weight decay, per-layer channel allocations, per-stage device counts and wall-clock are given in the archived configuration files, which are the authoritative record.}
  \label{tab:optim_schedule}
  \footnotesize
  \setlength{\tabcolsep}{6pt}
  \begin{tabular}{lccc}
    \toprule
    \textbf{Setting} & \textbf{Stage 1 (S1)} & \textbf{Stage 2 (S2)} & \textbf{Stage 3 (S3)} \\
    \midrule
    Peak learning rate        & $4\!\times\!10^{-4}$ & $2\!\times\!10^{-4}$ & $5\!\times\!10^{-6}$ \\
    Atom budget \texttt{bs\_atom} & $2{,}048$ atoms & $2{,}048$ atoms & $2{,}048$ atoms \\
    Total steps               & $1{,}000$k (of a $1{,}400$k horizon) & $1{,}400$k (its own run) & $400$k \\
    \bottomrule
  \end{tabular}
\end{table}

\begin{table}[H]
  \centering
  \caption{\textbf{Composition of the \emph{TZVP extreme-size} evaluation tier.} The released evaluation set
  (\S\ref{sec:eval_benchmarks}), enumerated from the archived LMDB shards. Every shard holds one system carried
  through one trajectory, so ``systems'' equals trajectories; atom count is constant within a trajectory. All
  systems are neutral, closed-shell explicitly solvated clusters, which is why H and O dominate the atom
  inventory. Elements are listed in decreasing abundance, with counts over one frame per system; the ions appear
  in single-digit numbers per category and are marked $^{\dagger}$. No configuration from this tier enters
  training at any stage.}
  \label{tab:extreme_tier}
  \footnotesize
  \setlength{\tabcolsep}{5pt}
  \resizebox{\textwidth}{!}{%
  \begin{tabular}{llccccl}
    \toprule
    \textbf{Category} & \textbf{Shard} & \textbf{Systems} & \textbf{Frames} & \textbf{Frames/system} & \textbf{Atoms} & \textbf{Elements} \\
    \midrule
    Protein, relaxation & \texttt{AFDB-1500}    & 10 & 1{,}010 & 101      & 1{,}501--1{,}555 & H, O, C, N, Cl$^{\dagger}$, S, Na$^{\dagger}$, Br$^{\dagger}$, Ca$^{\dagger}$, K$^{\dagger}$ \\
    Protein, MD         & \texttt{AFDB-1500-MD} & 18 & 875     & 35--51   & 1{,}408--1{,}450 & H, O, C, N, Na$^{\dagger}$, S, Cl$^{\dagger}$ \\
    DNA (10-mer)        & \texttt{DNA-10}       & 5  & 226     & 36--56   & 1{,}258--1{,}329 & H, O, C, N, Na$^{\dagger}$, P \\
    RNA (15-mer)        & \texttt{RNA-15}       & 5  & 505     & 101      & 1{,}413--1{,}527 & H, O, C, N, Na$^{\dagger}$, P \\
    Lipid               & \texttt{lipids}       & 5  & 501     & 100--101 & 1{,}215--1{,}282 & H, O, C, N, P, Cl$^{\dagger}$, Br$^{\dagger}$, Na$^{\dagger}$, Mg$^{\dagger}$, K$^{\dagger}$, Zn$^{\dagger}$ \\
    \midrule
    \textbf{Total}      &                       & \textbf{43} & \textbf{3{,}117} & --- & \textbf{1{,}215--1{,}555} & \\
    \bottomrule
  \end{tabular}}
\end{table}

\begin{table}[H]
  \centering
  \caption{\textbf{Equilibrium mass density from unbiased $NPT$ simulations.} Densities (g/cm$^3$) at 300~K, 1~bar over the equilibrated second half of each trajectory; deviation from experiment in parentheses (smaller is better). Baselines use the authors' recommended stress-based barostat and pure water only (``--''); UBio-MolFM uses a Monte-Carlo barostat with conservative forces. The pure-water column is the same measurement plotted in Extended Data Fig.~\ref{fig:density_baselines_si}.}
  \label{tab:density_benchmark}
  \footnotesize
  \setlength{\tabcolsep}{5pt}
  \resizebox{\textwidth}{!}{%
  \begin{tabular}{lccccc}
    \toprule
    \textbf{System}          & \textbf{Exp.}  & \textbf{UMA-S-1p2}  & \textbf{MACE-OMol}  & \textbf{DPA-4}                & \textbf{UBio-MolFM} \\
    \midrule
    Liquid water (300~K)     & 0.997          & 1.110 ($+11.3\%$)   & 1.089 ($+9.2\%$)    & 0.62 ($-38\%$, collapse)      & \textbf{0.987 ($-1.0\%$)} \\
    0.15\,mol/L NaCl(aq)     & $\approx$1.004 & --                  & --                  & --                            & \textbf{0.995 ($-0.9\%$)} \\
    0.15\,mol/L KCl(aq)      & $\approx$1.005 & --                  & --                  & --                            & \textbf{0.996 ($-0.9\%$)} \\
    \bottomrule
  \end{tabular}%
  }
\end{table}

\begin{table}[H]
  \centering
  \caption{\textbf{Complete supplementary-parameter file (\texttt{csa.frcmod}) for the Cyclosporine~A classical baseline}, as written by AmberTools \texttt{parmchk2} and used verbatim in every classical run of Fig.~\ref{fig:csa_pmf}c,d. The first line is \texttt{parmchk2}'s own default remark, reproduced as it stands rather than edited out, so that the block is byte-faithful to the file the runs loaded. The \texttt{MASS}, \texttt{BOND}, \texttt{ANGLE} and \texttt{NONBON} sections are empty: GAFF2 covered every mass, bond, angle and van der Waals term, and \emph{all} missing parameters were torsional. The ten proper dihedrals are centred without exception on \texttt{ns} (amide N carrying one H, assigned to the four residues with a free backbone N--H) and analogized from \texttt{n} (tertiary amide N) at penalty $0.0$; \texttt{n}$\to$\texttt{ns} differs only in H versus CH$_3$ on a planar amide nitrogen, so a zero penalty is credible. Of the five impropers, three take GAFF2's generic value, one is a general term at penalty $3.0$, and one---\texttt{c2-c3-c2-ha} at penalty $47.1$---substitutes an aromatic C--H out-of-plane term for an olefinic one on the MeBmt1 butenyl side chain, remote from both collective variables. Two of the four \texttt{ns} nitrogens are the $\text{HB}_1$ and $\text{HB}_2$ donors, so the torsions around the atoms that carry the latch coordinate are among the borrowed ones; \S\ref{subsec:csa_thermodynamics} and Methods state what this does and does not license us to conclude.}
  \label{tab:csa_frcmod}
  \tiny
\begin{verbatim}
Remark line goes here
MASS

BOND

ANGLE

DIHE
c3-c2-ns-c3   4    2.600       180.000           2.000      same as X -c2-n -X , penalty score=  0.0
o -c2-ns-c3   4    2.600       180.000           2.000      same as X -c2-n -X , penalty score=  0.0
c3-c3-ns-hn   6    0.000         0.000           2.000      same as X -c3-n -X , penalty score=  0.0
c2-c3-ns-hn   6    0.000         0.000           2.000      same as X -c3-n -X , penalty score=  0.0
h1-c3-ns-hn   6    0.000         0.000           2.000      same as X -c3-n -X , penalty score=  0.0
c3-c2-ns-hn   4    2.600       180.000           2.000      same as X -c2-n -X , penalty score=  0.0
o -c2-ns-hn   4    2.600       180.000           2.000      same as X -c2-n -X , penalty score=  0.0
c2-c3-ns-c2   6    0.000         0.000           2.000      same as X -c3-n -X , penalty score=  0.0
c3-c3-ns-c2   6    0.000         0.000           2.000      same as X -c3-n -X , penalty score=  0.0
h1-c3-ns-c2   6    0.000         0.000           2.000      same as X -c3-n -X , penalty score=  0.0

IMPROPER
c2-c3-ns-hn         1.1          180.0         2.0          Using the default value
c3-n -c2-o          1.1          180.0         2.0          Using the default value
c2-c3-n -c3         1.1          180.0         2.0          Using general improper torsional angle  X-c3- n-c3, penalty score=  3.0)
c3-ns-c2-o          1.1          180.0         2.0          Using the default value
c2-c3-c2-ha         1.1          180.0         2.0          Same as X -X -ca-ha, penalty score= 47.1 (use general term))

NONBON
\end{verbatim}
\end{table}

\begin{table}[H]
  \centering
  \caption{\textbf{Electrostatic-truncation control, per trajectory.} Adjacent-pair differential of
  $\Delta\mathbf{F}_\text{LR}=\mathbf{F}_\text{PME}-\mathbf{F}_{8\,\text{\AA}}$ at each trajectory's
  \emph{stiffest} axial K$^+$ pair, the only one to which linear response is applied; $30$ frames spanning
  $50$--$1000$~ps per run. Positive is expansive---restoring the missing term would widen that pair, so
  truncation had been compressing it; negative is compressive, which reverses the objection. The ratio of
  measured to required differential is \emph{sign-blind}: read it beside the sign, never instead of it.
  \textsc{nb} marks runs whose implied shift exceeds the $0.25$~\AA{} harmonic-validity limit; no
  quantitative claim is made for those and none of their shifts is quoted. Fourteen of the eighteen runs are
  compressive---twelve of them inside the harmonic range---including every UBio-MolFM replica that
  holds its contact column under either barostat.}
  \label{tab:lr_electrostatics}
  \footnotesize
  \begin{tabular}{llrrrl}
    \toprule
    \textbf{Trajectory} & \textbf{Pair} & \textbf{Signed mean} & \textbf{Implied shift} & \textbf{Ratio} & \textbf{Verdict} \\
     & & (eV/\AA) & (\AA) & & \\
    \midrule
    UBio-MolFM seed 42  & pair 1 & $-0.158$ & $-0.123$ & $0.58$ & compressive \\
    UBio-MolFM seed 84  & pair 1 & $-0.235$ & $-0.211$ & $0.70$ & compressive \\
    UBio-MolFM seed 168 & pair 1 & $-0.196$ & $-0.166$ & $0.60$ & compressive \\
    UBio-MolFM seed 210 & pair 1 & $-0.249$ & $-0.227$ & $0.86$ & compressive \\
    UBio-MolFM seed 126 & pair 0 & $+0.678$ & --- & $14.54$ & \textsc{nb} (vacancy-softened) \\
    \addlinespace
    released seed 124 & pair 2 & $+0.243$ & --- & $1.06$ & \textsc{nb} \\
    released seed 168 & pair 1 & $-0.170$ & $-0.151$ & $0.77$ & compressive \\
    released seed 210 & pair 1 & $-0.259$ & $-0.167$ & $0.63$ & compressive \\
    \addlinespace
    12-6 seed 1 & pair 1 & $-0.059$ & $-0.062$ & $0.53$ & compressive \\
    12-6 seed 2 & pair 1 & $-0.043$ & $-0.185$ & $2.96$ & compressive (sign-blind ratio) \\
    12-6 seed 3 & pair 1 & $-0.056$ & $-0.051$ & $0.58$ & compressive \\
    12-6 seed 4 & pair 1 & $-0.136$ & --- & $1.78$ & \textsc{nb} \\
    12-6 seed 5 & pair 1 & $-0.126$ & $-0.241$ & $1.20$ & compressive \\
    \addlinespace
    12-6-4 seed 1 & pair 1 & $+0.107$ & --- & $6.96$ & \textsc{nb} \\
    12-6-4 seed 2 & pair 1 & $-0.051$ & $-0.060$ & $0.64$ & compressive \\
    12-6-4 seed 3 & pair 1 & $-0.074$ & $-0.081$ & $0.70$ & compressive \\
    12-6-4 seed 4 & pair 1 & $-0.138$ & --- & $1.18$ & \textsc{nb} \\
    12-6-4 seed 5 & pair 1 & $+0.045$ & --- & $4.17$ & \textsc{nb} \\
    \bottomrule
  \end{tabular}
\end{table}

\end{document}